\documentclass[useAMS,usenatbib, usegraphicx]{mn2e}
\usepackage{amssymb} 

\newcommand{\rmd}{{\rm{d}}}
\newcommand{\rme}{{\rm{e}}}
\newcommand{\rmi}{{\rm{i}}}

\newcommand{\qe}{q_{\rm e}}

\newcommand{\beq}{\begin{equation}}
\newcommand{\eeq}{\end{equation}}
\newcommand{\barr}{\begin{eqnarray}}
\newcommand{\earr}{\end{eqnarray}}

\newcommand{\cm}{{\rm{cm}}}
\newcommand{\K}{{\rm{K}}}
\newcommand{\gr}{{\rm{gr}}}
\newcommand{\mH}{m_{\rm{H}}}
\newcommand{\nH}{n_{\rm{H}}}
\newcommand{\Am}{{\rm{\AA}}}
\newcommand{\erf}{{\,\rm{erf}\,}}
\newcommand{\C}{{\rm{C}}}
\newcommand{\g}{{\rm{g}}}
\newcommand{\debye}{{\rm{debye}}}
\newcommand{\eV}{{\rm{eV}}}

\newcommand{\ed}{{\rm{ed}}}
\newcommand{\mnras}{MNRAS}
\newcommand{\apj}{ApJ}

\newcommand{\apjl}{ApJL}
\newcommand{\aap}{AAP}

\begin{document}

\title[A refined model for spinning dust radiation]{A refined model for spinning dust radiation}
\author[Ali-Ha\"{\i}moud, Hirata \& Dickinson]{
       Yacine Ali-Ha\"{\i}moud$^{1}$\thanks{yacine@tapir.caltech.edu},
       Christopher M. Hirata$^{1}$\thanks{chirata@tapir.caltech.edu}, and
       Clive Dickinson$^{2}$\thanks{cdickins@ipac.caltech.edu}\\
$^{1}$California Institute of Technology, Mail Code 130-33, Pasadena, CA 91125, USA \\
$^{2}$Infrared Processing and Analysis Center, California Institute of Technology, M/S 220-6, Pasadena, CA 91125,
USA}

\date{Accepted 2009 February 4. Received 2009 February 4; in original form 2008 December 26}

\pagerange{\pageref{firstpage}--\pageref{lastpage}} \pubyear{2009}

\maketitle

\label{firstpage}

\begin{abstract}
We present a comprehensive treatment of the spectrum of electric dipole emission from spinning dust grains, updating the commonly used model of Draine \& Lazarian.  Grain angular velocity distributions 
are computed using the Fokker-Planck equation; we revisit the drift and diffusion coefficients for the major torques on the grain, including collisions, grain-plasma interactions, and infrared emission. 
We use updated grain optical properties and size distributions. The theoretical formalism is implemented in the companion code, {\sc SpDust}, which is publicly available. The effect of some environmental 
and grain parameters on the emissivity is shown and analysed.
\end{abstract}


\section{Introduction}

Observational cosmology has 
entered an area of high precision, exemplified by the most recent temperature results from sensitive cosmic microwave background (CMB) experiments \citep{Dickinson04,Readhead04,Kuo07,Hinshaw08}. However, foreground separation and 
removal remains a major challenge for any CMB measurement (e.g. \citealt{Eriksen08,Leach08}). In addition to 
the standard Galactic foregrounds, free-free, synchrotron and thermal dust 
emission, an unknown ``anomalous'' dust-correlated emission has been 
observed over the last decade, in the microwave region of the spectrum. 
The anomalous emissions was first interpreted as free-free emission from 
shock-heated gas by \citet{Leitch}, but \citet{DL98a} showed that this 
would require an extremely high plasma temperature and a corresponding 
unrealistic energy injection rate. They proposed instead two possible 
mechanisms to explain the anomalous microwave emission. One of them is the 
magnetic dipole emission from thermal fluctuations in the magnetisation of 
interstellar dust grains \citep{DL99}. The other possible mechanism, on 
which the present work focuses, is electric dipole radiation from the 
smallest carbonaceous grains, described in \citet{DL98b}, hereafter DL98b. 
The physical principle is quite straightforward: dust grains are 
presumably asymmetric, and thus will have a nonzero electric dipole 
moment.  These grains will spin due to interaction with the ambient 
interstellar medium (ISM) and radiation field, and thus radiate 
electromagnetic waves due to the rotation of their electric dipole moment. 
To get the electric dipole radiation spectrum, one thus needs three 
ingredients: the quantity of small grains, then their dipole moment, and 
finally their rotation rates.

Although the observational interest in electric dipole radiation from spinning dust
grains has only grown in the last decade, there is a long standing history of theoretical work
on the subject. \citet{Erickson} was the first to consider the possibility that rotating dust grains
could be the source of non-thermal radio-noise. \citet{hoyle} showed that this process was dominated by grains
with radius $a \lesssim 10^{-6} \ \cm$ and could lead to radio emission around $10$ GHz. \citet{Ferrara} estimated 
the spinning dust emissivity for thermally rotating grains. The first to provide a detailed treatment of rotational excitation
of small grains were \citet{Rouan}. They considered the effect of collisions with gas atoms and absorption and emission of radiation. \citet{AW93} evaluated the effect of collisions with ions and ``plasma drag'' (torques due to the electric field of passing ions).

DL98b provided the first comprehensive study of the rotational dynamics of small grains, including all the previous effects. They 
evaluated, as a function of grain radius and environmental conditions, rotational damping and excitation rates through collisions, 
``plasma drag'', 
infrared emission, emission of electric dipole radiation, photoelectric 
emission and formation of H$_2$ molecules.  The spectra they provided are 
now widely used in interpreting ISM microwave emission (e.g. \citealt{Finkbeiner04,Watson05,Casassus06,Casassus07,Casassus08,Dickinson07,Dickinson08,Dobler08}) and for 
CMB foreground analyses (e.g. \citealt{Banday03,Davies06,Bonaldi07,Hildebrandt07,Gold}).  Given that the DL98b 
models are now a decade old, and the recent surge in interest in anomalous 
emission, it is timely to revisit the theory of spinning dust emission, 
including the approximations made in DL98b.  This is the purpose of this 
paper.

As in DL98b, we concentrate on the rotation rate of the grains; the size 
distribution has been reconsidered by other authors, and the grain dipole 
moment distribution should be regarded as a model parameter since one 
cannot compute it from first principles. We first review and generalize 
DL98b rotational excitation and damping rates. We modify the rotational 
excitation and damping rates by collisions with neutral species, such that 
it respects detailed balance in the case where the evaporation 
temperature is equal to the gas temperature. We include the electric 
dipole potential when evaluating the effect of collisions with ions. Full 
hyperbolic trajectories and rotating grains are used when computing the 
effect of plasma drag. We correct the infrared emission damping rate which 
was underestimated for a given infrared spectrum. Finally, we use these 
excitation and damping rates to calculate the grain rotational 
distribution function by solving the Fokker Planck equation. Updated grain 
optical properties and size distribution are used throughout this 
analysis. An Interactive Data Language (IDL) code implementing the formulas in this paper, 
{\sc SpDust}, is available on the 
web\footnote{http://www.tapir.caltech.edu/$\sim$yacine/spdust/spdust.html}, and will 
hopefully allow for a more thorough exploration of the parameter space, 
as well as model fitting to observations.

The paper is organized as follows. In Section \ref{section : electric 
dipole} we remind the reader of the electric dipole radiation formula and 
give the resulting expected emissivity. In Section \ref{grain prop} we 
discuss the size distribution and dipole moments, along with other grain 
properties. We then turn to the main thrust of this study, which is the 
computation of the angular velocity distribution function. The theoretical 
formalism is exposed in Section \ref{Fokker}, which presents the 
Fokker-Planck equation. Sections \ref{collisions}--\ref{H2} discuss 
the various rotational damping and excitation processes : collisions with ions 
and neutral species, plasma drag, infrared emission, photoelectric 
emission, and random H$_2$ formation. The reader interested primarily in 
the predicted emission may wish to proceed directly to Section 
\ref{emissivity}, where we present the resulting emissivity and the effect 
of various parameters and environment conditions. Our conclusions are 
given in Section \ref{conclusion}. Appendix A exposes the techniques used to numerically evaluate 
integrals of rapidly oscillating functions involved in the plasma drag calculation. Appendix B presents an alternate, 
quantum mechanical derivation of the rotational damping rate through infrared emission.


\section{Electric dipole radiation} \label{section : electric dipole}

The power radiated by a dust grain spinning with an angular velocity $\bomega$, of electric dipole moment $\bmu$, with component $\bmu_{\bot}$ 
perpendicular to $\bomega$, is
\beq
P = \frac{2}{3}\ \frac{\mu_{\bot}^2\  \omega^4}{c^3}.
\eeq
This power is emitted at the frequency $\nu = \omega / 2\pi$.

To get the emissivity of electric dipole radiation per H atom, in erg$\,$s$^{-1}\,$sr$^{-1}\,$(H$\,$atom)$^{-1}$, one needs several ingredients:
\begin{itemize}
\item The grain size distribution function: $\nH^{-1}\rmd n_{\gr}/\rmd a$, which gives the number of dust grains per unit size per H atom.
\item The electric dipole moments as a function of grain size $a$: $\mu(a)$.
\item The angular velocity distribution function, $f_a(\omega)$, which depends upon the grain radius and environmental conditions. It depends on the 
angular velocity modulus only in a perfectly isotropic environment, with no strong electromagnetic fields forcing the dipole moments to align in some 
particular direction.
\end{itemize}
One then readily gets the emissivity of spinning dust grains per H atom:
\beq
\frac{j_{\nu}}{n_H} = \frac{1}{4\pi}\ \int_{a_{\rmn{min}}}^{a_{\rmn{max}}} \rmd a\ \frac{1}{\nH}
\frac{\rmd n_{\gr}}{\rmd a}\ 4\pi \omega^2 f_a(\omega) \  2\pi \ \frac{2}{3}\frac{\mu_{a \bot}^2
\omega^4}{c^3},
\eeq
where $\omega = 2 \pi \nu$.

\section{Dust grains properties} \label{grain prop}

\subsection{Grain shapes}

The grains are characterized by their volume-equivalent radius $a$, such that the grain volume is $4 \pi a^3/3$. The radius $a$ is in fact a measure 
of the number of C atoms in the grain, which we assume to be
\beq
N_{\C} = \frac{4 \pi a^3 \rho_{\C}}{3 m_{\C}} \approx 468 \ a_{-7}^3  \label{N_C}
\eeq
where $\rho_{\C} = 2.24\g\ \cm^{-3}$ is the density of ideal graphite and $a_{-7} \equiv a/(10^{-7} \ \cm)$.

We follow \citet{DL01}, hereafter DL01, for the number $N_{\rmn{H}}$ of H-atoms in the grains (see their Eq.~8). 
Following DL98b, we account for the fact that the smallest grains may be sheetlike\footnote{DL98b allow for a possible population of linear grains, 
although they do not actually use them.}, as expected for polycyclic aromatic hydrocarbons (PAHs). We assume that this is the case for $a < a_2 = 6 \ $\AA \ (this corresponds to $N_{\rmn C} \approx 100$ 
carbon atoms, the size of a large PAH). We model them as disks of thickness $d = 3.35 \ 
$\AA, the interlayer separation in graphite. In many cases, these grains will be rotating primarily around the axis of largest moment of inertia
\citep{Purcell79},
which is perpendicular to the plane of the grain. When computing various rates, we will usually assume a spherical geometry, with a 
``surface-equivalent'' radius $a_s$ or a ``cylindrical excitation-equivalent'' radius $a_{cx}$, defined as :
\beq
4 \pi a_s^2 \equiv \oint \rmd S
{\rm ~~~~and~~~~}
4 \pi a_{cx}^4 \equiv \frac{3}{2} \oint \rho^2 \rmd S, \label{acx}
\eeq
where $\rho \equiv r \sin \theta$ is the distance to the axis of symmetry and $\rmd S$ is the surface area element.

\begin{figure} 
  \includegraphics[width = 88mm]{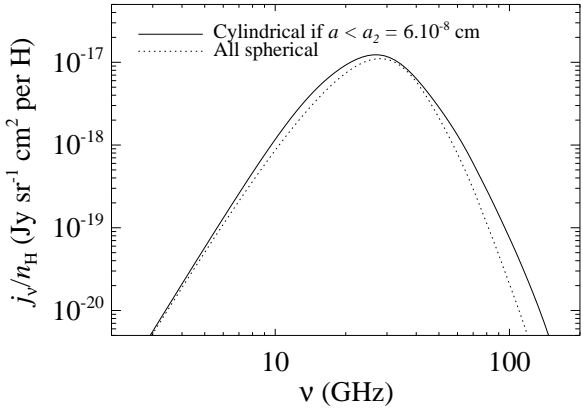}
\caption{Effect of relaxing the assumption of cylindrical grains on the spectrum, for a fiducial Cold Neutral Medium environment (CNM, defined in Eq. (\ref{eq : CNM})). At equal radius, spherical grains have a smaller moment of inertia than the cylindrical ones, which are rotating primarily about their axis of largest moment of inertia. They will thus radiate at slightly higher frequencies. For the CNM, we find an increase of peak frequency $ \Delta \nu_{\rmn{peak}}/\nu_{\rmn{peak}} \approx 6 \%$. The high-frequency tail of the spectrum is due to the smallest dipole moments of the assumed Gaussian distribution for the intrinsic dipole moments (see Section \ref{Section dipole} and Fig.\ref{figure : dipole_effect}). For a spherical distribution of dipole moments, there are fewer grains with a low dipole than for a planar distribution. This explains the decrease in power at high frequencies. For the CNM, this results in a decrease of total emitted power $ \Delta j_{\rmn{tot}}/j_{\rmn{tot}} \approx -16 \%$.}
\label{spherical}
\end{figure}

Although the assumption of cylindrical grains for $a < a_2$ is not critical, it does have an effect on the spectrum, which is shown in Fig.~\ref{spherical}. 

\subsection{Size distribution} \label{size_dist}

Following \citet{WD01a}, hereafter WD01a, we adopt the following size distribution for carbonaceous dust, for grain radii $ a_{\min} 
= 3.5 \ $\AA$ < a < a_{\max} = 100 $\AA:
\barr
\frac{1}{\nH} \frac{dn_{\gr}}{da} &=& D(a) + \frac{C}{a} \Bigg( \frac{a}{a_t} \Bigg)^{\alpha} F(a ; \beta, a_t)
\nonumber \\
&&\times
\left\{
\begin{array}{lc}
1, & a_{\rm min} < a < a_t\\
\rme^{-[(a-a_t)/a_c]^3}, & a > a_t
\end{array}
\right.,
\earr
where
\beq
F(a;\beta, a_t) = \left\{
\begin{array}{lc}
1+\beta a/a_t, & \beta \geq 0\\
(1 - \beta a/a_t)^{-1}, & \beta < 0 \end{array}
\right..
\eeq
The function $D(a)$ describes truncated (at 3.5 \AA) lognormal grain populations,
\beq
D(a) = \sum_{i=1}^2 \frac{B_i}{a} \exp \Big{\{}- \frac{1}{2}\Big[\frac{\ln(a/a_{0,i})}{\sigma} \Big]^2 \Big{\}},
\eeq
with the normalizations $B_i$ defined to place a total number $b_{\C,i}$ of carbon atoms per hydrogen nucleus in the i$^{\rmn{th}}$ lognormal population. Here $b_{\C,1} = 0.75 b_{\C}$, $b_{\C,2} = 0.25 b_{\C}$, $b_{\C}$ being the total carbon abundance per hydrogen nucleus in the lognormal populations, 
$a_{0,1} = 3.5 \ $\AA, $a_{0,2} = 30 \ $\AA, and $\sigma = 0.4$. This size distribution has a total of six adjustable parameters $(b_C, C, a_t, a_c, 
\alpha, \beta)$. For a given choice of $b_C$, the other five parameters can be found in WD01b, Table 1.


We consider only carbonaceous grains in this work. The abundance of small silicate grains in the diffuse phases is indeed limited by the absence of the 10$\ \mu$m band in emission, as discussed in WD01b. Note, however, that \citet{LD01_Si} found that as much as $\sim 10 \% $ of the interstellar silicate mass could be in the form of ultrasmall particles ($ a \lesssim 15 \Am$) without violating any existing observational constraints. While consistent with observations, our assumption is thus not required by them.

\subsection{Dipole moments} \label{Section dipole}

Although only the component of the dipole moment perpendicular to $\bomega$ is of importance for the electric dipole emission, the total dipole moment 
will be needed in coming calculations. In case of spherical grains, we assume the dipole moment and rotation axis are randomly oriented. For 
cylindrical grains, the dipole moment is mainly oriented in the plane of the grain, perpendicular to the rotation axis.

The dipole moments have two components. First, an intrinsic part $\bmu_i$, which results from the addition of dipole moments from individual 
molecular bonds. For a given grain radius, we assume a multivariate Gaussian distribution, with variance proportional to the 
number of atoms in the grain, $N_{\rmn{at}} = N_{\rmn C} + N_{\rmn H}$:
\beq
P(\mu_i)  \propto \left\{\begin {array}{lll}
\mu_i^2 \rme^{-3 \mu_i^2/2 \langle\mu_i^2\rangle} & & \textrm{spherical grains} \\
\mu_i \rme^{-\mu_i^2/\langle\mu_i^2\rangle} & & \textrm{disklike grains}\end{array}\right., \label{Pmui}
\eeq
with
\beq
\langle\mu_i^2\rangle = N_{\rmn{at}}\beta^2.
\eeq
These are appropriate assumptions if the dipole moments of bonds add in a random-walk fashion, although we caution that this need not be the case.
Counterexamples could include dipole moments dominated by a single feature, e.g. a PAH with a single OH group.
The formula given above is in that case intended to give an estimate of the total dipole moment, with the value of $\beta$ tuned to reproduce approximately observed dipole moments for laboratory 
molecules (see e.g. DL98b, Table 3).

The rms dipole moment per atom, $\beta$, is poorly known. Following DL98b, we will take $\beta = 0.38 \ \debye$ as a fiducial value, corresponding to 
\beq
\langle \mu_i^2 \rangle|_{a = 10^{-7}\ \cm} = (9.3 \ \debye)^2.
\eeq
In addition, for grains with charge $Ze$, a displacement $d$ between charge centroid and center of mass (e.g. due to asymmetric grain shape or isotopic substitution) may add another, uncorrelated 
component.  We 
assume that the displacement is proportional to the excitation equivalent radius: $d = \epsilon a_{cx}$, where $\epsilon =0.01$ (DL98b).
In most cases this is negligible compared to the intrinsic component, so we model it as a single 
value for the sake of simplicity.
The total dipole moment is thus given by
\beq
\mu^2 = \mu_i^2 + (\epsilon Z \qe a_{cx})^2,
\eeq
where $\qe$ is the elementary charge.

\subsection{Grain charge}
\label{s:charge}

The rotational damping and excitation rates will be dependent on the grain charge. DL98b showed that the characteristic timescale for changes in charge 
is much shorter than the characteristic rotational damping time. We will therefore average the damping and excitation rates over grain charges, as well 
as the electric dipole moment when computing the power radiated. We therefore need the charge distribution function\footnote{We use the same notation 
for different distribution functions. The context and their argument should make their meaning unambiguous.} of the grains as a function of their 
radius and environmental conditions, $f_a(Z)$.

There are three main processes contributing to grain charging: collisional charging by electrons and ions,
which rates we denote $J_e(Z, a)$ and $J_i(Z, a)$ respectively, and photoelectric
emission of electrons caused by the impinging radiation, which
rate is $J_{pe}(Z, a)$. For every grain radius, the steady state charge distribution function
is obtained by solving recursively the following equations:
\beq
\left[J_i(Z, a) + J_{pe}(Z, a) \right]f_a(Z) = J_e(Z+1, a) f_a(Z+1).
\eeq
We use the equations of \citet{DraineSutin} for collisional processes, updated with the \citet{WD01b} electron sticking coefficients, for $J_i$ and 
$J_e$.  The photoelectric emission rate is computed according to WD01b.
The radiation field is taken to be a multiple $\chi$ of the average interstellar radiation field $u_{\rmn{ISRF}}$, as estimated by \citet{Mezger} and 
\citet{Mathis}.

\section{The Fokker-Planck equation} \label{Fokker}

\subsection{Form of the equation in spherical polar coordinates}

The stationary angular velocity distribution function $f_a(\bomega)$ (such that
$f_a(\bomega) \rmd^3\bomega$ is the probability of the grain's
angular velocity being $\bomega$ within $\rmd^3\bomega$) is
determined from the stationary Fokker-Planck equation.  We differ here from DL98b who assumed the distribution was Maxwellian and calculated its 
approximate rms grain rotation rate $<\omega^2>^{1/2}$.  The Fokker-Planck equation is valid in the limit of continuous torques, i.e. if every 
interaction changing the rotation rate of the grain does so by a small amount $\delta \omega \ll \omega$. This is, therefore, accurate for the largest 
grains, which have large moments of inertia. But it fails to describe precisely the smallest ones ($a \ \lesssim 7 \ \Am$), for which DL98b showed that impulsive torques are 
important (see their Section 7 and Fig. 7). However, we believe that the actual distribution function would differ from the one we calculate only at very high frequencies, where the 
dust emissivity is dominated by the vibrational emission. Indeed, the occasional impulsive torques on the grains enhance the distribution function for high values of the rotation rate, where the solution of the Fokker-Planck equation predicts an exponential cutoff, as we shall see later. The peak of the distribution will not be affected significantly, as the variations of the rotation rate of a grain within the peak are not impulsive.

The stationary Fokker-Planck equation is given by
\beq
\frac{\partial}{\partial \omega^i}\left[D^i(\bomega) f_a(\bomega)
\right]  + \frac{1}{2} \frac{\partial^2}{\partial \omega^i \partial \omega^j}
\left[ E^{ij}(\bomega) f_a(\bomega) \right]
 = 0.
\eeq
The coefficients are defined as:
\beq
D^i(\bomega) \equiv - \lim_{\delta t \rightarrow 0} \frac{\langle\delta  \omega^i \rangle}{\delta t}
{\rm ~~~and~~~~}
E^{ij}(\bomega) \equiv \lim_{\delta t \rightarrow 0} \frac{\langle \delta \omega^i \delta \omega^j \rangle}{\delta t}.
\eeq
We assume that the medium is isotropic, and there are no
physical processes that allow for a preferred direction, such as a
magnetic field. As a consequence, the rotational distribution function only depends upon the magnitude $\omega$ of $\bomega$. Moreover, in a local 
orthonormal frame $(\hat{\bmath e}_{\omega} , \hat{\bmath e}_{{\theta}}, \hat{\bmath e}_{\phi})$, where $\omega, \theta$ and
$\phi$ are the usual spherical polar coordinates defining $\bomega$, the excitation
coefficient take up the following form :
\beq
E^{\hat\omega\hat\omega} = E_{\parallel}(\omega)
\eeq
accounts for fluctuations along $\hat\bomega$, and
\beq
E^{\hat{\theta} \hat{\theta}} = E^{\hat \phi \hat \phi} =
E_{\bot}(\omega)
\eeq
accounts for fluctuations perpendicular to $\bomega$.
The components in the coordinate basis are thus:
\beq
E^{\omega \omega } = E_{\parallel}(\omega), \ \ E^{\theta \theta } = \frac{E_{\bot}(\omega)}{\omega^2}
, \ \ E^{\phi \phi} = \frac{E_{\bot}(\omega)}{\omega^2 \sin^2\theta}.
\eeq
Moreover, we assume there are no systematic torques, so the damping coefficient is directed along $\bomega$ and we have
\beq
{\bmath D}( \bomega) = D(\omega) \hat{\bmath e}_{\omega}.
\eeq
In the spherical polar coordinate basis, the Fokker-Planck equation then becomes:
\barr
\frac{1}{\omega^2} \frac{\rmd}{\rmd \omega} \left[ \omega^2 D(\omega)
f_a(\omega) \right]  
&& \nonumber\\
+ \frac{1}{2\omega^2} \frac{\rmd^2}{\rmd \omega^2} \left[ \omega^2 E_{\parallel}(\omega) f_a(\omega) \right]
&& \nonumber\\
- \frac{1}{\omega^2} \frac{\rmd}{\rmd \omega} \left[ \omega \ E_{\bot}(\omega)f_a(\omega) \right]  &=& 0.
\earr
Integrating once, we get the following first order differential
equation:
\beq
\frac{\rmd f_a}{\rmd \omega} + 2 \frac{\tilde{D}}{E_{\parallel}} f_a = 0,  \label{FP}
\eeq
where
\beq
\tilde D \equiv D + \frac{1}{\omega}(E_{\parallel} - E_{\bot}) + \frac{1}{2}\frac{\rmd E_{\parallel}}{\rmd \omega}.
\eeq
Note that $\tilde D$ is simply equal to $D$ if the fluctuations are
isotropic and independent of $\omega$.

The coefficients $D$, $E_\parallel$, $E_{\bot}$, and therefore $\tilde D$ from various independent rotational damping and excitation processes are 
additive.

A given process is said to respect detailed balance, when, if that process were the only one taking place, the grain would rotate thermally, i.e. 
$f_a(\omega) \propto \exp(-I \omega^2/2 k T) $. As one can see from the Fokker-Planck equation, this implies that this process 
must satisfy :
\beq
\tilde D = \frac{I \omega}{2 k T} E_{\parallel}. \label{detailed balance}
\eeq
Excitation rates are often easier to calculate than damping rates, since they are positive definite and do not rely on near-cancellation of processes 
that increase versus decreasing $\omega$.  Thus in some cases, we will make use of detailed balance 
(i.e. the fluctuation-dissipation theorem), to obtain the damping rate, knowing the excitation rate.

\subsection{Normalized damping and excitation coefficients}

We will see in the next section that for collisions with neutral H atoms, at a temperature $T$, for a spherical dust grain at the same temperature $T$, 
the 
damping and parallel excitation coefficients have the following form:
\beq
\tilde D_H = \frac{\omega}{\tau_{\rm H}}
{\rm ~~~~and~~~~}
E_{||, H}  = E_{\bot, H} = \frac{2 k T}{I \tau_{\rm H}},
\eeq
where
\beq
\tau_{\rm H} \equiv \left[ \nH m_{\rm H} \left(\frac{2 k T}{\pi m_{\rm H}} \right)^{1/2} \frac{4 \pi a_{cx}^4}{3 I} \right]^{-1}
\label{tauH}
\eeq
is the characteristic rotational damping timescale for collisions with neutral H atoms. Note that they respect the detailed balance condition.

We normalize the damping and excitation coefficients of each process to those of collisions with H atoms. Taking DL98b notation, we define, for each 
process $X$ :
\beq
F_X(\omega) \equiv \frac{\tau_{\rm H}}{\omega} \tilde D_X  \label{F_X}
\eeq
\beq
G_X(\omega) \equiv \frac{I \tau_{\rm H}}{2 k T} E_{\parallel,X}(\omega)  \label{G_X}
\eeq
A special case is made of the rotational damping through electric dipole radiation (subscript $_{\ed}$), because of its specific $\omega^3$ dependence:
\beq
\left.\frac{\rmd}{\rmd t}\left(\frac{1}{2}I \omega^2 \right)\right|_{\ed} = \frac{2}{3}\frac{\mu_{\bot}^2 \omega^4}{c^3},
\eeq
so
\beq
\left. \frac{\rmd \omega}{\rmd t}\right|_{\ed} = - D_{\ed}(\omega) =  - \frac{2}{3} \frac{\mu_{\bot}^2  \omega^3}{I c^3}  = - \frac{I \omega^3}{3 k T} \frac{1}{\tau_{\ed}}.  \label{D_ed}
\eeq
Here we define, following DL98b:
\beq
\tau_{\ed} \equiv \frac{I^2 c^3}{2 k T \mu_{\bot}^2}
\eeq
Using Eqs.~(\ref{F_X}), (\ref{G_X}) and (\ref{D_ed}) in Eq.~(\ref{FP}), the final equation for the distribution function is
\beq
\frac{\rmd f_a}{\rmd \omega} + \left[\frac{I \omega}{k T} \frac{F}{G}  +  \frac{\tau_{\rmn H}}{\tau_{\ed}}  \frac{1}{3 G} \frac{I^2 
\omega^3}{(k T )^2} \right] f_a = 0,
\eeq
where
\beq F \equiv \sum_X F_X {\rm ~~~~and~~~~} G \equiv \sum_X G_X. \eeq
One can see that the conditions to get a thermal, Maxwellian distribution $f_a(\omega) \propto \exp(-I \omega^2/2 k T)$ are:
\beq
F = G = \textrm{constant}
{\rm ~~~~and~~~~}
\frac{\tau_{\rm H}}{\tau_{\ed}} \rightarrow 0.
\eeq
Otherwise, the general solution to this equation is :
\barr
f_a(\omega) &\propto&  \exp \Bigl\{-\int_0^{\omega}  d \omega' \Bigl[\frac{I \omega'}{k T}   \frac{F(\omega')}{G(\omega')}  
\nonumber \\
&& +  \  \frac{\tau_H}{3 
\tau_{\ed}G(\omega')}   \frac{I^2 \omega'^3}{ (k T )^2} \Bigr] \Bigr\}.
\label{equation : f_a(omega)}
\earr
If all $F_X$'s and $G_X$'s are constant, this has a simple form :
\beq
f_a(\omega) \propto \exp \left[ - \frac{F}{G} \frac{I \omega^2}{2 k T} - \frac{\tau_H}{\tau_{\ed}} \frac{1}{3 G} \Big(\frac{I \omega^2}{2 k T} \Big)^2 
\right].
\eeq
Note that the damping through electric dipole radiation causes the distribution to be non-Maxwellian.

In the general case, some $F_X$'s and $G_X$'s may depend upon $\omega$ and one has to compute numerically the resulting distribution function, using 
Eq.~(\ref{equation : f_a(omega)}).

We now turn to the calculation of the various damping and excitation coefficients, due to collisions, plasma drag, infrared emission, photoelectron emission, and random $\textrm{H}_2$ formation. In the following microphysics sections that form the heart of the paper, we compute excitation and damping coefficients as a function of grain radius and environmental conditions. We evaluate them numerically for a fiducial Cold Neutral Medium (CNM) environment, defined explicitly in Eq.~(\ref{eq : CNM}).

\section{Collisional damping and excitation} \label{collisions}

In this section we correct the results of DL98b, Appendix B, which did not take into account the fact that not all neutrals escape the grain surface when computing the damping rate.\\
The microphysics of collisions is
complex and beyond the scope of this study (for a discussion of the physics and chemistry of PAHs and their relation with the interstellar gas see for example \citet{Omont}). We therefore
use the following simplifying assumptions:
\begin{itemize}
\item The grain is in a stationary state: the rate at
which species collide with it is equal to the rate at which they
leave its surface.
\item We assume that all species (neutrals and ions) colliding with the grain
stick and that they depart the grain as neutrals. In extremely dense environments, the colliding species may bouce off the grain surface instead of sticking. This case is discussed at the end of Section \ref{section : Tev}.
\item Even if the impacting species may not collide
equiprobably everywhere on the grain's surface (e.g. if the grain
is non spherical or if it has a dipole moment), we assume they
somehow get re-distributed on the grain surface and leave it equiprobably
from any point.
\item We assume, as in DL98b, that neutrals leave the grain
surface with a thermal velocity distribution in the grain's frame, with a temperature $T_{ev}$ of the order of the infrared emission characteristic 
temperature. Unlike DL98b, we estimate $T_{ev}$ as a function of grain radius and ambient radiation field (see section \ref{section : Tev}).
\end{itemize}
Using those assumptions, one can compute the rate of collisional
damping and excitation. We will perform the calculations for a
spherical grain in the general case. To find the relevant equivalent radius to use for a cylindrical grain, we will carry out the explicit calculation 
in the case of collisions of a neutral grain with neutral H atoms.
Note that as pointed in DL98b, the rotational excitation in
case of collisions has two origins: the random excitation by
incoming particles (superscript $^{(in)}$), as well as the random
excitation by ``evaporating'' neutrals (superscript $^{(ev)}$).

\subsection{General considerations: spherical grain}

We use the usual spherical polar coordinates around the spherical grain, taking the rotation axis as a reference.
The local phase-space density at the grain surface is:
\beq
f_{ev}({\bmath v}, \theta) = K(\theta) \exp\Big[-\frac{m({\bmath v} - {\bmath v}_0)^2}{2 k T_{ev}} \Big]
\eeq
with the local velocity
\beq
{\bmath v}_0 \equiv \bomega \times {\bmath r} = a \omega \sin \theta \hat{\bmath e}_{\phi}.
\eeq
The normalization constant $K(\theta)$ is found by imposing that, locally, the flux of evaporating (and escaping) particles is equal to the flux of 
colliding particles. Except for the case of ions interacting with the electric dipole of the grain, the flux of colliding particles will be homogenous 
on the grain surface. If it is not the case, we approximate the local flux by the total rate of collisions $\rmd N_{coll}/\rmd t$ divided by the grain area:
\beq
\frac{1}{4 \pi 
a^2}\frac{\rmd N_{coll}}{\rmd t} =
K \int v_r \exp \left[-\frac{m({\bmath v} - {\bmath v}_0)^2}{2 k T_{ev}} \right] P_{\rm esc}\, \rmd^3{\bmath v},
\label{eq:knorm}
\eeq
where $P_{\rm esc}=1$ for velocities at the grain surface leading to escape, and $0$ otherwise.

All particles evaporating from the grain are neutrals. They interact with the grain through the induced dipole potential (we neglect the dipole-induced dipole interaction with the dipole moment of the 
grain):
\beq U(r) = - \frac{1}{2}\alpha  \frac{Z_g^2 \qe^2}{r^4}, \eeq
where $\alpha$ is the polarizability of the escaping neutral and $\qe$ is the elementary charge.
The polarizability of hydrogen is a standard result in nonrelativistic quantum mechanics and is $\frac92a_0^3=0.67\,$\AA$^3$ where $a_0$ is the Bohr radius \citep{LLQ}.
We also take $\alpha=0.20$\AA$^3$ for helium\footnote{We assume that all the helium is neutral and $n_{\rmn{He}}/\nH = 1/12$}  \citep{TH72}, and $\alpha=1.54$\AA$^3$ for carbon \citep{MK72}, which is important since C$^+$ is often the dominant ion if the hydrogen is self-shielded. For molecular hydrogen H$_2$, we take $\alpha=0.79$\AA$^3$ \citep{Marlow65}. 


\subsubsection{Computation of $P_{esc}$}

The radial coordinate of the escaping neutral is the solution of the following equation:
\beq
\dot r^2 + V_{\rm eff}(r) \equiv \dot r^2 + \frac{a^2}{r^2} v_{\parallel}^2 - \frac{a^4}{r^4} v_a^2 = \frac{2 E}{m},
\eeq
where $v_{\parallel}$ is the modulus of the tangential velocity at the grain surface and
\beq
v_a^2 \equiv \frac{Z_g^2 \qe^2 \alpha}{m a^4}. \label{va}
\eeq
The effective potential has a maximum at the radius
\beq
r_a = \sqrt 2 \ a \frac{v_a}{v_{\parallel}}; \ \ \ \ V_{eff}(r_a) = \frac{v_{\parallel}^4}{4 v_a^2}.
\eeq
To escape, a neutral needs to have either $a > r_a$ or\\ 
$2 E/m > V_{\rm eff}(r_a)$. These two conditions can be combined to get:
\barr
P_{\rm esc} \!\!\!\! &=& \!\!\!\! 1  \ \textrm{if}\left\{
\begin{array}{lcl} v_r > v_a  & {\rm or} & \\
 0 < v_r < v_a & {\rm and} & v_{\parallel} > \sqrt{2 v_a(v_a - v_r)} \end{array} \right.,
\nonumber \\ &&
\earr
where $v_r$ is the radial velocity at the grain surface.

\subsubsection{Computation of $K(\theta)$}

Following DL98b, we define $\epsilon_e^2 \equiv m v_a^2/2 k T_{ev}$, which describes whether the typical evaporating atom has enough energy to 
overcome the induced dipole attraction to the grain ($\epsilon_e<1$) or not ($\epsilon_e>1$).
We also define the ratio of rotational velocity to thermal velocity at the grain surface, which is small compared to unity:
\beq
\Omega \equiv a \omega \sqrt{\frac{m}{2 k T_{ev}}} \sim \Big( \frac{m}{m_{\rm{grain}}} \frac{T_{\rm{rot}}}{T_{ev}} \Big)^{1/2} \ll 1.
\eeq
In terms of those dimensionless quantities, we can find the normalization constant $K$.  The right-hand side of Eq.~(\ref{eq:knorm}) can be expanded 
using the substitution
\beq
(v_r,v_\theta,v_\phi) = \sqrt{\frac{2kT_{ev}}m}\,(u_r,u\cos\psi,u\sin\psi)
\eeq
to yield
\barr
\frac{1}{4 \pi
a^2}\frac{\rmd N_{\rm coll}}{\rmd t}
\!\!\!\!&=&\!\!\!\!K\left(\frac{2 k T_{ev}}{m} \right)^2 \frac{\pi}{2} \Bigg[  \rme^{- \epsilon_e^2}
\nonumber \\
&&\!\!\!\! + \rme^{-(\Omega \sin \theta)^2}
\int_{0}^{\epsilon_e}  2u_r \rmd u_r\, \rme^{-u_r^2} 
\nonumber \\
&&\!\!\!\!\times\!
\int_{\sqrt{2 \epsilon_e(\epsilon_e-u_r)}}^{\infty} \!\!\!\!\! 2 u \, \rmd u \,  \rme^{-u^2}
I_0(2u\Omega \sin \theta) 
\Bigg],
\earr
where
\beq
I_0(X) = \frac{1}{2 \pi}\int_0^{2 \pi} e^{X \sin \psi} d \psi = 1 + \frac14X^2 + ...
\eeq
is a modified Bessel function of the first kind.

Expanding to second order in $\Omega$, we get :
\beq
K = \left(\frac{2 k T_{ev}}{m} \right)^{-2} \frac{2}{\pi} \frac{e^{\epsilon_e^2}}{e^{-\epsilon_e^2} + \sqrt{\pi} \epsilon_e \erf(\epsilon_e)}
\frac{1}{4 \pi a^2} \frac{d N_{coll}}{dt}
\eeq
up to corrections of order $\mathcal O (\Omega^2) $.

\subsubsection{Damping and excitation rates} \label{collision damping rate}

Each escaping neutral particle takes away an angular momentum
\beq
{\bmath L} = m a (v_{\theta}  \hat{\bmath e}_{\phi} - v_{\phi} \hat{\bmath e}_{\theta}).
\eeq
As $P_{esc}$ is an even function of $v_{\theta}$, the average of $v_{\theta}$ vanishes. The loss of angular momentum along the z-direction per unit 
time per unit area is given by
\barr
\frac{\rmd L_z}{\rmd t\, \rmd S} &=&
- m a \sin \theta \ K \int v_r v_{\phi} \exp \left[-\frac{m({\bmath v} - {\bmath v}_0)^2}{2 k T_{ev}} \right]
\nonumber \\
&& \times P_{esc} \,\rmd v_r \,\rmd v_{\theta} \,\rmd v_{\phi}.
\label{Pesc}
\earr
Here we differ from DL98b as we take into account the fact that not all particles escape from the grain. Expanding in $\Omega$ and using the expression 
for $K$ we get, up to corrections of order $\mathcal O (\Omega^2) $:
\beq
\frac{\rmd L_z}{\rmd t\, \rmd S} = - \frac{1}{4 \pi} m \sin^2 \theta\, \frac{\rme^{-\epsilon_e^2} + 2 \epsilon_e^2}{e^{-\epsilon_e^2} + \sqrt \pi 
\epsilon_e \erf\epsilon_e}  \frac{\rmd N_{\rm coll}}{\rmd t} \omega.
\eeq
Integrating over the whole grain surface, we find the damping rate
\beq
D(\omega) = - \frac{1}{I}\frac{\rmd L_z}{\rmd t} =  \ \frac{\rme^{-\epsilon_e^2} + 2 \epsilon_e^2}{\rme^{-\epsilon_e^2} + \sqrt \pi \epsilon_e 
\erf\epsilon_e} \frac{2 m a^2}{3 I } \frac{\rmd N_{\rm coll}}{\rmd t} \omega.
\eeq
A similar calculation leads to the excitation rate through evaporating particles:
\barr
{E_{\parallel}}^{(ev)}(\omega) &=& \frac{1}{I^2} {\frac{\rmd \Delta L_z^2}{\rmd t}}^{(ev)}
\nonumber \\ &=&
\frac{\rme^{-\epsilon_e^2} + 2 \epsilon_e^2}{\rme^{-\epsilon_e^2} + \sqrt \pi \epsilon_e \erf\epsilon_e}\frac{2 m a^2}{3 I^2}\frac{\rmd 
N_{\rm coll}}{\rmd t} \ k T_{ev}
\nonumber \\
&=& \frac{k T_{ev}}{I \omega} \ D(\omega),
\earr
up to terms quadratic in $\Omega$.

This implies the remarkable relation
\beq
G_{\rmn{coll}}^{(ev)} = \frac{T_{ev}}{2 T} F_{\rmn{coll}}.  \label{G_coll_ev}
\eeq
Physically, this occurs because if $T_{ev}=T$ then the collisions with neutrals satisfy detailed balance, Eq.~(\ref{detailed balance}).  The factor of 
2 arises since in this case there is an equal contribution to the excitation from incoming and evaporating particles.

We derive a stronger damping rate due to evaporating atoms than DL98b: for $\epsilon_e\ll
1$ this results in no change, but for $\epsilon_e\gg 1$ we find much stronger damping.  The physical origin of this is that atoms that evaporate
with prograde velocities relative to the local grain surface ($v_\phi>v_{0\phi}$) typically have more angular momentum than atoms that evaporate with 
retrograde velocities.  Therefore the centrifugal potential helps them to escape the grain.  DL98b neglected this effect, but for $\epsilon_e\gg 1$ it 
is dominant.

The excitation rate through incoming particles will be calculated for each case.

\subsubsection{Evaporation temperature $T_{ev}$} \label{section : Tev}

DL98b assume that the evaporating temperature is a constant, independent of grain size. This accurately describes the largest grains, for which the temperature may be approximated as a constant, obtained from equating the absorbed and emitted energy (DL98b):
\beq 
T_c = \frac{h c}{k} \left[ \frac{\langle Q \rangle_{*} u_{*}}{8 \pi h c Q_0 \lambda_0^{\alpha} \Gamma(\alpha + 4) \zeta(\alpha +4)} \right]^{1/(\alpha + 4)}
\eeq
where in the infrared the grain absorption efficiency is assumed to be a power-law
\beq 
Q_{\nu} = Q_0 \left(\frac{\nu}{\nu_0}\right)^{\alpha} \ \ \ , \ \ \lambda_0 = \frac{c}{\nu_0}
\eeq 
with typically $\alpha = 2$ and $\langle Q \rangle_{*} u_{*} \equiv \int \rmd \nu Q_{\nu} u_{\nu}$. Note that we have $T_c \propto \chi^{1/6}$ and a weak dependence on grain radius as the absorption efficiencies cancel out.

\begin{figure}
  \includegraphics[width = 88mm]{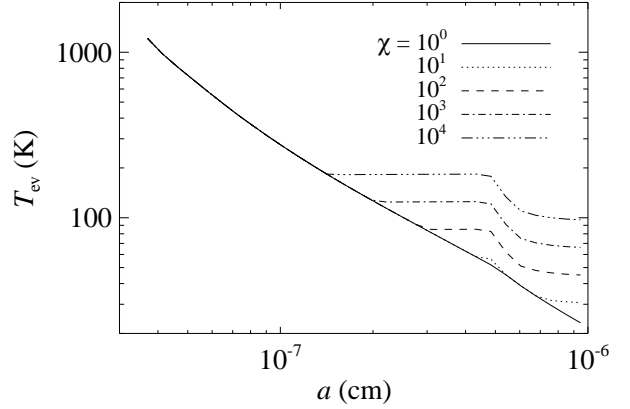}
\caption{Evaporation temperature $T_{ev}$ as a function of grain radius $a$, for various values of the ambient radiation field $u$, parameterized by $\chi = u/u_{\rmn{ISRF}}$. The curves join at small radii, for which the grains undergo temperature spikes. The kink at $ a = 50 \Am$ results from the DL01 prescription for PAH-graphite optical properties.}
  \label{figure : Tev}
\end{figure}

\begin{figure}
  \includegraphics[width = 88mm]{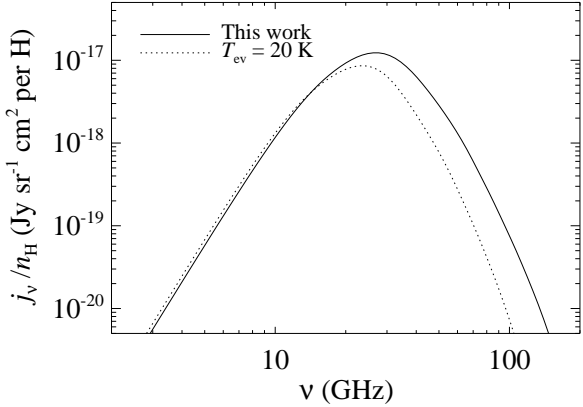}
\caption{Effect of the evaporation temperature model on the spinning dust spectrum for the Cold Neutral Medium (CNM, Eq. (\ref{eq : CNM})). Our prescription results in a much higher evaporation temperature for the smallest grains, compared to DL98b, who assume a constant $T_{ev} = 20 \ \K$ for all grain sizes. This leads to a decreased damping rate (see discussion at the end of section \ref{collision damping rate}) and an increased excitation rate through collisions, and therefore increases the peak frequency of the spectrum.}
  \label{CNM Tev effect}
\end{figure}

However, the smallest grains undergo sudden thermal spikes after each photon absorption, followed by long intervals during wich the grain drops to its vibrational ground state. The neutrals or ions that 
have stuck to the grain after a collision cannot be thermally ejected from a grain in the ground state so we assume ejection during thermal spikes.  A simple assumption is that in this case ejection 
occurs after a photon absorption and thermalization of the photon's energy.  We take
\beq 
E_{\gamma} = \frac{\int Q_{\nu} u_{\nu} \rmd \nu }{\int Q_{\nu} \frac{u_{\nu}}{h \nu} \rmd \nu }
\eeq
 as the typical energy of an absorbed photon. Typically, $E_{\gamma} \approx 5 \ \eV$. We then calculate the corresponding grain temperature following DL01: we solve for $T_q$ such that $\bar{E}(T_q) = E_{\gamma}$, where 
\beq
\bar E(T) = \sum_{j=1}^{N_m} \frac{\hbar \omega_j}{\exp(\hbar \omega_j/k T) - 1} 
\eeq
is the expectation value of the energy of the grain, and the sum runs over its $N_m$ vibrational degrees of freedom.
We take $T_{ev} =\max(T_c, T_q)$ as the evaporation temperature. The result is shown in Fig.~\ref{figure : Tev}. One can see that we obtain much higher evaporation temperatures than the ones used by DL98b\footnote{The mechanism we describe for atomic ejection from grains is called photo-thermo-dissociation (PDT). \citet{Rouan} also mention another possible mechanism, photo-dissociation (PD), which is an atomic ejection following the direct interaction of a UV photon with a given C-H bond. PD may lead to even higher ejection temperatures, of order $10~000 \ \K$}. The effect may be significant on the final spectrum, as can be seen from Fig.~\ref{CNM Tev effect}.\\[5pt]
\emph{High density, low radiation field case}\\[3pt]
The previous treatment is valid only if the rate of photon absorption is high enough to eject all stuck species before all available sites on the grain are occupied.  We approximate the number of available sites on the grain by the number of superficial C-atoms :
\beq 
N_{\rmn{sites}} = \left\{
\begin{array}{lc}
N_{\rmn C}(a) & \textrm{ for cylindrical grains}\\
\frac{3 d}{a} N_{\rmn C}(a)  &  \textrm{for spherical grains}\end{array}
\right.,
\eeq
where $N_{\rmn C}(a)$ was defined in Eq. (\ref{N_C}) and $ d = 3.35$\AA \ is the interlayer separation in graphite. The ratio of collision rate to photon absorption rate is given by :
\beq
R_{\rmn{coll/abs}} = \frac{\nH \sqrt{8 k T/\pi m_{\rmn{H}}}}{\int Q_{\nu} \frac{u_{\nu}}{h \nu} \rmd \nu  \ c}
 \approx 0.1 \times \frac{\nH}{30 \cm^{-3}} T_2^{1/2} \chi^{-1} a_{-7}^{-1}.
\eeq 
In most environments, $R_{\rmn{coll/abs}}  \ll N_{\rmn{sites}}$ so there is no accumulation of stuck species. In very dense and dark clouds however, the rate of collisions may become so high compared to the rate of photon absorption that all the sites are occupied. In that case, the assumption that incoming species stick to the grain is no more valid. They will instead bounce off the irregular grain surface. From the fluctuation-dissipation theorem, one expects that, for collisions with neutral species, $F_n = G_n$. Thus, we set the effective evaporation temperature equal to the gas temperature in that case (see Eq. (\ref{Fn Gn}) and discussion below):
\beq
T_{ev} = T \ \ \ \textrm{if} \ \ \ R_{\rmn{coll/abs}} > N_{\rmn{sites}}.
\eeq
The actual transition from sticking to elastic collisions should of course be smooth, unlike the discontinuous step we assume here. Our treatment should approximately reflect the physics of collisions except near the transition regime $R_{\rmn{coll/abs}} \sim N_{\rmn{sites}}$.

\subsection{Collision with neutral H atoms: neutral grain, general grain shape}

We assume that the grain is neutral, and has no dipole moment, so there is no interaction whatsoever between the grain and the neutral H atoms (purely geometric cross-section).
The phase-space density of incoming H atoms at the grain surface is simply
\beq
f_{in}({\bmath v}) = \nH \left(\frac{m_{\rmn H}}{2 \pi k T} \right)^{3/2} \rme^{-m_{\rmn H}v^2/2 k T},
\eeq
from which one can easily get the excitation rate through incoming particles :
\beq
\frac{\rmd \Delta L_z^2}{\rmd t\, \rmd S}^{(in)} = \int v_n (m_{\rmn H} \rho v_{\phi})^2 f_{in}({\bmath v}) \rmd^3{\bmath v}
\eeq
where $v_n$ is the component of the velocity normal to the grain surface.  This evaluates to
\beq
\frac{\rmd \Delta L_z^2}{\rmd t\, \rmd S}^{(in)} = \nH \mH^2 \rho^2 \frac{\pi}{4} \left(\frac{2 k T}{\pi \mH} \right)^{3/2}.
\eeq
Integrating over the grain surface, we get
\beq
\frac{\rmd \Delta L_z^2}{\rmd t}^{(in)} = k T \nH \mH \left( \frac{2 k T}{\pi \mH}\right)^{1/2} \frac{4 \pi a_{cx}^4}{3},
\eeq
where $a_{cx}$ was defined in equation (\ref{acx}). For a spherical grain, $a_{cx} = a$. For a disklike grain of thickness $d$ and radius $b$, 
spinning around its axis of symmetry, we have
\beq
a_{cx} = \left[\frac{3}{8} b^3 \ (2 d + b)\right]^{1/4} \eeq
We can write the excitation rate by incoming H atoms as
\beq
E_{\parallel, \rmn H}^{(in)}  = \frac{k T}{I \tau_{\rmn H}}
\eeq
where $\tau_{\rmn H}$ was defined in equation (\ref{tauH}).

The case of evaporating particles is very similar. Assuming the grain surface is at the same temperature $T$ as the gas, the phase-space density of 
evaporating particles is
\beq
f_{ev}({\bmath v}) = \nH \left(\frac{m_{\rmn H}}{2 \pi k T} \right)^{3/2} \exp \left( - \frac{m_{\rmn H} ({\bmath v} - {\bmath v}_0)^2}{2 k T} \right).
\eeq
In that case $P_{\rm esc}= 1$ for all outgoing particles.
The same calculation therefore leads to
\beq
E_{\parallel, \rmn H}^{(ev)}  = \frac{k T}{I \tau_{\rmn H}}
\eeq
up to terms of order $\mathcal O(\Omega^2)$, which comes from the fact that we did not take into account the slight change of $\omega$ after the 
particle has collided (we assumed the same $\omega$ for the incoming and the outgoing particle).
Detailed balance ensures that
\beq
\tilde D_{\rmn H} = \frac{\omega}{\tau_{\rmn H}}.
\eeq
Therefore, for non spherical grains, we will compute collision rates assuming a spherical geometry with radius $a_{cx}$. We just showed that this is 
an exact result for collisions with neutral H atoms. The collision rates are indeed proportional to the area of the grain, but the angular momentum 
gained depends on $\langle\rho^2\rangle$, so $a_{cx}$ will approximately reflect both dependencies.

\subsection{Collisions with neutral atoms: charged grains}

In that case, the incoming neutrals interact with the same potential as the outgoing particles:
\beq
U(r) = -\frac{1}{2}\alpha \frac{Z_g^2 \qe^2}{r^4}.
\eeq
We use the same notation as DL98b and define
\beq
\epsilon_n \equiv \sqrt{\frac{m v_a^2}{2 k T}}
{\rm ~~~~and~~~~}
b_0(v) \equiv a \sqrt{\frac{2v_a}{v}},
\eeq
where $v_a$ was defined in Eq. (\ref{va}).\\
We recall, from DL98b, that a trajectory with impact parameter $b$ and velocity at infinity $v$ leads to a collision if
\beq
b \leq b_{\max}(v) = \left\{ \begin{array}{lc}
b_0(v)  & \rmn{if} \ \   v \leq v_a \\
a\sqrt{1 + v_a^2/v^2}  & \rmn{if}  \ \  v \geq v_a \end{array} \right..
\eeq
We compute the collision rate
\barr
\frac{\rmd N_{\rm coll}}{\rmd t} \!\!&=& \!\!n_n \int_0^{\infty} \rmd v \, 4 \pi v^3
\pi b_{\max}^2(v) \left(\frac{m_n}{2 \pi k T}\right)^{3/2} \rme^{-m v^2/2 k T}\nonumber\\
&=&\!\! n_n 2 \pi a^2 \left( \frac{2 k T}{\pi m} \right)^{1/2} \left[ e^{- \epsilon_n^2} + \sqrt \pi \epsilon_n \erf\epsilon_n \right].
\earr
We can now get the normalized damping and excitation rates for collisions with neutrals:
\barr
F_n &=& \frac{n_n}{n_H} \sqrt{\frac{m_n}{m_{\rm H}}} \frac{ e^{- \epsilon_n^2} + \sqrt \pi \epsilon_n \erf\epsilon_n}{e^{-\epsilon_e^2} + \sqrt 
\pi \epsilon_e \erf\epsilon_e} \left(e^{-\epsilon_e^2} + 2 \epsilon_e^2 \right),
\nonumber \\
G_n^{(ev)} &=& \frac{T_{ev}}{2 T} F_n, {\rm ~~~and}
\nonumber \\
G_n^{(in)} &=& \frac{n_n}{2 n_H} \sqrt{ \frac{m_n}{m_{\rm H}}} \left(e^{-\epsilon_n^2} + 2 \epsilon_n^2 \right),  \label{Fn Gn}
\earr
where the result for $G_n^{(in)}$ is identical to that of DL98b.
Note that when $T = T_{ev}$, $G_n^{(ev)} = G_n^{(in)} = F_n/2$ so the principle of detailed balance holds. Moreover, in the case of a neutral grain, if the only rotational excitation and damping process were collisions with neutral species, then the rotational distribution function would be a Maxwellian. In that case the rotational temperature would be given by $T_{\rmn{rot}} = G_n/F_n \times T = \frac{1}{2}(T + T_{ev})$, the arithmetic mean of the gas and evaporation temperatures, as was already shown by \citet{Purcell79}. 

This is the contribution of an individual neutral, for a given grain charge. To get the total contribution, one must average over all grain charges 
(DL98b showed that the charging timescale is much shorter than the collision timescale) and sum over all neutral species, which we take to be atomic and molecular hydrogen, and helium\footnote{Collisions with neutral helium have little effect on the spectrum: the helium contribution dominates $F_n$ and $G_n$ only in the case where the medium is strongly ionized, i.e. when the dominant rotational excitation and damping mechanisms are rather collisions with ions or plasma drag. We include them for completeness.} (with $n_{\rmn{He}}/\nH = 1/12$). 

\subsection{Collisions with ions: charged grains} \label{section : collisions ions charged grain}

The ion interacts with the grain through the Coulomb, electric dipole, and ``image charge'' potentials. The latter dominates over the Coulomb potential only in the immediate vicinity of the grain surface, so we will neglect it for charged grains. Properly accounting for it would result in a slight increase in both damping and excitation rates as this is an attractive potential. The general solution for this problem, with a rotating electric dipole moment, is still not analytical. Thus, for simplicity, we will only consider the case where the electric dipole moment can be considered as non-rotating, i.e. when the timescale of the collision is short compared to the rotation period of the grain. This is justified as, when the ion reaches the vicinity of the grain surface, the ratio of collision timescale to rotation timescale is approximately $ \omega a/v \sim \sqrt{m_i/m_{gr}} \ll 1$.  We will assume 
that the grain is spherical, so that the electric dipole moment is randomly oriented relative to the rotation axis (for cylindrical grains this is not 
the case but we will assume so for simplicity). Note that when the grain rotates rapidly, the component of the dipole moment perpendicular to the 
rotation axis averages out, but not the parallel component. Although this problem will be different in nature as this alignment creates anisotropic 
excitation by collisions, the magnitude of the non rotating part of the dipole moment will remain of the same order (a factor $1/\sqrt3$ smaller only), 
so our approximation should give a decent idea of what the effect of the dipole moment is on the trajectory.\\

We assume a spherical geometry with radius $a_{cx}$. Taking $\bmu$ as the polar axis for spherical polar coordinates, the interaction potential of the ion in the Coulomb and dipole field of the grain is given by
\beq
V(r, \theta) = \frac{Z_g Z_i \qe^2}{r} + \frac{Z_i \qe \mu \cos \theta}{r^2}.
\eeq
The motion in this potential has two obvious constants: the energy $E$ and the
angular momentum along the $z$-axis (along $\bmu$), $L_z$.  For this special potential, however, there exists a third constant of the motion.
The torque $\dot{\bmath L}$ exerted on the ion comes entirely from the second term in the potential and is
\beq
\dot{\bmath L} = -{\bmath r}\times\nabla V({\bmath r}) = \frac{Z_i \qe\mu\sin\theta}{r^2}\hat{\bmath e}_\phi.
\eeq
Since the azimuthal component of angular momentum is
${\bmath L}\cdot\hat{\bmath e}_\phi = mr^2\dot\theta$, we can then determine the overall rate of change of the angular momentum,
\beq
\frac{\rmd}{\rmd t}(L^2) = 2m_iZ_i\qe\mu\sin\theta\,\dot\theta
= -2m_iZ_i\qe\mu\frac\rmd{\rmd t}\cos\theta.
\eeq
Therefore we find the constant of the motion\footnote{This may also be derived by the Hamilton-Jacobi method in spherical polar coordinates.}
\beq
\mathcal A \equiv L^2 + 2m_iZ_i\qe\mu\cos\theta.
\eeq
Its value can be determined by taking the incoming trajectory at infinity with approach angle $\theta_\infty$,
\beq \mathcal A = (m_i b v)^2 + 2 m_i Z_i \qe \mu \cos \theta_{\infty}. \eeq
The energy of the trajectory is
\beq
\frac{1}{2} m_i \dot r^2 + V_{\rm eff}(r) = E
\eeq
where $V_{\rm eff}(r)$ is the sum of the potential $V(r,\theta)$ and the tangential kinetic energy $L^2/2m_ir^2$:
\beq
V_{\rm eff}(r) \equiv \frac{Z_g Z_i \qe^2}{r} + \frac{m_i
(b v)^2 + 2 Z_i \qe \mu \cos \theta_{\infty}}{2 r^2}.
\eeq
It is easier to work with the following dimensionless parameters 
\beq
\psi \equiv \frac{Z_g Z_i \qe^2}{a_{cx} k T} \ \ , \ \ \tilde \mu \equiv \frac{Z_i \qe \mu}{a_{cx}^2 k T}. \label{psi-mutilde}
\eeq
Their physical meaning is as follows : $|\psi|\ll 1$ when the thermal energy of the ion dominates over the
electrostatic interaction energy, and $|\psi|\gg 1$ when the electrostatic interaction dominates.  The sign determines whether the interaction is attractive ($\psi<0$) or repulsive ($\psi>0$). $\tilde \mu$ is the equivalent quantity for the dipole interaction. Note that we consider only positively charged ions so $\tilde \mu >0$. We also work with the dimensionless variables
\beq
c = \frac{b}{a_{cx}} \ \ ,  \ \ u = \sqrt{\frac{m_i}{2 k T}} v . \label{c-u}
\eeq
The effective potential can now be written
\beq
V_{\rm eff}(r) = k T \left[\psi \ \frac{a_{cx}}{r} +\left( u^2 c^2 + \tilde \mu \cos \theta_{\infty} \right) 
\left(\frac{a_{cx}}{r}\right)^2 \right]
\eeq
A study of this potential leads to the following condition for collision :
\beq
\cos \theta_{\infty} < X_{\max}(c, u) \equiv {\tilde \mu}^{-1}\left(u^2 -u^2 c^2 - \psi \right)
\eeq
Note that if $X_{max} < -1$, then there is never collision,
for any angle. If $X_{max} > 1$, then all angles lead to a
collision. We define
\beq
X(c,u) \equiv \max\left\{-1, \min\left[ 1, X_{\max}(c,u)\right] \right\}.  \label{def : X}
\eeq
Now, we can compute the collision rate
\barr
\frac{\rmd N_{\rm coll}}{\rmd t} &=&
n_i \int 2 \pi v^3 \rmd v \, 2 \pi b \rmd b\, \Big(\frac{m_i}{2 \pi k T} \Big)^{3/2}
\nonumber \\ && \times \rme^{- m_i v^2/2 k T}
(X+1) \nonumber \\
&=& 2 n_i \sqrt{\frac{2 \pi k T}{m_i}} a_{cx}^2 \nonumber \\ && \times \int 2 u^3 \rme^{-u^2} \rmd u \ 2 c \ \rmd c  \ \frac{X + 1}{2} .
\label{collision rate}
\earr
We can also get the excitation rate by incoming ions
\barr
\frac{\rmd\Delta L_z^2}{\rmd t} &=&
n_i \int \frac{(m b v)^2}{3} 2 \pi v^3 \rmd v  \, 2 \pi b \rmd b \, \Big(\frac{m_i}{2 \pi k T} \Big)^{3/2}
\nonumber \\ && \times \rme^{- m_i v^2/2 k T}
(X+1) \nonumber \\
&=& \frac{2 n_i m_i^2 a_{cx}^4}{3 \pi} \left(\frac{2 \pi k T}{m_i}\right)^{3/2} \nonumber\\
&& \times \int u^5 \rme^{-u^2} \rmd u \ 4 c^3 \rmd c \ \frac{X + 1}{2}.
\label{excitation rate}
\earr
These integrals can be evaluated explicitely and one then gets, for the charged grains $Z_g \neq 0$ 
\barr
F_i(Z_g \neq 0) &=& \frac{n_i}{n_{\rm H}} \sqrt{\frac{m_i}{m_{\rm H}}} \frac{\rme^{-\epsilon_i^2} + 2 \epsilon_i^2}{\rme^{-\epsilon_i^2} + \sqrt \pi \epsilon_i 
\erf\epsilon_i} \ g_1(\psi, \tilde \mu),
\nonumber \\
G_i^{(ev)}(Z_g \neq 0) &=& \frac{T_{ev}}{2 T} F_i(Z_g \neq 0), {\rm ~~~~and}
\nonumber \\
G_i^{(in)}(Z_g \neq 0) &=& \frac{n_i}{2 n_{\rm H}} \sqrt{\frac{m_i}{m_{\rm H}} } \ g_2(\psi, \tilde \mu),
\earr
where we have defined $ g_1(\psi, \tilde \mu) = $
\barr
 \left\{ \begin{array}{ll} 1- \psi  & \psi < 0 \\
\rme^{-\psi}\sinh \tilde \mu / \tilde \mu & \psi > 0 \end{array} \right.\ \ \ \ \ \ \ \ \ \ \ \ \ \  , \ \tilde \mu \le |\psi| \nonumber \\
 \frac{ 1 - \rme^{-(\psi + \tilde \mu)} + \tilde \mu - \psi + \frac{1}{2} (\tilde \mu -\psi)^2 }{2 \tilde \mu} \ \ \  , 
\ \tilde \mu > |\psi| , \label{g1}
\earr
and $ g_2(\psi, \tilde \mu) = $
\barr
\left\{ \begin{array}{ll} 1- \psi + \psi^2/2 + {\tilde \mu}^2/6 & \psi < 0 \\
\rme^{-\psi}\sinh \tilde \mu / \tilde \mu & \psi > 0 \end{array} \right.\ \ \ \ \ \ \ \ \ \ \ \ \   , \ \tilde \mu \le |\psi| \nonumber \\
\frac{ 1 - \rme^{-(\psi + \tilde \mu)} + \tilde \mu - \psi + \frac{1}{2} (\tilde \mu -\psi)^2 + \frac{1}{6}(\tilde \mu -\psi)^3 }{2 \tilde \mu} \ \ \  , 
\ \tilde \mu > |\psi| . \label{g2}
\earr
Note that these functions coincide with the functions $g_1(\psi) \ , \ g_2(\psi)$ defined in DL98b for $\tilde \mu =0$. We also defined
$ \epsilon_i^2 \equiv Z_g^2 \qe^2 \alpha_i/2 a^4 k T_{ev}$ (here $\alpha_i$ is the polarizability of species $i$ after it neutralizes on the grain surface, e.g. when considering collisions with the C$^+$ ion we 
take the polarizability of 
the neutral C atom).  Note that even when $T_{ev} = T$, $F_i \neq G_i$ as the incoming and outgoing particles are in different ionization states; 
detailed balance does not apply since realistic ISM phases are not in Saha equilibrium.
Numerically, one has (with $T_2 \equiv T/100 \rmn K$)
\barr
\psi &\approx& 170 \ Z_g \  a_{-7}^{-1} \ T_2^{-1}\\
\tilde \mu &\approx& 
30 \ \frac{\langle \mu^2 \rangle^{1/2}|_{10^{-7} \cm}}{9.3 \ \debye} \ a_{-7}^{-1/2} \ T_2^{-1} \label{mu-numerics}
\earr
From these values, one can see that in general the effect of the dipole moment cannot be neglected a priori, as $\tilde \mu$ is not small compared to unity. However, in general $\tilde \mu < |\psi|$. This 
implies that, for negatively charged grains, the dipole moment has little or no effect on the excitation and damping rate. For positively charged grains, the damping and excitation rate are both 
increased by the huge factor $\sinh \tilde \mu / \tilde \mu$, but still remain extremely small due to the Coulomb repulsion, which shows in the factor $\rme^{-\psi}$.

We therefore conclude that DL98b approximation of neglecting the effect of the electric dipole moment on the trajectory of ions, is essentially valid in the case of collisions with charged grains. It only has a significant effect for positively charged grains, for which the coulomb repulsion implies an extremely small rate of collisions with ions anyway. We still account for the electric dipole moment for the sake of completeness.

\subsection{Collisions with ions, neutral grain}

\begin{figure*}
  \includegraphics[width = 88mm]{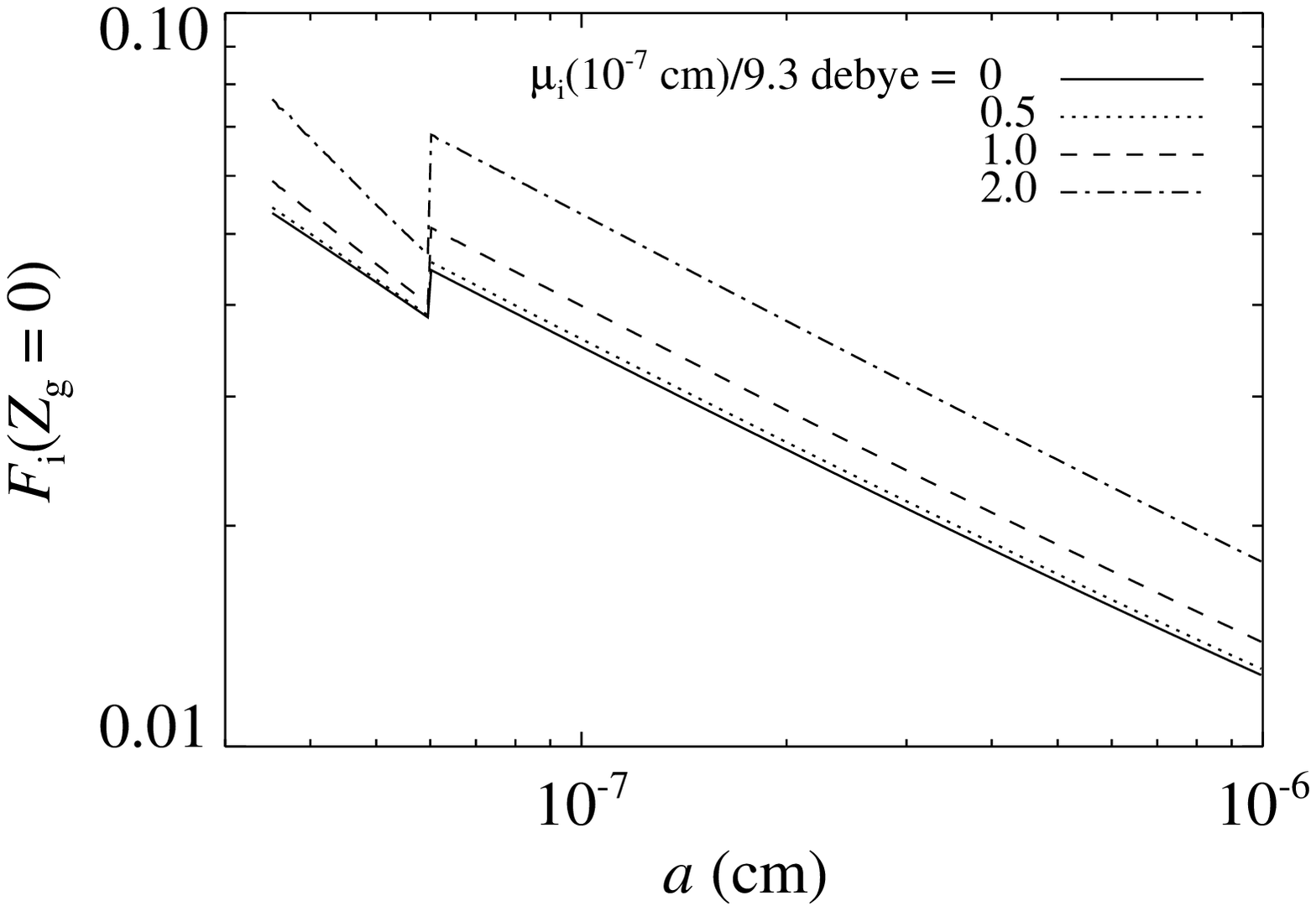}
  \includegraphics[width = 88mm]{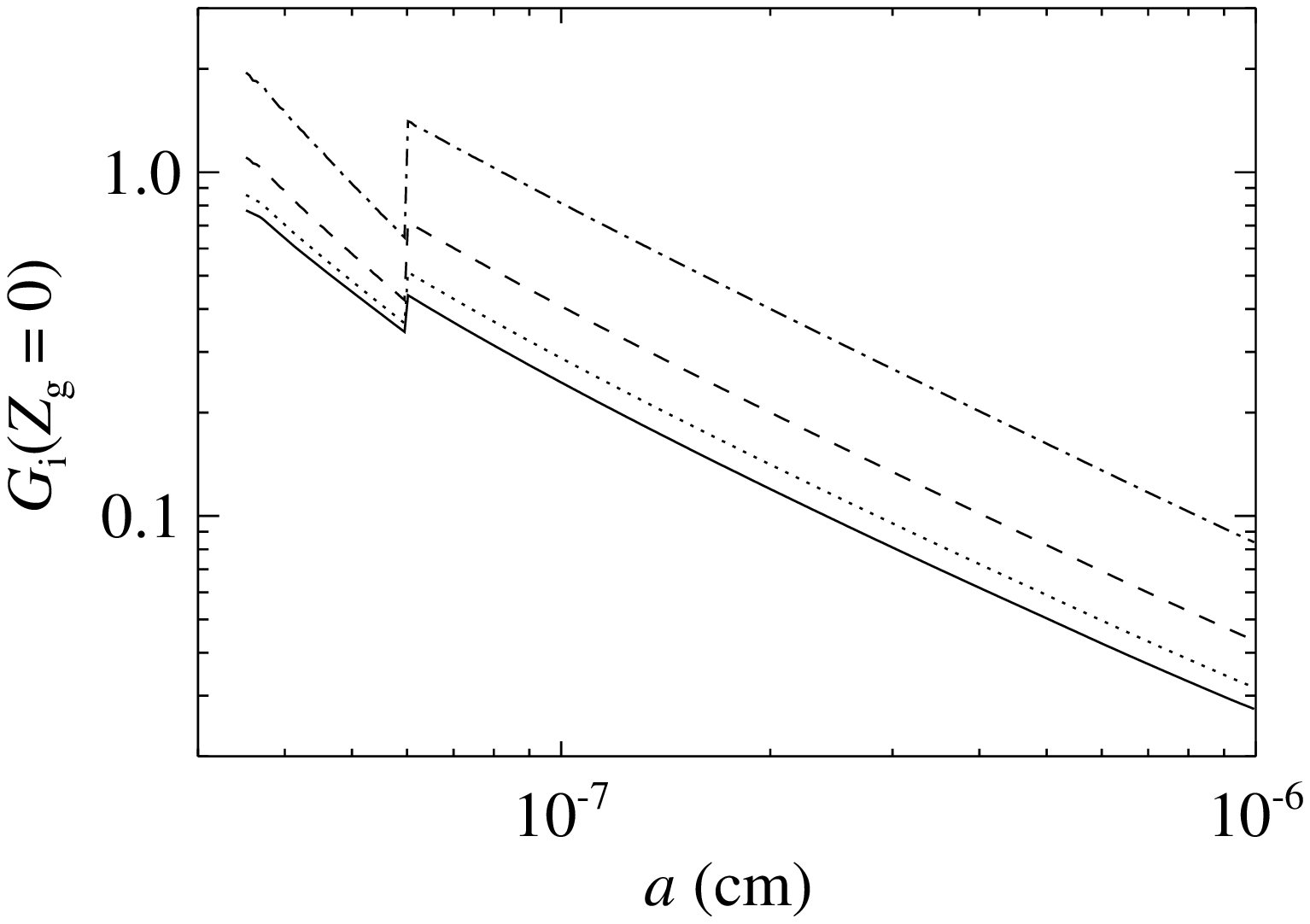}
\caption{$F_i(Z_g=0)$ (left panel) and $G_i(Z_g=0)$ (right panel) for several values of the electric dipole moment, in CNM conditions, Eq. (\ref{eq : CNM}).}
  \label{figure : FGi neutral}
\end{figure*}

In that case the Coulomb potential vanishes, and the ``image charge '' potential has to be taken into account. We carry the calculation using the same assumptions as in the previous section : slowly rotating spherical grain, with radius $a_{cx}$. 
Taking $\bmu$ as the polar axis for spherical polar coordinates, the interaction potential of the ion in the dipole and induced dipole field of 
the grain is given by
\beq
V(r, \theta) = -\frac{Z_i^2 \qe^2 a_{cx}^3}{2r^2(r^2-a_{cx}^2)} + \frac{Z_i \qe \mu \cos \theta}{r^2}.
\eeq
The considerations that lead to the third constant of motion $\mathcal A$ hold again.
The energy of the trajectory is
\beq
\frac{1}{2} m \dot r^2 + V_{\rm eff}(r) = E
\eeq
where $V_{\rm eff}(r)$ is given by
\beq
V_{\rm eff}(r) \equiv - \frac{Z_i^2 \qe^2 a^3}{2 r^2(r^2-a^2)} + \frac{m
(b v)^2 + 2 Z_i \qe \mu \cos \theta_{\infty}}{2 r^2}
\eeq
Following DL98b, we define the dimensionless parameter
\beq
\phi^2 \equiv \frac{2 Z_i^2 \qe^2}{a_{cx} k T}, \label{def : phi}
\eeq
which describes whether the image charge attraction dominates over the thermal energy ($\phi\gg 1$) or the thermal energy dominates ($\phi\ll 1$). The effective potential can be written
\beq
V_{\rm eff}(r) = k T \left[ - \frac{\phi^2 a_{cx}^4}{4 r^2(r^2 - a_{cx}^2)} + \left( u^2 c^2 + \tilde \mu \cos \theta_{\infty} \right) 
\left(\frac{a_{cx}}{r}\right)^2 \right]
\eeq
where $\tilde \mu \ , \ c$, and $u$ were defined in equations (\ref{psi-mutilde}) and (\ref{c-u}).\\
A study of this potential leads to the following condition for collision :
\beq
\cos \theta_{\infty} < X_{\max}(c, u) \equiv {\tilde \mu}^{-1}\left(u^2 -u^2 c^2 + \phi u \right) \label{Xmax neutral}
\eeq
The collision and excitation rates are obtained as in Eq. (\ref{def : X}), (\ref{collision rate}) and (\ref{excitation rate}). One can then obtain the normalized damping and excitation rates for collisions of ions with a neutral grain:
\barr
F_i(Z_g = 0) &=& \frac{n_i}{n_H}\sqrt{\frac{m_i}{m_{\rm H}}} \ h_1(\phi, \tilde \mu) \\
G_i^{(ev)}(Z_g = 0) &=& \frac{T_{ev}}{2 T} F_i(Z_g = 0)\\
G_i^{(in)}(Z_g = 0) &=& \frac{n_i}{2 n_H}\sqrt{\frac{m_i}{m_{\rm H}}} \ h_2(\phi, \tilde \mu)   \label{eq : Gi neutral}
\earr
where we have defined
\barr
h_1(\phi, \tilde \mu) &\equiv&  \frac{1}{2} + \frac{\tilde \mu}{4} + \frac{2 + \phi^2}{4 \tilde \mu}\left(1- \rme^{-u_0^2}\right)- \frac{\phi u_0}{4 \tilde \mu} \rme^{-u_0^2}\nonumber\\ && + \frac{\pi^{1/2} \phi}{2}\left(1+ \frac{3 - 2\tilde \mu}{4 \tilde \mu} \erf u_0\right)  \label{h1}\\
h_2(\phi, \tilde \mu) &\equiv& \frac{1}{2} + \frac{3 \pi^{1/2}}{4}\ \phi + \frac{\phi^2}{4} + \frac{\tilde \mu^2}{12} + \frac{\tilde \mu}{4}  \nonumber\\
&& + \frac{1 + \phi^2}{2 \tilde \mu}(1 - \rme^{-u_0^2}) + \frac{2 \tilde \mu \phi^2 + \phi(2 \tilde \mu -7)u_0}{16 \tilde \mu} \rme^{-u_0^2}\nonumber\\
&& + \frac{\pi^{1/2}\phi}{32  \tilde \mu}\left(4 \tilde \mu^2 - 12 \tilde \mu +15 + 2\phi^2 \right) \erf u_0 \label{h2}\\
u_0 &\equiv& \frac{- \phi + \sqrt{\phi^2 + 4 \tilde \mu}}{2} 
\earr
Note that in the limit $\tilde \mu \rightarrow 0$ we recover DL98b result, as 
\barr
h_1(\phi, \tilde \mu \rightarrow 0) &=& 1 + \frac{\pi^{1/2}}{2} \phi + \mathcal O(\tilde \mu^3)\\
h_2(\phi, \tilde \mu \rightarrow 0) &=& 1 + \frac{3 \pi^{1/2}}{4}\ \phi + \frac{\phi^2}{2} +  \mathcal O(\tilde \mu^2)
\earr
However, the parameter $\tilde \mu$ is not small in general, as we saw in Eq. (\ref{mu-numerics}), so the effect of the dipole moment on the trajectory cannot be neglected. Note that we also have $ \phi \approx 18 \ a_{-7}^{-1/2}T_2^{-1/2}$.
The net effect of the dipole moment is to increase the collision and excitation rates, as can be seen from Fig.~\ref{figure : FGi neutral}. In contrast to the case of charged grains, the electric dipole moment does have a significant effect and cannot be discarded. \\

The effect of the dipole moment is always to increase both the collision and excitation rates, for both charged and neutral grains. This can be understood as follows. When the dipole moment vanishes, ions with a given velocity at infinity $v$ collide with the grain if their impact parameter is such that $ b < b_{\max}(v)$. The effect of the dipole moment is to make a smooth transition from non-colliding to colliding trajectories : all ions with impact parameter $ b < b_1(v)$ collide with the grain, a fraction $(X(b,v)+1)/2$ of those for which $b_1(v) < b < b_2(v)$ do collide, and none of the ions with $b > b_2(v)$ collide. $b_1$ and $b_2$ are such that $b_1 < b_{\max} < b_2$. As a result, a fraction of trajectories for which $b_1(v) < b < b_{\max}(v)$ do not lead to collision anymore (compared to the vanishing dipole case), and a fraction of trajectories $b_{\max}(v) < b < b_2(v)$ now lead to collision. The suppressed colliding trajectories have a lower rate of collision and angular momentum than the added colliding trajectories. Thus the net effect of the dipole moment is to increase the collision and rotational excitation rates.

\section{Plasma drag}

DL98b computed the effect of torques from passing ions on the electric dipole moments of the dust grains, which they named ``plasma drag''. They computed this effect for straight-line trajectories (the 
``Born approximation'').  Here we include the full hyperbolic trajectory in the case of charged grains. We also account for the rotation of the grain explicitly. Moreover, we do not include trajectories 
leading to collisions, as they will give away their entire angular momentum through collision, which we already accounted for. A precise calculation is important because plasma drag is one of the major 
excitation processes in some environments.  Note that treatments of the plasma drag effect that treat the plasma as a linear dielectric medium \citep{ragot} with the drag due to the imaginary part of the 
dielectric constant $\Im \epsilon(\omega,k)$ implicitly assume the Born approximation and do not capture the effects considered here.

We will find that the straight-line approximation usually overestimates the plasma drag.  In the case of positively charged grains, there is a range of impact parameters where the ion trajectory is 
deflected away from the grain, thereby suppressing angular momentum transfer.  For negatively charged grains, ions can be focused by electrostatic attraction.  \citet{AW93} 
argued that this is not a significant correction because the increased torque during close approach balances the shorter interaction time since the ion gains kinetic energy as it is attracted to the 
grain; however, we will see that in these cases there is a cancellation of angular momentum transfer in different parts of the trajectory that leads to reduced drag.  For very special cases, the grain 
can corotate with the ion during close approach leading to an enhancement of the plasma drag, but this occurs for only a narrow range of impact parameters and does not compensate for the reduction of 
plasma drag that we find in other regimes.

As in DL98b, we find it easiest to directly compute the plasma excitation $G_p$ and use the fluctuation-dissipation theorem to infer $F_p=G_p$.

\subsection{Charged grain}

We consider the trajectories of positively charged ion (charge
$Z_i > 0$) in the electric potential of a charged dust grain
(charge $Z_g \neq 0$). The trajectories are not strictly hyperbolic due to the presence of the electric dipole potential (see Section \ref{section : collisions ions charged grain}). However, we saw that it has little influence on collisions and we will neglect its effect on the trajectory here, assuming they are hyperbolic and determined by the Coulomb potential only. The eccentricity of the hyperbolic trajectory of the ion will be denoted $e$ (as opposed to the elementary charge $\qe$).

Let the ion trajectory (a hyperbola) be in the $(\hat{\bmath e}_y, \hat{\bmath e}_z)$ plane,
symmetric about the $\hat{\bmath e}_y$ axis. The ion position is given by
\beq
{\bmath r} = r\hat{\bmath e}_r = r(\alpha)\left(\cos\alpha\,\hat{\bmath e}_y + \sin\alpha\,\hat{\bmath e}_z\right).
\eeq
The hyperbolic trajectory of impact parameter $b$ and velocity at infinity $v$ can be described in polar coordinates as
\barr
r(\alpha) = \frac{p}{e \cos \alpha - 1} &  \alpha \in (-\alpha_e, \alpha_e) & (Z_g > 0), \nonumber\\
r(\alpha) = \frac{p}{1- e \cos \alpha} & \alpha \in (\alpha_e, 2\pi-\alpha_e) & (Z_g < 0);
\label{eq:ralpha}
\earr
the eccentricity and semilatus rectum\footnote{The positive and negative cases of Eq.~(\ref{eq:ralpha}) could have been unified by taking the negative branches of the square root in Eq.~(\ref{eq:e}), 
however we have not taken this route.} of the trajectory are
\beq
e = \sqrt{1 + \left(\frac{m_i b v^2}{Z_i Z_g \qe^2} \right)^2} {\rm ~~~and~~~}  p = b \sqrt{e^2 - 1}.
\label{eq:e}
\eeq
The range of longitudes $\alpha$ of the trajectory are determined by the limiting case
\beq
\alpha_e \equiv \arccos\frac1e.
\eeq
The longitude can be related to the true anomaly $f$ familiar from planetary dynamics by $\alpha=f$ for repulsive ($Z_g>0$) cases and $\alpha=\pi+f$ for attractive ($Z_g<0$) cases.
We will need the following expression for the time $t(\alpha)$, valid in both cases (for the case of an attractive potential, see e.g. \citet{orbits}, Eq. (2.4.12)):
\barr
t(\alpha) &=& \frac{b}{v} \frac{1}{e+1}
\Bigg[ \sqrt{\frac{e+1}{e-1}}\ln \left|\frac{\tan
\frac{\alpha}{2} + \sqrt{\frac{e-1}{e+1}}}{\tan
\frac{\alpha}{2} - \sqrt{\frac{e-1}{e+1}}}\right|
\nonumber \\ &&
-\frac{2 e \tan \frac{\alpha}{2}}{\tan^2 \frac{\alpha}{2} -
\frac{e-1}{e+1}} \Bigg].
\earr

In order to characterize the torque on the grain, we must first take the unit vector in the direction of grain rotation,
\beq
\hat{\bmath e}_\omega = \sin\theta\cos\phi\,\hat{\bmath e}_x + \sin\theta\sin\phi\,\hat{\bmath e}_y + \cos\theta\,\hat{\bmath e}_z,
\eeq
so that $\bomega = \omega\hat{\bmath e}_\omega$.  We use $(\theta,\phi)$ to parameterize the (general) direction of rotation.  We define the other two axes:
\barr
\hat{\bmath e}_\theta &=& \cos\theta\cos\phi\,\hat{\bmath e}_x + \cos\theta\sin\phi\,\hat{\bmath e}_y - \sin\theta\,\hat{\bmath e}_z
{\rm ~~~and}\nonumber \\
\hat{\bmath e}_\phi &=& -\sin\phi\,\hat{\bmath e}_x + \cos\phi\,\hat{\bmath e}_y.
\earr
In this system the electric dipole moment of the grain is
\beq
\bmu = \mu_\parallel \hat{\bmath e}_{\omega} + \mu_{\bot} \left[
\cos(\omega t + \chi) \hat{\bmath e}_{\theta} + \sin(\omega t +
\chi) \hat{\bmath e}_{\phi} \right],
\eeq
where $t=0$ is taken to be the time when the ion is at the
closest approach (i.e. ${\bmath r}\parallel\hat{\bmath e}_y$) and $\chi \in
[0, 2\pi)$ is the random angle that $\bmu_{\bot}$ makes with $\hat{\bmath e}_{\theta}$ at that time.

The ion electric field exerts a torque on the grain dipole moment:
\beq
I \frac{\rmd\bomega }{\rmd t} =
\bmu \times {\bmath E}  = - I \frac{Z_i \qe}{r^2} \bmu \times \hat{\bmath e}_r.
\eeq
Using the conservation of angular momentum, $r^2 \dot \alpha = bv$, we can rewrite:
\beq \frac{\rmd\bomega}{\rmd \alpha} = - \frac{Z_i \qe}{I b v} \bmu \times
\hat{\bmath e}_r.
\eeq
We project that along the direction of $\hat{\bmath e}_\omega$:
\barr
\frac{\rmd\omega_{\parallel} }{\rmd \alpha}  &=& - \frac{Z_i \qe \mu_{\bot}}{I b v} \big[ \cos(\omega t + \chi) \cos \alpha \cos \phi
\nonumber \\ &&
- \sin(\omega t + \chi)\, (\cos \alpha \cos \theta \sin \phi - \sin \alpha \sin \theta ) \big].
\earr
Expanding the sines and cosines, we integrate over the trajectory.  We keep only the parts of the integral for which the inbound and outbound parts do not cancel, i.e. those which are even under 
$\alpha\rightarrow-\alpha$ ($Z_g>0$) or $\alpha\rightarrow2\pi-\alpha$ ($Z_g<0$); note that $t(\alpha)$ is even.  We are then left with
\barr
\delta \omega_\parallel &=& \frac{Z_i \qe \mu_{\bot}}{I b v}
\Big[(\sin
\chi \cos \theta \sin \phi - \cos \chi \cos \phi)
\nonumber \\ && \times
 \int \cos
\omega t  \cos \alpha \rmd \alpha
\nonumber \\ &&
- \cos \chi \sin \theta \int
\sin \omega t \sin \alpha  \rmd \alpha \Big].
\earr

In order to find the plasma excitation coefficient, we need to sum $\delta \omega_\parallel^2$ over collisions.
We begin by averaging $\delta \omega_\parallel^2$ over solid angles for
$(\theta, \phi)$ and over angles for $\chi$. The result is
\beq
\langle\delta \omega_\parallel^2\rangle = \frac{1}{3}
\left( \frac{2 Z_i \qe \mu_{\bot}}{I b v} \right)^2 \mathcal I\left( \frac{\omega b}{v}, e, Z_g \right). \label{delta omega^2}
\eeq
We have defined the integral
\barr
\mathcal I\left( \frac{\omega b}{v}, e, Z_g \right) &\equiv&
\left( \int \cos \omega t \cos \alpha
\rmd \alpha \right)^2
\nonumber \\ &&
 + \left(\int \sin \omega t \sin \alpha \rmd\alpha \right)^2,
\earr
where the integration limits are given by $0<\alpha<\alpha_e$ ($Z_g>0$) or $\alpha_e<\alpha<\pi$ ($Z_g<0$).  Note that $\mathcal I$ only integrates over the inbound part of the trajectory; the outbound 
part is equal by symmetry.

The excitation rate due to plasma drag is then given by:
\barr
\frac{\rmd \Delta\omega_\parallel^2}{ \rmd t} &=& \int_0^{\infty} \rmd v
\int_{b_{\max}(v)}^{\infty} 2 \pi b \,\rmd b\,n_i \, 4 \pi v^3
\nonumber \\ &&
 \times \left(\frac{m_i}{2 \pi k T} \right)^{3/2} \rme^{-m_i v^2/2 k T}
 \frac{1}{3}
\left( \frac{2 Z_i \qe
\mu_{\bot}}{I b v} \right)^2
\nonumber \\ && \times
 \mathcal I\left( \frac{\omega b}{v}, e, Z_g \right),
\earr
where $b_{\max}(v)$, the maximum impact parameter for collision to occur, is defined as  
\beq
b_{\max}(v) = \left\{ \begin{array}{ll} 0  & m v^2/2kT \le \psi \\
a_{cx} \sqrt{1 - (2 k T/mv^2)\psi} & m v^2/2kT > \psi \end{array} \right.
\eeq
where $\psi = Z_g Z_i \qe^2/a_{cx}kT$.\\ 
Note that technically the integration over impact parameters should stop at the Debye length 
\beq
\lambda_D = \sqrt{\frac{k T}{4\pi n_e \qe^2}} \approx 398 \left( \frac{T_2}{n_e/0.03 \cm^{-3}} \right)^{1/2}
\eeq 
However, we will see below that the integrand vanishes exponentially for 
\beq
b > v / \omega \approx 4.5 \times 10^{-6} a_{-7}^{5/2} \sqrt{\frac{m_{\rm H}}{m_i}} \frac{v}{v_{\rmn{th}}} \frac{\omega_{\rmn{th}}}{\omega} \ \cm
\eeq
which is much smaller than the Debye length.\\
Converting this into an excitation coefficient, we find
\beq
G_p = \frac{n_i}{n_{\rm H}} \sqrt{\frac{m_i}{m_{\rm H}}} \left( \frac{Z_i \qe \mu_{\bot}}{a_{cx}^2 k T} \right)^2 
\times g_p\left(\psi, \sqrt{\frac{m_i a_{cx}^2}{2 k T}} \omega \right) 
\eeq
where
\beq
g_p \left(\psi, \Omega \right) \equiv 
\int_0^{\infty} 2 u \rme^{-u^2} \rmd u \
\int_{\frac{b_{\rm max}}{a_{cx}}}^{\infty} \frac{\rmd c}{c} \,\mathcal I\left(\frac{\Omega c}{u}, e, Z_g \right), \label{Gp}
\eeq
where the eccentricity is given by 
\beq
e = \sqrt{1 + \left(\frac{2 c u^2}{\psi}\right)^2} .
\eeq
Note that we recover DL98b result\footnote{DL98b include a term due to the parallel component of $\bmu$ which 
is not relevant as it only leads to excitation perpendicular to $\bomega$.} in the limit $\mathcal I = 1$.

This expression has to be averaged over the grain charge and summed over all present ions.

\subsubsection*{Straight line limit for $\mathcal I$}

In the limit $e \rightarrow \infty$, it is easier to express the integrals as a function of time, using
\barr
\cos \alpha &=& \frac{y}{\sqrt{y^2 + z^2}} = \frac{b}{\sqrt{b^2 + (v t)^2}}, \nonumber\\
\sin \alpha &=& \frac{v \ t}{\sqrt{b^2 + (v t)^2}}, {\rm ~~and}\nonumber\\
\rmd\alpha &=& \frac{1}{1 + \left(\frac{v t}{b} \right)^2} \frac{v }{b} \rmd t.
\earr
In this case, the first integral for $\mathcal I$ reduces to
\barr
\int\cos\omega t\cos\alpha\,\rmd\alpha &=& \frac vb\int_0^\infty\frac{ \cos\omega t\,\rmd t}{[1+(vt/b)^2]^{3/2}}
\nonumber \\ &=& \frac{\omega b}vK_1\left(\frac{\omega b}v \right),
\earr
where $K_1$ is a modified Bessel function of the second kind.  [Here we used Eq.~(9.6.25) of \citet{as} with $\nu=1$, $z=1$, and $x=\omega b/v$.]
The other integral is
\barr
\int\sin\omega t\sin\alpha\,\rmd\alpha &=&
\frac vb\int_0^\infty \frac{vt}b \frac{ \sin\omega t}{[1+(vt/b)^2]^{3/2}}\rmd t
\nonumber \\
&=& \int_0^\infty \frac{\tau\sin x\tau\,\rmd\tau}{(1+\tau^2)^{3/2}},
\earr
where $x=\omega b/v$.  Since $\tau(1+\tau^2)^{-3/2}$ is the derivative of $-(1+\tau^2)^{-1/2}$, we can integrate by parts and find
\beq
\left.\frac{-\sin x\tau}{\sqrt{1+\tau^2}}\right|_0^\infty + \int_0^\infty \frac{x\cos x\tau\,\rmd\tau}{\sqrt{1+\tau^2}}.
\eeq
The boundary terms evaluate to zero, and the second integral can again be evaluated to $xK_0(x)$ using Eq.~(9.6.25) of \citet{as} with $\nu=0$ and $z=1$.
Thus we have
\beq
\mathcal I
= x^2[K_0^2(x)+K_1^2(x)], {\rm ~~~~} x=\frac{\omega b}v. \label{straight line}
\eeq
Note that when $\omega \rightarrow 0$ we recover DL98b result, i.e.
\beq
\mathcal I \left( \frac{\omega b}{v} = 0, e\rightarrow \infty , Z_g \right) =1.
\eeq
We moreover have an exact functional shape for the cutoff at large rotation rates.

\subsubsection*{Non rotating grain limit for $\mathcal I$}

It is straightforward to show that
\beq
\mathcal I \left(\frac{\omega b}{v} = 0, e \right) = 1 - \frac{1}{e^2}
\eeq
for both positively and negatively charged grains. Thus, the nearly parabolic trajectories $e-1 \ll 1$ are suppressed by a factor $\sim 2 (e-1)$.\\

\begin{figure*}
  \includegraphics[width = 88mm]{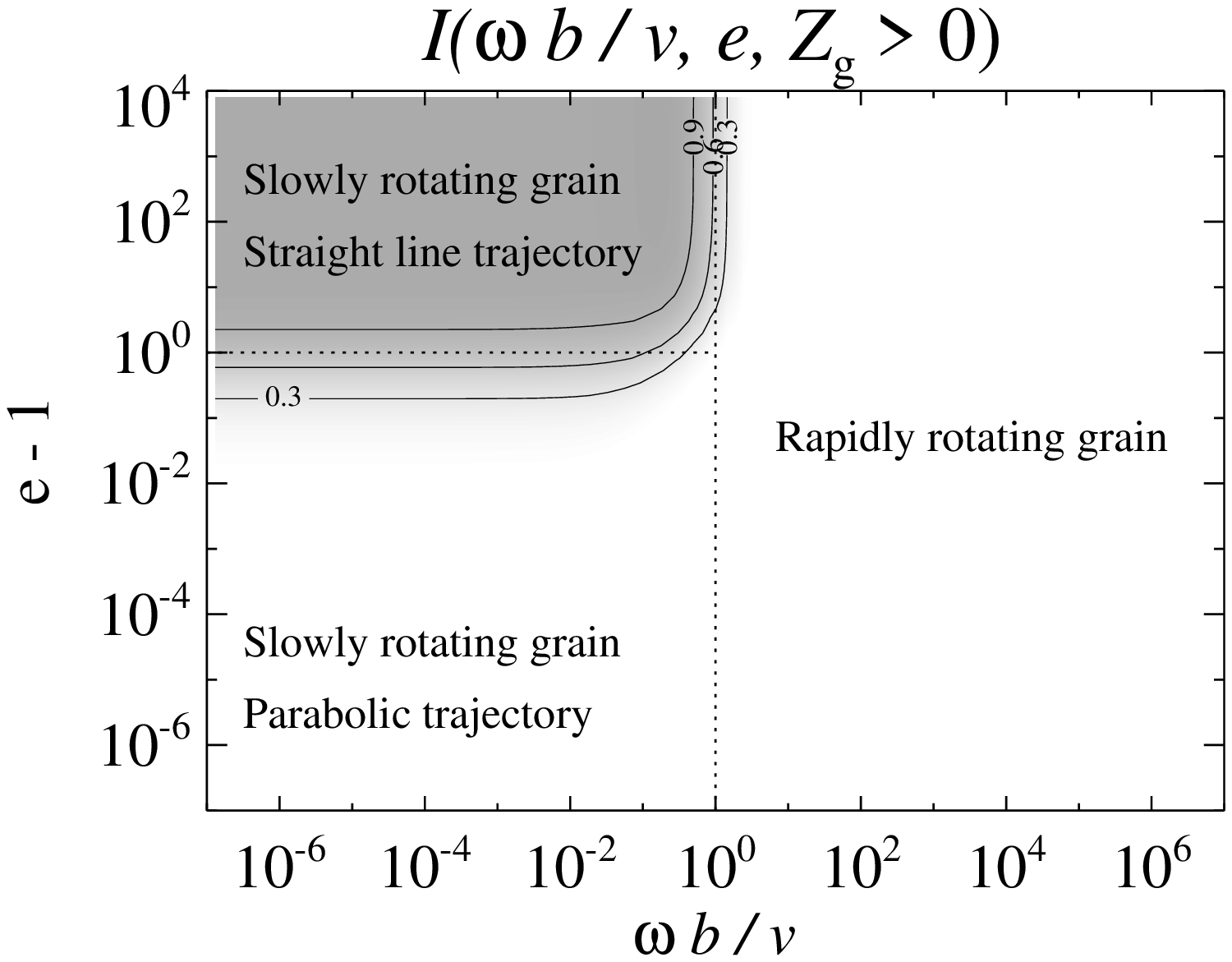}
  \includegraphics[width = 88mm]{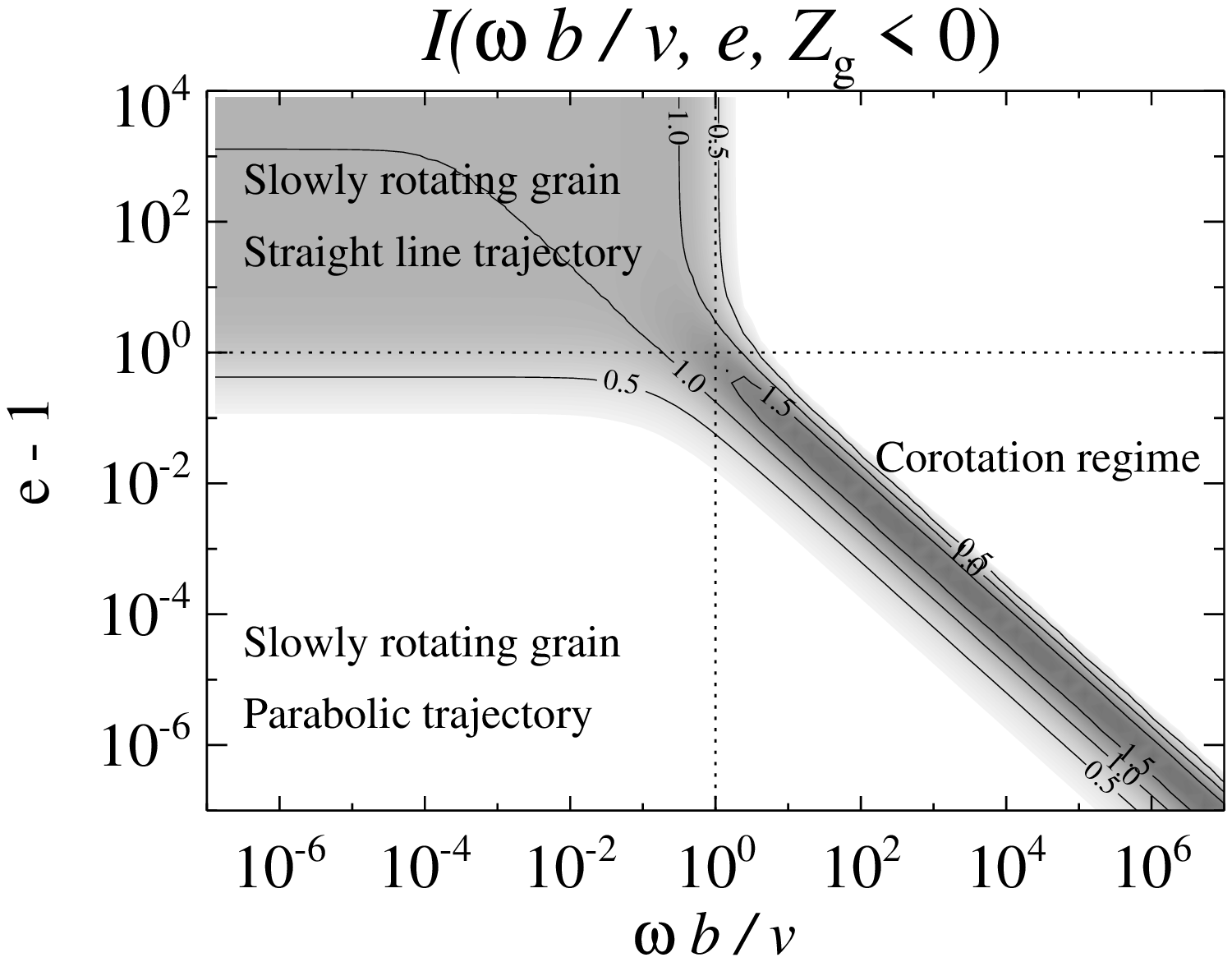}
  \caption{Contour levels of $\mathcal I(\frac{\omega b}{v}, e, Z_g > 0)$ (left panel) and $\mathcal I(\frac{\omega b}{v}, e, Z_g < 0)$ (right panel). Both show that $\mathcal I$ goes to unity for slowly rotating grain, straightline trajectories, and vanishes for rapidly rotating grains or nearly parabolic trajectory. In the case of negatively charged grains, though, there is a visible corotation regime, where $e-1 \ll 1$ and $\omega b / v (e-1) \sim 1$, for which the ion and the grain approximately corotate, enhancing the torque given to the grain.}
  \label{figure : Iplasma}
\end{figure*}

The numerical calculation of $\mathcal I$ in the general case is tricky because it involves integrating an oscillating function which frequency goes to infinity at one limit of the integral, as $t(\alpha \rightarrow 
\alpha_e) \rightarrow \infty$. We refer the reader to Appendix~\ref{app:i} for the description of the method used for numerical computation. Fig.~\ref{figure : Iplasma} shows the resulting 
dimensionless 
torques. An important feature is that for negatively charged grains, ions with nearly parabolic trajectories may corotate with the grain which results in an enhanced torque.

\subsection{Neutral grain}

The exact calculation of the trajectory in the electric dipole potential and the ``image charge potential'' is untractable analytically, and would require a heavy numerical calculation. Therefore, we will make the following simplifications. First, we neglect the effect of the electric dipole moment on the trajectory. This assumption is somewhat cavalier, as we saw previously that the electric dipole moment may significantly affect the ion trajectory in the case of a neutral grain. Furthermore, although trajectories in the ``image charge potential'' will be curved in general, we will consider them to be straight lines. Thus, we will approximate the torque given to the grain by Eq. (\ref{delta omega^2}), where $\mathcal I$ is given by Eq. (\ref{straight line}). Colliding trajectories should not be taken into account for the plasma drag excitation rate. Thus, we integrate the torque only over trajectories with impact parameter $b > b_{\max}(v)$, with 
\beq
b_{\max} = a_{cx} \sqrt{1 + \frac{\phi}{u}}
\eeq
(see DL98b Eq. (B24) and the definition of $\phi$ Eq. (\ref{def : phi})).\\

\begin{figure}
  \includegraphics[width = 88mm]{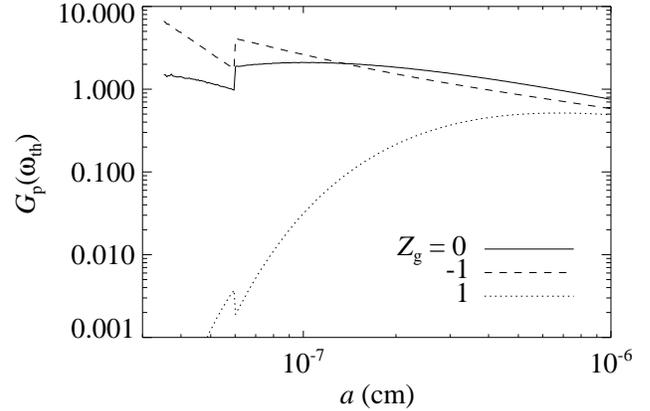}
  \caption{Normalized excitation rate due to plasma drag $G_p(\omega_{\rmn{th}})$ for a neutral grain, a positively charged grain, and a negatively charged grain in CNM conditions (Eq. (\ref{eq : CNM})), evaluated at 
the "thermal rotation rate" $\omega_{\rmn{th}} = \left(2 k T/I \right)^{1/2}$. All of them are lower than estimated by DL98b. It is clear that the positively charged grains are much less excited than the neutral and
negatively charged grains. The kink at $6 \Am$ is due to the change of grain shape.}
  \label{figure : Gp}
\end{figure}

Therefore, in the case of neutral grains, we have
\beq
G_p(Z_g = 0) = \frac{n_i}{n_{\rm H}} \sqrt{\frac{m_i}{m_{\rm H}}} \left( \frac{Z_i \qe \mu_{\bot}}{a_{cx}^2 k T} \right)^2 
\times \tilde{g_{p}}\left(\phi, \sqrt{\frac{m_i a_{cx}^2}{2 k T}} \omega \right) 
\eeq
where
\beq
\tilde{g_{p}} \left(\phi, \Omega \right) \equiv 
\int_0^{\infty} 2 u \rme^{-u^2} \rmd u \
\int_{\frac{b_{\rm max}}{a_{cx}}}^{\infty} \frac{\rmd c}{c} \,\mathcal I\left(\frac{\Omega c}{u}, e\rightarrow \infty \right) \label{Gp neutral}
\eeq
The normalized excitation rate for plasma drag for Cold Neutral Medium conditions, Eq. (\ref{eq : CNM}), is shown in Fig.~\ref{figure : Gp}.

\section{Infrared emission}
\label{s:ir}

A dust grain absorbs light and reemits it in the infrared. A
rotating grain will also
radiate angular momentum, which damps its rotation.

DL98b compute this damping rate by modelling the grain as composed
of six rotating dipoles. We give here a more accurate calculation,
using the correlation functions of the dipole moment in the grain
frame. Our result is a factor of two greater than that of DL98b.
We present a classical calculation in this section; the quantum calculation is presented in Appendix~\ref{app:q} and gives the same result.

The rates of emission of energy and of angular
momentum by a varying electric dipole moment:
\beq
\dot E = \frac{2}{3c^3} \ddot{\bmath p}^2 {\rm ~~~and~~~}
\dot{\bmath L} = \frac{2}{3c^3} \dot{\bmath p} \times
\ddot{\bmath p}.
\eeq
We denote the coordinates of the dipole moment in the frame
corotating with the grain with unprimed indices, and the ones in
the "lab frame" with primed indices. Take a grain rotating around
the $z$-axis, without precession, with angular frequency $\omega$. We have
\barr
{p_x}' &=& \cos \omega t\ p_x - \sin\omega t\ p_y, \nonumber \\
{p_y}' &=& \sin \omega t\ p_x + \cos\omega t\ p_y, {\rm ~~~and}\nonumber \\
{p_z}' &=& p_z.
\earr
A straightforward calculation leads to the following expressions in the lab frame :
\barr
{\ddot {\bmath p}}^{2}
&=& \ddot p_x^2 + \ddot p_y^2 + \ddot p_z^2 + 4 \omega\left(\dot p_x \ddot p_y - \dot p_y \ddot p_x \right)\nonumber\\
&&+ \omega^2\left[ 4\left(\dot p_x^2 + \dot p_y^2 \right) - 2 \left( p_x \ddot p_x + p_y \ddot p_y\right) \right]\nonumber\\
&&+ 4 \omega^3 \left(p_x \dot p_y - p_y \dot p_x \right) + \omega^4 \left( p_x^2 + p_y^2\right) \label{ddotp}
\earr
and
\barr
\left( \dot{\bmath p} \times \ddot{\bmath p} \right)_z
\!\!\!\!&=&\!\!\!\! \dot p_x \ddot p_y - \dot p_y \ddot p_x + \omega \left[ 2 \left(\dot p_x^2 +  \dot p_y^2 \right) - \left( p_x \ddot p_x + p_y \ddot p_y\right) \right]\nonumber \\
&&\!\!\!\! + 3 \omega^2 \left(p_x \dot p_y - p_y \dot p_x \right) + \omega^3 \left( p_x^2 + p_y^2\right).
\earr

Since we are interested in the statistical properties of the emission, we define the unequal-time dipole moment correlation function in grain coordinates,
\beq
C_{ij}(\tau) \equiv \left\langle(p_i(t)- \langle p_i\rangle) (p_j(t+\tau) - \langle p_j\rangle)\right\rangle,
\eeq
where $\langle p_i\rangle = \mu_i$ is just the constant dipole moment of the grain.
We further assume statistical spherical symmetry of the dipole moment in the grain coordinates, i.e. $C_{ij} = C\delta_{ij}$.
(For a planar grain, the values of the correlation functions depend on the in-plane or out-of-plane character of the vibrational modes and may be anisotropic.  However if the 
infrared emission arises during thermal spikes when the grain is not rotating around its axis of greatest angular momentum, we expect the isotropic analysis to be a good approximation.)
The average values of the previous formulae then become\footnote{Expectation values of derivatives such as $\langle \dot p_x^2\rangle$ can be expressed in terms of correlation functions via integration 
by parts.
In this case, $\langle \dot p_x^2\rangle = \langle (p_x\dot p_x)\dot{}\rangle - \langle p_x\ddot p_x\rangle$.  The first term vanishes for a stationary process, and the second is $-C''(0)$.}
\beq
\langle\ddot{\bmath{p}}^{2}\rangle = 3 C''''(0) -12 \omega^2 C''(0) + 2
\omega^4 C(0) \label{eqn:powerisot}
\eeq
and
\beq
\left\langle\dot{\bmath p} \times \ddot{\bmath p} \right\rangle_z = -6
\omega C''(0) + 2 \omega^3 C(0), \label{eqn:angular momisot}
\eeq
where $'$ denotes the derivative of the correlation function with respect to $\tau$.

The Wiener-Khintchine Theorem relates the correlation functions to
the Spectral density $S_{\nu}$,
$C(\tau) = \int_0^{\infty} S_{\nu} \cos(2 \pi \nu \tau) \rmd \nu$.
Plugging back into Eqs.~(\ref{eqn:powerisot}) and (\ref{eqn:angular
momisot}), we get
\beq
\langle\ddot{\bmath{p}}^{  2}\rangle = \int_0^{\infty} \left[ 3 (2\pi \nu)^4 + 12 \omega^2 (2 \pi \nu)^2 + 2 \omega^4 \right] S_{\nu} \,\rmd \nu
\eeq
and
\beq
\langle\dot{\bmath p} \times \ddot{\bmath p} \rangle_z =
\int_0^{\infty} \left[ 6 \omega(2\pi \nu)^2 + 2 \omega^3 \right]
S_{\nu}\, \rmd\nu.
\eeq
Now making use of the assumption that the grain rotates
slowly, i.e. that $\nu_{rot} \equiv \omega/2\pi \ll \nu_0 \equiv $ typical
frequency of emission, in the infrared, we get, at the
lowest order in $\nu_{rot} / \nu_0$, the
average total power and average total rate of radiation of angular
momentum:
\beq
\left\langle\frac{\rmd E}{\rmd t}\right\rangle = \frac{2}{3 c^3}
\langle\ddot{\vec{p}}^{  2}\rangle = \frac{2}{c^3} \int_0^{\infty}
(2\pi \nu)^4  S_{\nu} \,\rmd \nu
\eeq
and
\beq
\left\langle\frac{\rmd L_z}{\rmd t}\right\rangle= \frac{2}{3 c^3}
\left\langle\dot{\bmath p} \times \ddot{\bmath p} \right\rangle_z
=\frac{4 \omega}{c^3} \int_0^{\infty} (2\pi \nu)^2 S_{\nu}\,\rmd\nu.
\eeq

If one knows the infrared power radiated per steradian per frequency interval $F_{\nu}$, such that
\beq
\left\langle\frac{\rmd E}{\rmd t}\right\rangle = 4 \pi \int_0^{\infty}
F_{\nu} \, \rmd\nu,
\label{eq:y-power}
\eeq
one can deduce the rate of angular momentum loss through infrared emission:
\beq
\left\langle\frac{\rmd L_z}{\rmd t}\right\rangle = \frac{2 \omega}{\pi} \int_0^{\infty} \frac{F_{\nu}}{\nu^2} \, \rmd\nu.
\label{eq:y-dlz}
\eeq
This result is twice as big as the one given in DL98b.
[The difference occurs because DL98b modeled the dipole fluctuations with six uncorrelated rotating dipoles, one rotating each direction in the $xy$, $yz$, and $xz$ planes.  They counted the radiated 
power from 
all six of these, but only considered the angular momentum loss from two of them (in the $xy$ plane).  The dipoles rotating in the $xz$ and $yz$ planes containing the rotation axis also emit net angular 
momentum however, and if they are considered one recovers the factor of 2.]

This classical treatment does not predict the rotational excitation from the recoil given by individual photons, which is a quantum effect. As in DL98b, we set
\beq
\left\langle\frac{\rmd\Delta L^2}{\rmd t}\right\rangle = \frac{\rmd N_{\rmn{phot}}}{\rmd t}\ \hbar^2 = \frac{h}{\pi} \int_0^{\infty} \frac{F_{\nu}}{\nu} \,\rmd\nu.
\eeq

\begin{figure*}
  \includegraphics[width = 88mm]{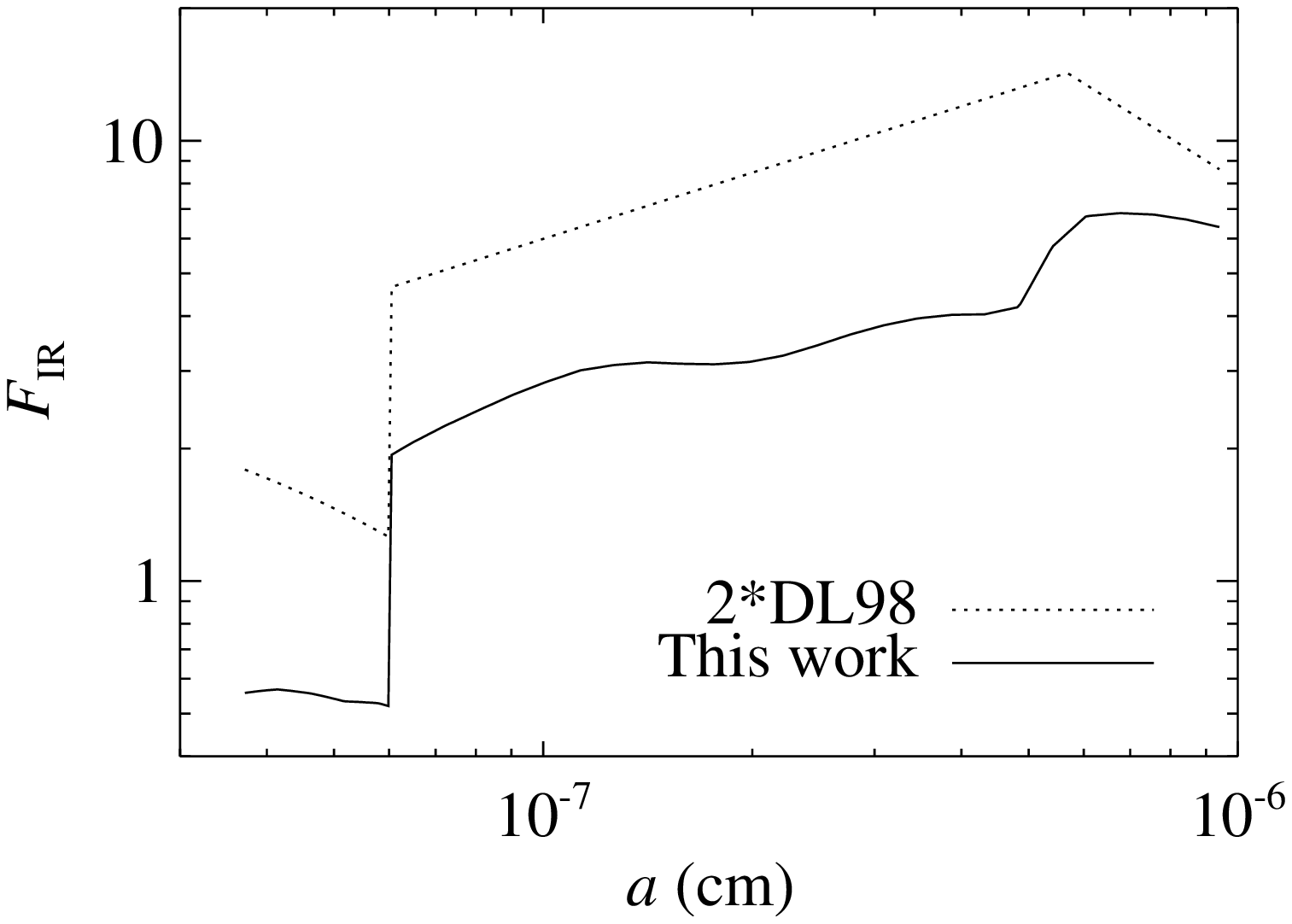}
  \includegraphics[width = 88mm]{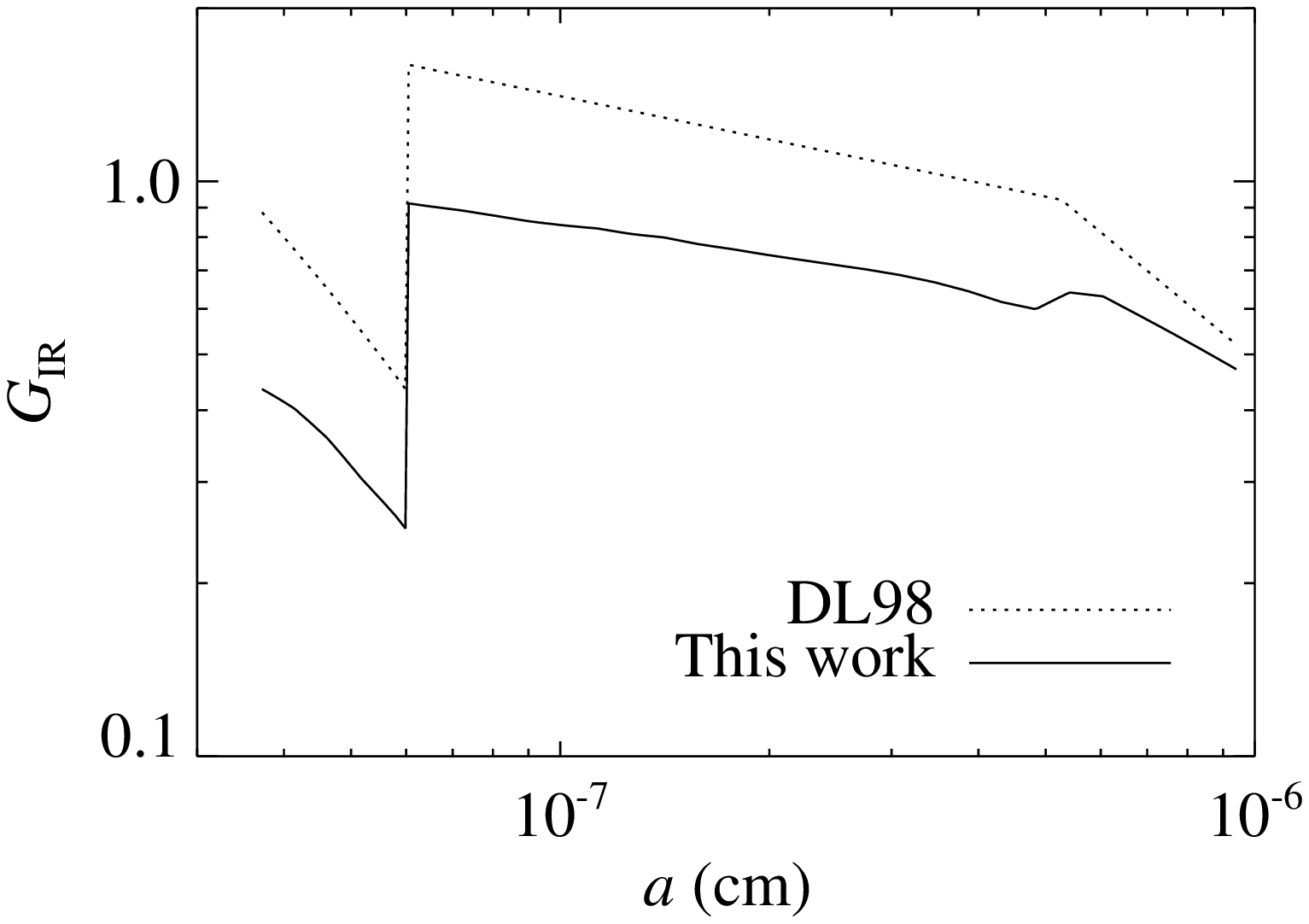}
  \caption{Infrared emission damping and excitation coefficients $F_{\rmn{IR}}, \ G_{\rmn{IR}}$, in CNM conditions (Eq. (\ref{eq : CNM})), compared with DL98b result. The difference is mainly due to differences in grain absorption efficiencies and the calculation of the infrared spectrum. We used Li $\&$ Draine (2001) absorption efficiencies and DL01 model to compute the infrared emissivity. The kink at $50 \ \Am$ in our result is due to a change in optical properties of dust grains. The kink around $50 \ \Am$ in DL98b result is due to the change from constant temperature limit (larger grains) to thermal spikes limit (smaller grains). The fact that they coincide is purely chance, and would not be necessarily the case for other environment conditions. The discontinuity at $6 \Am$ is due to the change in grain shape. }
  \label{figure : FGIR}
\end{figure*}

The normalized damping and excitation rates are then
\barr
F_{\rmn{IR}} &=& \frac{2 \tau_{\rmn H}}{\pi  I} \int_0^{\infty} \frac{F_{\nu}}{\nu^2} \ d\nu {\rm ~~and}\nonumber \\
G_{\rmn{IR}} &=& \frac{h }{6 \pi  I } \frac{\tau_{\rmn H}}{k T} \int_0^{\infty} \frac{F_{\nu}}{\nu} \, \rmd\nu.
\earr
We calculate the infrared emissivity of PAHs and small carbonaceous grains using the "thermal continuous" approximation, described in DL01. They indeed show that this treatment leads to spectra very 
close to those predicted by the exact statistical treatment, and has the advantage of being computationally much faster. We obtain the steady-state energy distribution function and then get the infrared 
emissivity, as explained in DL01.\\
We checked numerically that we recover DL98 result for low values of the radiation field intensity : $F_{\rmn{IR}}$, $G_{\rmn{IR}} \propto \chi$. However, their result for high values of the radiation field ($F_{\rmn{IR}} \propto \chi^{2/3}$, $G_{\rmn{IR}} \propto \chi^{5/6}$) relies on the fact that the absorption efficiency $Q_{\nu} \propto \nu^{2}$ at the characteristic frequencies of infrared emission. This is not valid anymore for high radiation fields, which offset the emission spectrum to higher frequencies, where the absorption efficiency has not a simple dependence on frequency anymore. We show the resulting infrared emission and damping coefficients in Fig. \ref{figure : FGIR}.

\section{Photoelectric emission}

An electron ejected from the grain carries away an angular momentum along the rotation axis (z-axis) equal to :
\beq
\Delta L_z = m_e \rho \left( v_{\phi}' - \rho \omega \right),
\eeq
where $v_{\phi}'$ is its tangential velocity in the grain frame.
From this we deduce that
\beq
F_{pe} = \frac{m_e}{m_{\rmn{H}}} \frac{J_{pe}}{2 \pi a_s^2 \nH \sqrt{{2 k T}/{\pi m_{\rmn{H}}}}},
\eeq
where $J_{pe}$ is the photoemission rate and was described in section \ref{s:charge}.
The excitation rate can be obtained by first noticing that the rotational velocity is much smaller than the velocity of ejected electrons:
\beq
a \omega \ll v_{\phi}'
\eeq
so that we have, up to small corrections
\beq
\Delta L_z^2 = m_e^2 \rho^2 v_{\phi}'^2.
\eeq
We assume a cosine-law directional distribution for escaping electrons, so that $\langle v_{\phi}'^2\rangle = \frac{1}{4}  v_e^2$, where we denote $v_e$
the average velocity of the electron at the grain surface. The latter satisfies
\beq
\frac{1}{2} m_e v_e^2 - \frac{(Z_g + 1) \qe^2}{a_s} = E_{pe},
\eeq
where $E_{pe}$ is the average energy at infinity of the photoejected electron.
We finally get
\beq
\langle \Delta L_z^2\rangle = m_e^2 \frac{2}{3} \frac{a_{cx}^4}{a_s^2} \frac{1}{4}v_e^2 = \frac{m_e}{3}  \frac{a_{cx}^4}{a_s^2}\left[E_{pe} + \frac{(Z_g + 1) \qe^2}{a_s} \right].
\eeq
So the normalized excitation rate is
\beq
G_{pe} = \frac{m_e}{4 \nH \left(8 \pi m_{\rmn H} k T \right)^{1/2} a_s^2 k T } \left[ \Gamma_{pe} + \frac{(Z_g + 1)\qe^2}{a_s} J_{pe} \right],
\eeq
where $\Gamma_{pe}$ is the heating rate due to photoemission of electrons, obtained from WD01b.

\section{Random H$_2$ formation} \label{H2}

DL98b showed that the random formation of H$_2$ molecules on the grain surface does not make a major contribution to rotational excitation. We use their result:
\beq
G_{\rmn{H}_2} = \frac{\gamma}{4}(1 - y)\frac{E_f}{k T} \left[ 1 + \frac{\langle J(J+1)\rangle \hbar^2}{2 m_{\rmn{H}} E_f a_x^2} \right],
\eeq
where $\gamma$ is the efficiency of H$_2$ formation, $y = 2 n_{\rmn H_2}/\nH$, $E_f \approx 0.2 \eV$ is the average translational kinetic energy of the nascent H$_2$, and $\langle J(J+1)\rangle \approx 10^2$ gives its average angular momentum.

\section{Resulting emissivity and effect of various parameters} \label{emissivity}

Throughout this section and unless otherwise stated, we will take as a fiducial environment the CNM parameters specified by
\barr
\nH & = & 30 \ \cm^{-3} \ \ \ , \ \ \ T = 100 \ \K  \nonumber \\
 x_{\rmn{H}} &\equiv& n(\rmn{H}^+)/\nH  =  10^{-3} \ \ \ , \ \ \ x_{\rmn{C}} \equiv n(\rmn{C}^+)/\nH = 3 \times 10^{-4} \nonumber \\
\chi &\equiv& u/u_{\rmn{ISRF}}= 1\ \ \ , \ \ \ \gamma = 0.
\label{eq : CNM}
\earr
We also take the rms intrinsic dipole moment to be 
\beq
\langle\mu_i^2\rangle^{1/2}\left(a = 10^{-7}{\rm cm}\right) = 9.3 \ \debye.
\eeq
For the size distribution parameters, we use those given by WD01a for a ratio of visual extinction to reddening $R_V = 3.1$, and a carbon abundance in 
the log-normal distributions $b_C = 6 \times 10^{-5}$.

This section is intended to give some intuition into the effect of various parameters on the spinning dust spectrum. However, the reader should keep in mind that environment parameter space is 
many-dimensional, and changing several parameters at once may lead to modifications that are not superpositions of the effects described here.

\subsection{General shape of the rotational distribution function}

\begin{figure}
  \includegraphics[width = 88mm]{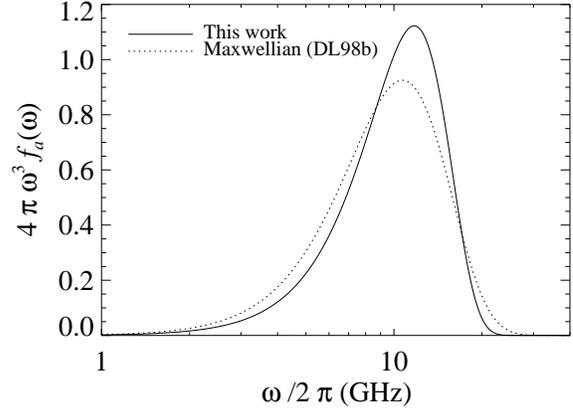}
  \caption{Rotational distribution function for a grain radius $ a = 7 \ \Am$, in CNM conditions, for a single value of the dipole moment $\mu_i(10^{-7} \cm) = 9.3 \ \debye$. The plot compares the solution of the Fokker-Planck equation with the DL98b Maxwellian approximation (DL98b Eq.~(57) used with our $F$, $G$). Note that DL98b prescription gives $(<\omega^2>)^{1/2} = 2 \pi \times 10.7 \ \rmn{GHz} $, which is in excellent agreement with the value we get, $(<\omega^2>)^{1/2} = 2 \pi \times 10.9 \ \rmn{GHz}$. However, the shape of the distribution function is significantly different.  }
  \label{figure : f_rot}
\end{figure}

The rotational distribution function is obtained as described in section \ref{Fokker}. We remind the reader that the Fokker-Planck equation is not stricltly valid for the smallest grains, for which impulsive torques are important. It however still describes their rotational distribution function with more accuracy than a simple Maxwellian. Moreover, DL98 showed that impulsive torques may be neglected for grain radii $ a \geq 7 \Am$. In Fig.~\ref{figure : f_rot}, we show that the rotational distribution function obtained by the Fokker-Planck equation differs significantly from a Maxwellian. It has a sharper cutoff at high frequencies due to the proper accounting for rotational damping through electric dipole radiation.

In what follows we will analyze the effect of various parameters on the spinning dust emissivity. As can be seen from equation (\ref{equation : f_a(omega)}) and the expressions derived next for the normalized damping and excitation rates, the rotational distribution function has complex dependencies on all grain and environment parameters. To get some intuition on the physics of spinning dust and the influence of each parameter, we will rely on a simplified expression for the rotational distribution function in the following sections:
\beq
f_a(\omega) \propto \exp\left(-\frac{F}{G}\frac{I \omega^2}{2 k T} - \frac{\tau_{\rmn H}}{3 G \tau_{\rmn{ed}}} \left(\frac{I \omega^2}{2 k T}\right)^2 \right) \label{equation : f_a approx}
\eeq
where we approximate the plasma drag excitation rate (which is in principle a function of $\omega$) by the constant
\beq 
G_p \approx G_p(\omega_{\rmn{th}}) \ \ \ , \ \ \omega_{\rmn{th}} \equiv \left(\frac{3 k T}{I}\right)^{1/2} \label{equation : Gp approx}
\eeq
In our analysis we will also neglect the charge-displacement induced dipole moment as it has a minor contribution. Of course the actual rotational distribution function and emissivity are computed using the exact equations developed in this paper.

For a given grain radius $a$ and intrinsic electric dipole moment $\mu_i$, the power radiated is $P_{\nu}\left(a ; \mu_i\right) \propto \nu^6 f_a(2 \pi \nu ; \mu_i)$. It is straightforward, from Eq. 
(\ref{equation : f_a approx}), to show that the peak frequency is given by
\beq
\nu_{\rmn{peak}} \approx \left(\frac{2}{1 + \sqrt{1 + \xi}} \frac{G}{F} \right)^{1/2} \frac{1}{2 \pi} \sqrt{\frac{6 k T}{I}},
\eeq
where we defined the parameter
\beq 
\xi \equiv  \frac{8 G}{F^2}\frac{\tau_{\rmn H}}{\tau_{\rmn{ed}}} \label{xi}
\eeq
which denotes the non-Maxwellian character of the distribution function.\\ 
For $\xi \ll 1$, the distribution is nearly Maxwellian, 
\beq
f_a(\omega)\propto \exp\left(-\frac{F}{G}\frac{I \omega^2}{2 k T}\right), 
\eeq
and the peak frequency is given by
\beq 
\nu_{\rmn{peak}} \approx \left(\frac{G}{F}\right)^{1/2}  \frac{1}{2 \pi}  \sqrt{ \frac{6 k T}{I}}   \ \ \  \ \ (\xi \ll 1).
\label{nu_peak, xi ll 1}
\eeq
Moreover, the total power emitted by a single grain\\ 
$j_a \propto  \mu^2 \int \omega^6 f_a(\omega) \rmd \omega$ has the following dependence:
\beq
j_a \propto \mu^2 \left(\frac{G}{F}\right)^2 T^2 \ \ \  \ \ (\xi \ll 1). \label{power, xi ll 1}
\eeq
For $\xi \gg 1$ the distribution is strongly non-Maxwellian\footnote{Interestingly, \citet{Erickson} had already obtained a result similar to Eq.~(\ref{f_a, xi gg 1}) with a Fokker-Planck equation.},
\beq
f_a(\omega)\propto \exp\left(- \frac{\tau_{\rmn H}}{3 G \tau_{\rmn{ed}}} \left(\frac{I \omega^2}{2 k T}\right)^2 \right), \label{f_a, xi gg 1}
\eeq
and the peak frequency is given by
\beq 
\nu_{\rmn{peak}} \approx \left(\frac{G \tau_{\rmn{ed}}}{2 \tau_{\rmn H}}\right)^{1/4} \frac{1}{2 \pi} \sqrt{ \frac{6 k T}{I}}    \ \ \ \ \ (\xi \gg 1). \label{nu_peak, xi gg 1}
\eeq
The total power is then given by
\beq
j_a \propto \mu^2 \ \frac{G \tau_{\rmn{ed}}}{\rmn{\tau_{\rmn H}}} \ T^2 \ \ \ \ \ (\xi \gg 1). \label{power, xi gg 1}
\eeq

\begin{figure}
  \includegraphics[width = 88mm]{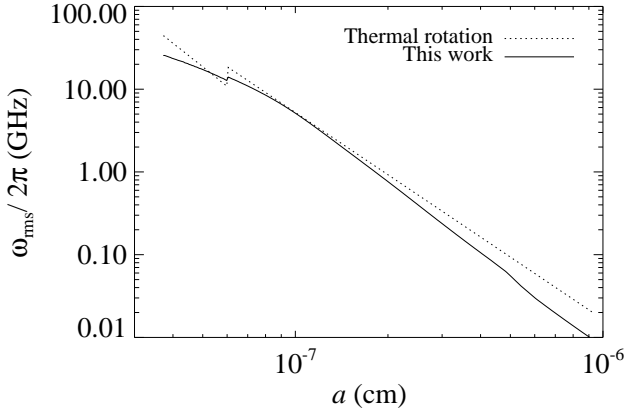}
  \caption{rms rotation rate $\omega_{\rmn{rms}} \equiv  <\omega^2>^{1/2} $ as a function of grain radius $a$, for CNM conditions. The rotation rate the grain would achieve if it were rotating thermally (in that case $\omega_{\rmn{rms}} = \sqrt{3 k T/I}$) is also shown. It can be seen that the grains rotate subthermally. The kink at $ 6 \ \Am$ is due to the change in grain geometry.} 
  \label{figure : omega_rms}
\end{figure}

In Fig.~\ref{figure : omega_rms} we show the rms rotation rate $ <\omega^2>^{1/2}$ as a function of grain radius. As can be expected, the smallest grains are rotating with the greatest angular velocity, as they have the smallest moment of inertia. Consequently, they radiate at the highest frequencies, and constitute the peak of the spectrum. Therefore, we will use Eqs. (\ref{equation : f_a approx}) to (\ref{power, xi gg 1}) for a grain of radius $a_{\rmn{min}} = 3.5 \ \Am$ to evaluate the effect of various parameters on the emissivity.  

We finally remind the reader with the dependencies of characteristic timescales :
\beq
\tau_{\rmn H} \propto \nH^{-1} T^{-1/2} \ \ \ , \ \ \ \tau_{\rmn{ed}} \propto \mu^{-2} T^{-1} \label{eq : tau_dependence}
\eeq

\subsection{Emissivity} \label{section : Emissivity}

Once the rotational distribution function is known, as a function of the intrinsic electric dipole moment, $f_a(\omega ; \mu_i)$, one can get the power radiated by a grain of radius $a$ by averaging over the intrinsic dipole moments gaussian distribution $P(\mu_i)$ defined in equation (\ref{Pmui}). One gets :
\beq
P_{\nu}(a) = \int d \mu_i P(\mu_i)  \ \frac{2}{3} \ \frac{\mu_{\bot}^2 \omega^6}{c^3} \ 2 \pi \ f_a(\omega ; \mu_i)
\eeq
where $\mu_{\bot}^2 = \frac23 \mu^2$ for spherical grains, and $\mu_{\bot}^2 =  \mu^2$ for cylindrical grains.

\begin{figure} 
  \includegraphics[width = 88mm]{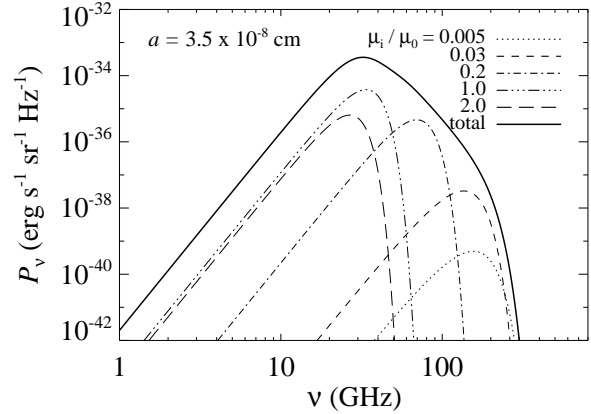}
  \caption{Power radiated by one grain of radius $ a = 3.5 \ \Am$ in CNM conditions. The dotted and dashed lines show the contributions of various values of the intrinsic dipole moment, which is assumed to have a gaussian distribution with rms value $ \ \  \ \ \ \  \ \ \ \ \ \ \mu_0 \equiv <\mu_i^2>^{1/2}(a)$. The solid line is the total power.}
  \label{figure : dipole_effect}
\end{figure}

The overall effect of averaging over the dipole moments distribution is to broaden the spectrum, as can be seen in Fig.~\ref{figure : dipole_effect}. The peak frequency remains approximately equal to 
that of $P_{\nu}(\mu_i = \langle\mu_i^2\rangle^{1/2})$. We will discuss the effect of the rms intrinsic dipole moment in section \ref{section : mu_effect}.

\begin{figure}
  \includegraphics[width = 88mm]{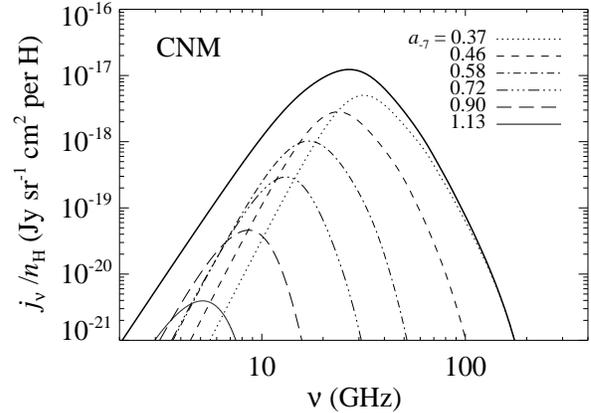}
  \caption{Spinning dust emissivity for CNM environment. Contributions from grains of various sizes are shown ($a_{-7} \equiv a/(10^{-7} \ \cm)$). The grain size distribution parameters are taken from WD01a with $R_V = 3.1$ and $b_c = 6 \times 10^{-5}$.}
  \label{figure : emissivity}
\end{figure}

The emissivity per H atom is then obtained by integrating the power radiated by each grain over the grain size distribution function, described in section \ref{size_dist}. The emissivity for the CNM environment is shown in figure \ref{figure : emissivity}. Note that the grain size distribution directly weights the spectrum, and thus needs to be known with accuracy, which is not quite the case yet for the very small grains.

\subsection{Effect of the rms intrinsic dipole moment $\langle\mu_i^2\rangle^{1/2}$} \label{section : mu_effect}

\begin{figure*}
  \includegraphics[width = 80mm]{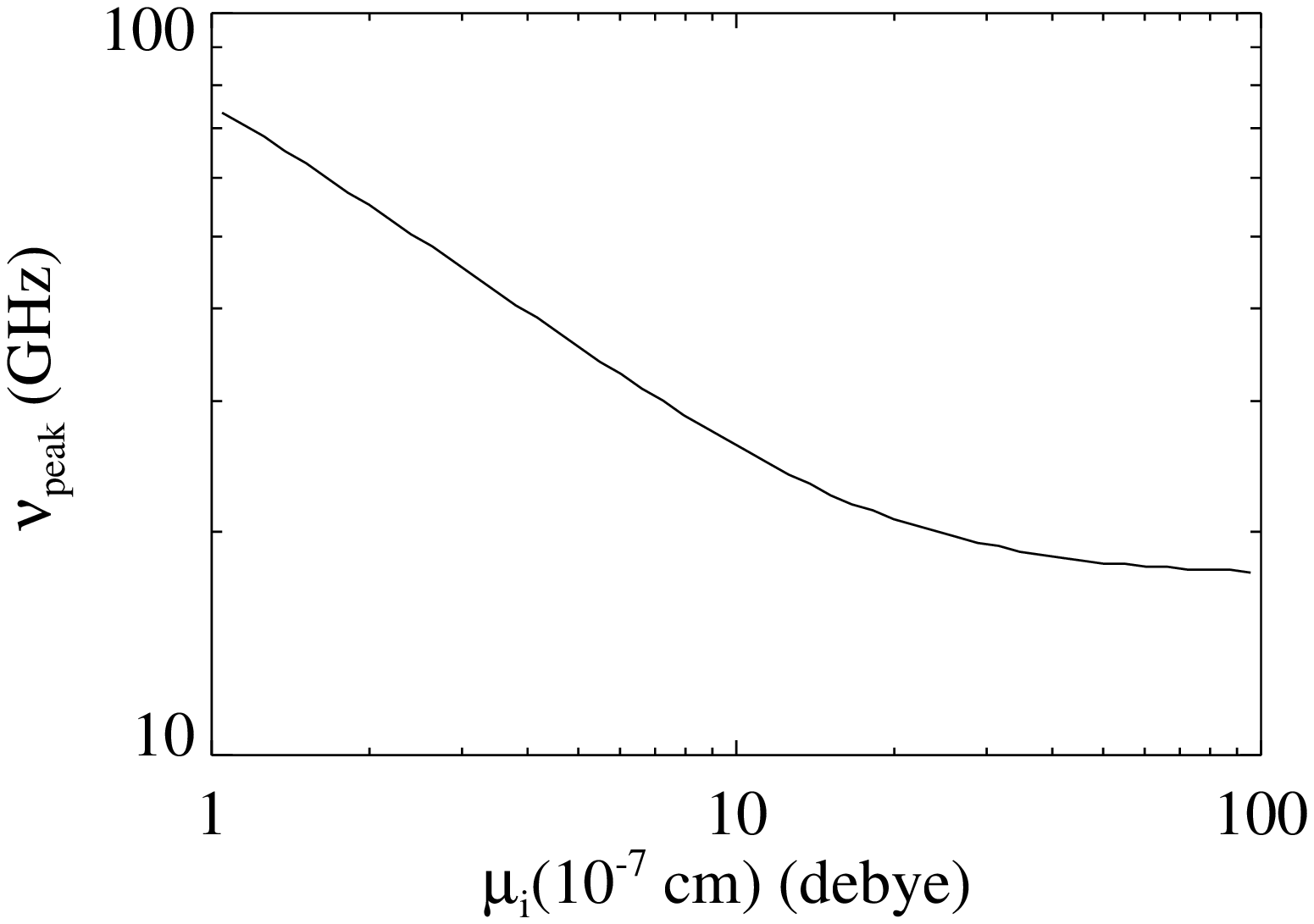}
  \includegraphics[width = 80mm]{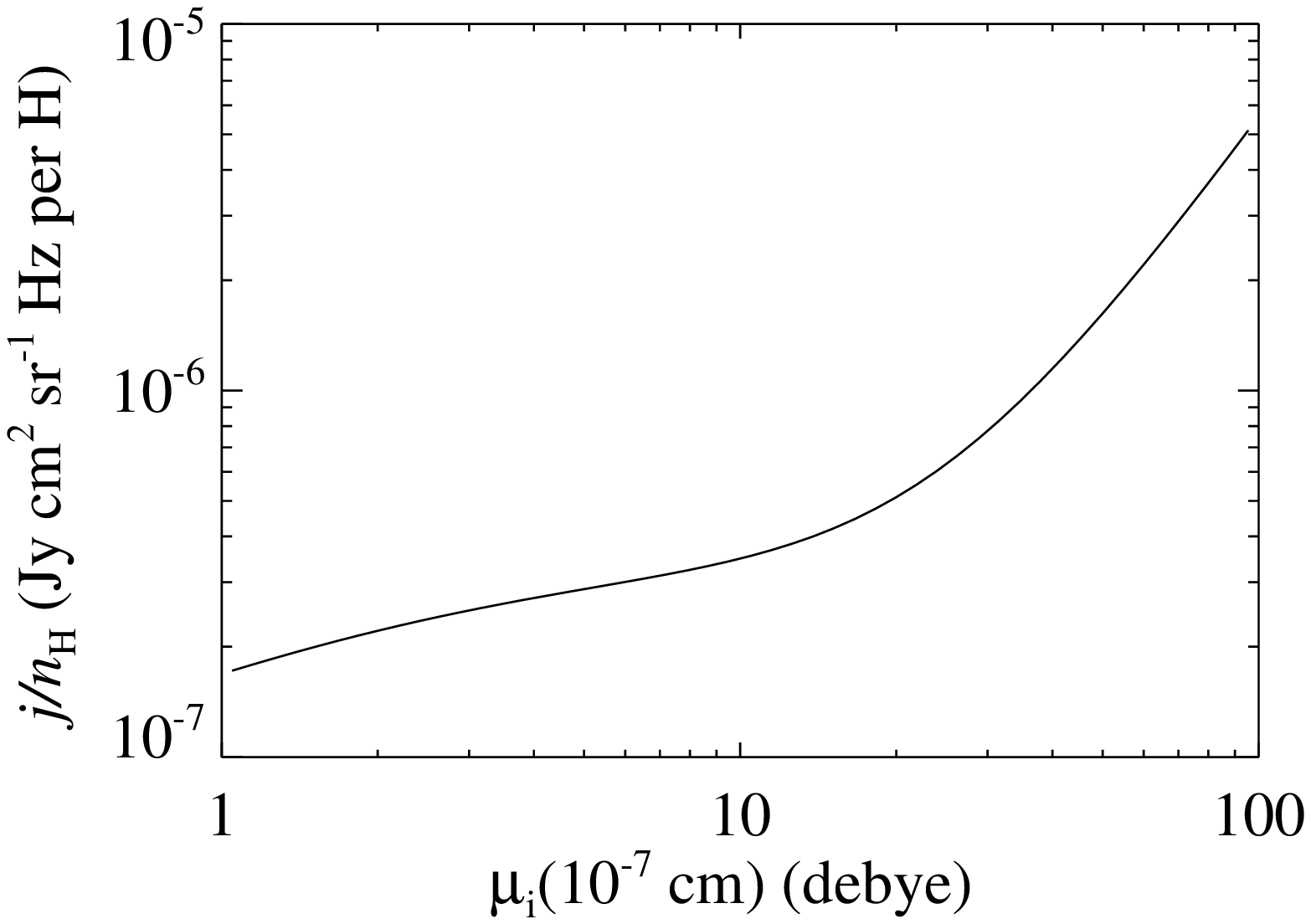}
  \caption{Effect of the intrinsic electric dipole moment on the peak frequency and the total spinning dust emission. Environment parameters are set to CNM conditions defined in equation (\ref{eq : CNM}). Increasing the electric dipole moment decreases the peak frequency and increases the total power radiated.}
  \label{figure : mu_effect}
\end{figure*}

Varying the rms intrinsic dipole moment affects the spectrum in three main ways. First, it affects the total power radiated, as $P_{\nu} \propto \mu^2$. Then, it affects the non-Maxwellian character of the distribution function, as $\tau_{\rmn{ed}} \propto \mu^2$. Finally, it affects the rotational damping and excitation rates essentially through plasma drag, which has $G_p \propto \mu^2$ (the effect on $G_i$ is not as important).\\
Throughout the range of values considered, 
\beq
1 \ \debye < \mu_i(10^{-7}\ \cm) < 100 \ \debye,
\eeq
and in CNM conditions, the distribution function remains strongly non-Maxwellian : $\xi \gtrsim 60$. Therefore, we can use the strongly non-Maxwellian limit Eq. (\ref{nu_peak, xi gg 1}) to evaluate the peak frequency.\\[5pt]
\emph{Low dipole moment limit} \\ 
For low values of the electric dipole moment, plasma drag has little effect on both the rotational damping and excitation. Therefore, $F$ and $G$ are roughly independent of $\mu$, and, from Eqs. (\ref{nu_peak, xi gg 1}), (\ref{power, xi gg 1}) and $\tau_{\rmn{ed}} \propto \mu^2$, we get
\beq
\nu_{\rmn{peak}}\left(\mu_i \rightarrow 0 \right) \propto \mu_i^{-1/2}  \label{nu_peak low mu}
\eeq
\beq
j/\nH \left(\mu_i \rightarrow 0 \right) \rightarrow \rmn{constant} \label{j low mu}
\eeq  
One can see in Fig.~\ref{figure : mu_effect} that Eq. (\ref{nu_peak low mu}) is quite accurately satisfied. The total power has a weak dependence on $\mu_i$ for low values of the intrinsic dipole moment, but is not strictly independent of it, which comes from the multiple approximations made in this analysis (neglecting the charge displacement-induced dipole moment, and using Eq. (\ref{power, xi gg 1}) for the total power, after integration over the size distribution, instead of the total power radiated by a single grain).\\[5pt]
\emph{High dipole moment limit} \\ 
For high values of the electric dipole moment, plasma drag dominates both rotational damping and excitation. Therefore, $G \approx G_p \propto \mu^2$ so we get
\beq
\nu_{\rmn{peak}}\left(\mu_i \rightarrow \infty \right) \rightarrow \rmn{constant}  \label{nu_peak high mu}
\eeq
\beq
j/\nH \left(\mu_i \rightarrow \infty \right) \propto \mu_i^2 \label{j high mu},
\eeq  
which describe approximately the behavior observed in Fig.~\ref{figure : mu_effect}.

\subsection{Effect of the number density $\nH$} \label{section nH}

The main effects of the number density are :
\begin{itemize}
\item
Changing the relative contribution of gas-induced and radiation-induced rotational damping and excitation. For very low number densities, $F_{\rmn{IR}}$ and $G_{\rmn{IR}} \propto \tau_{\rmn H} 
\propto\nH^{-1}$ dominate over other rotational damping and excitation rates. For high densities, plasma drag and collisions are dominant. Note that the charge distribution is also modified as the 
higher the density, the more important is collisional charging compared to photoemission. As a consequence, the grains are positively charged at low densities, and tend to be negatively charged at high densities, due to the higher rate of collisions with electrons.\\
\item
Influencing the non-Maxwellian character of the rotational distribution function. The higher the number density, the closer is the distribution function to a Maxwellian. Numerical calculation shows 
that starting from CNM conditions and varying only $\nH$, we transition to the Maxwellian regime ($\xi \lesssim 1$) if $\nH \gtrsim 10^5\ $cm$^{-3}$.
\end{itemize}

\emph{Low density limit} \\
For very low number densities, the distribution is highly non-Maxwellian and we can use Eqs. (\ref{nu_peak, xi gg 1}) and (\ref{power, xi gg 1}), with $G = G_{\rmn{IR}}$, to estimate the peak frequency and total power. As $G_{\rmn{IR}}/\tau_{\rmn H}$ is independent of $\nH$, both the number density and total power should asymptote to a constant value. We can estimate numerically the peak frequency in CNM conditions and get:
\beq
\nu_{\rmn{peak}}\left(\nH \rightarrow 0 \right) \approx 13 \ \rmn{GHz}  
\eeq
\beq
j/\nH\left(\nH \rightarrow 0 \right) \rightarrow \rmn{constant}
\eeq
which is in good agreement with Fig.~\ref{figure : nh_effect}.\\[5pt]
\emph{Intermediate densities}\\
Over the range $ 10^{2} \ \cm^{-3} \lesssim \nH \lesssim 10^{4} \ \cm^{-3}$, gas processes are dominant over infrared emission, so $F$, $G$ are roughly independent of $\nH$. In addition, the distribution is still strongly non-Maxwellian. Using Eqs. (\ref{nu_peak, xi gg 1}) and (\ref{power, xi gg 1}), we thus find
\beq
\nu_{\rmn{peak}}\left(10^{2} \ \cm^{-3} \lesssim \nH \lesssim 10^{4} \ \cm^{-3}\right) \propto \nH^{1/4} 
\eeq
\beq
j/\nH\left(10^{2} \ \cm^{-3} \lesssim \nH \lesssim 10^{4} \ \cm^{-3} \right) \propto \nH
\eeq
The kink around $\nH \sim 3 \times 10^3 \ \cm^3$ is due to our discontinous treatment of the evaporation temperature for high densities (see section \ref{section : Tev}), and to the replacement of the integration over all grain radii by a discrete summation when numerically computing the spectrum. Therefore the spectra should not be considered as very accurate in that region.
 
\emph{High density limit}\\
For very high number densities, the excitation and damping is dominated by gas processes, and the electric dipole damping becomes negligibly small, so that the rotational distribution function is 
actually a Maxwellian, although not thermal. Using Eqs. (\ref{nu_peak, xi ll 1}) and (\ref{power, xi ll 1}), we find
\beq
\nu_{\rmn{peak}}\left(\nH \rightarrow \infty \right) \approx 150 \ \rmn{GHz}  
\eeq
\beq
j/\nH\left(\nH \rightarrow \infty \right) \rightarrow \rmn{constant}
\eeq

\subsection{Effect of the gas temperature $T$}
\begin{figure*}
  \includegraphics[width = 80mm]{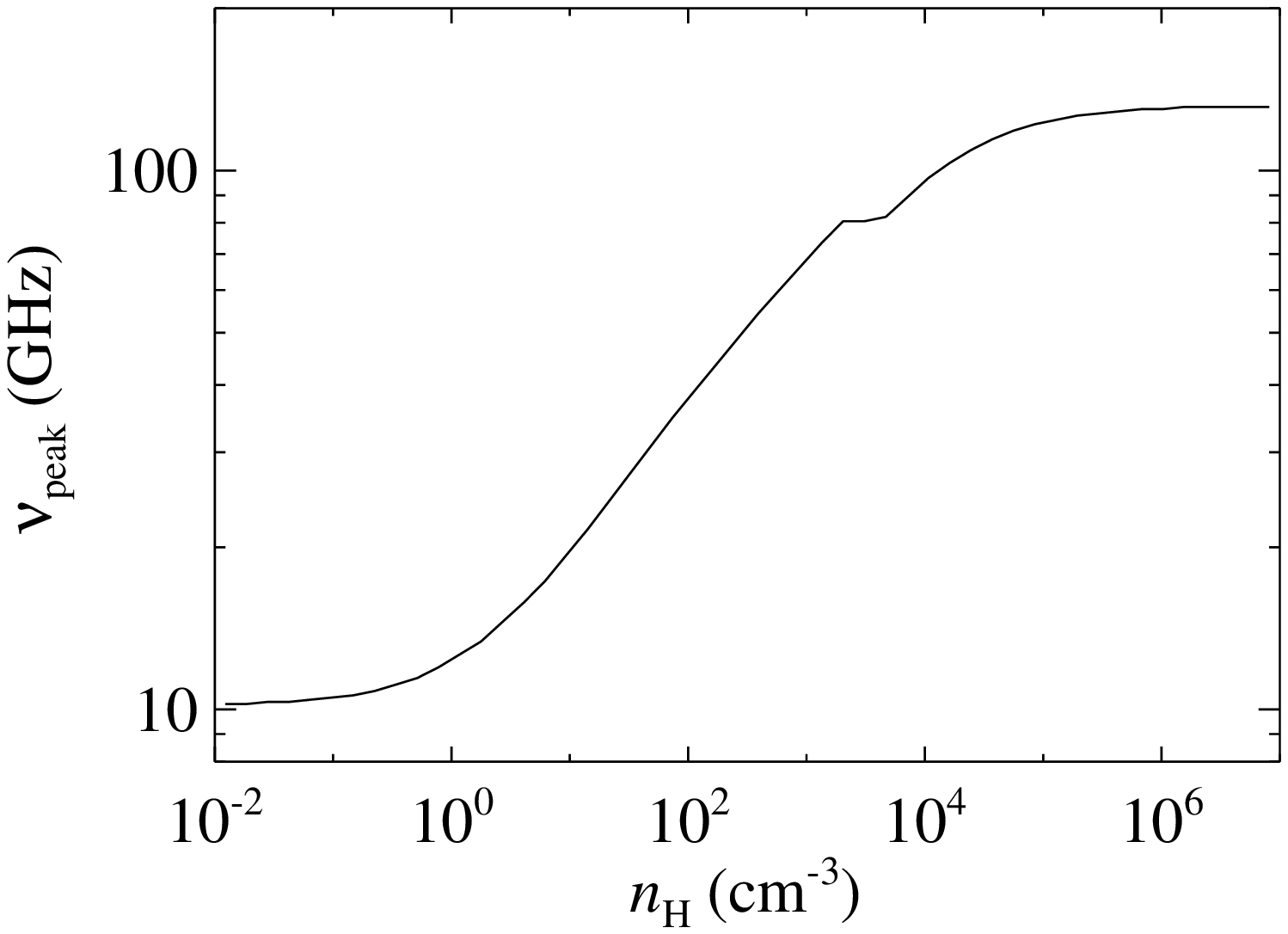}
  \includegraphics[width = 80mm]{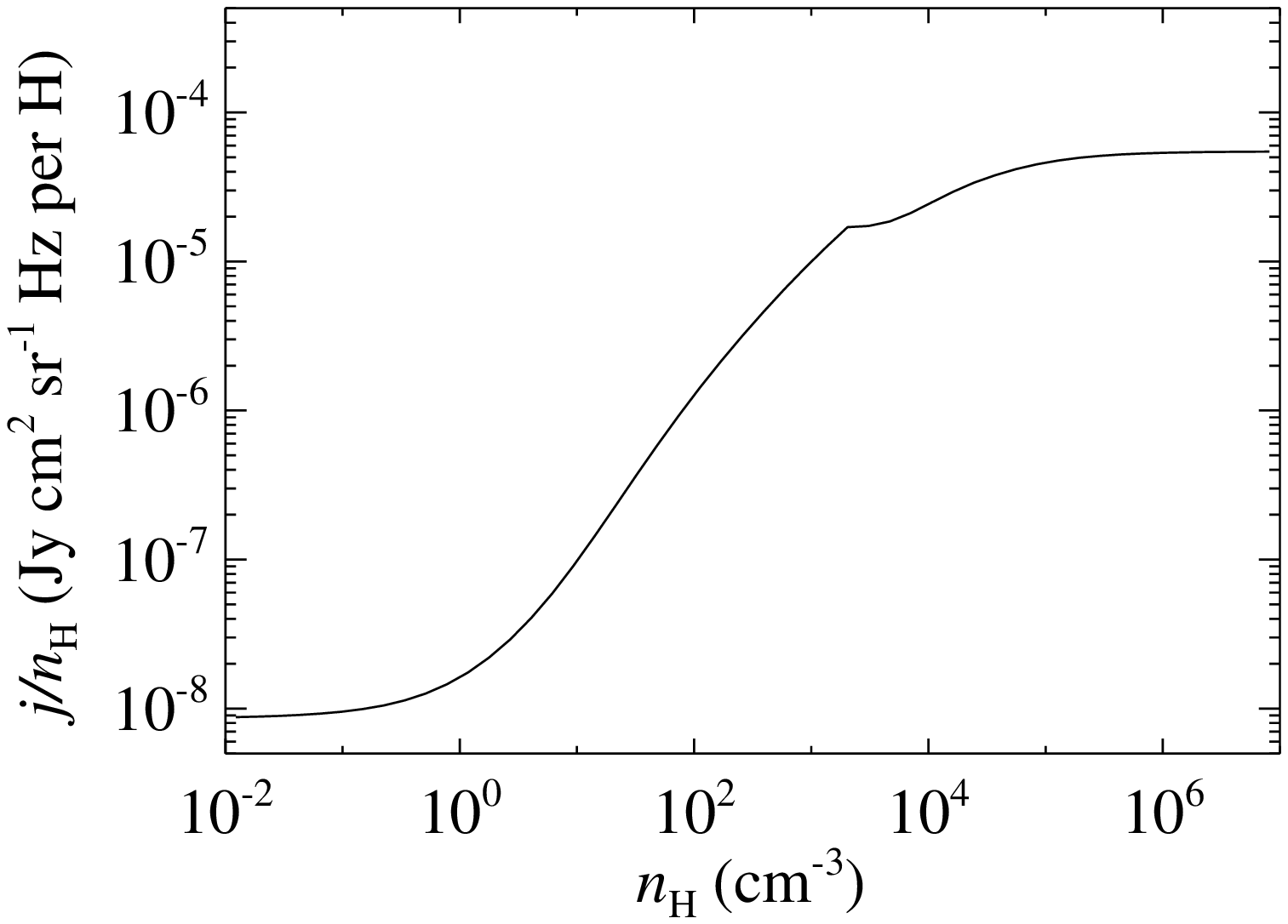}
  \includegraphics[width = 80mm]{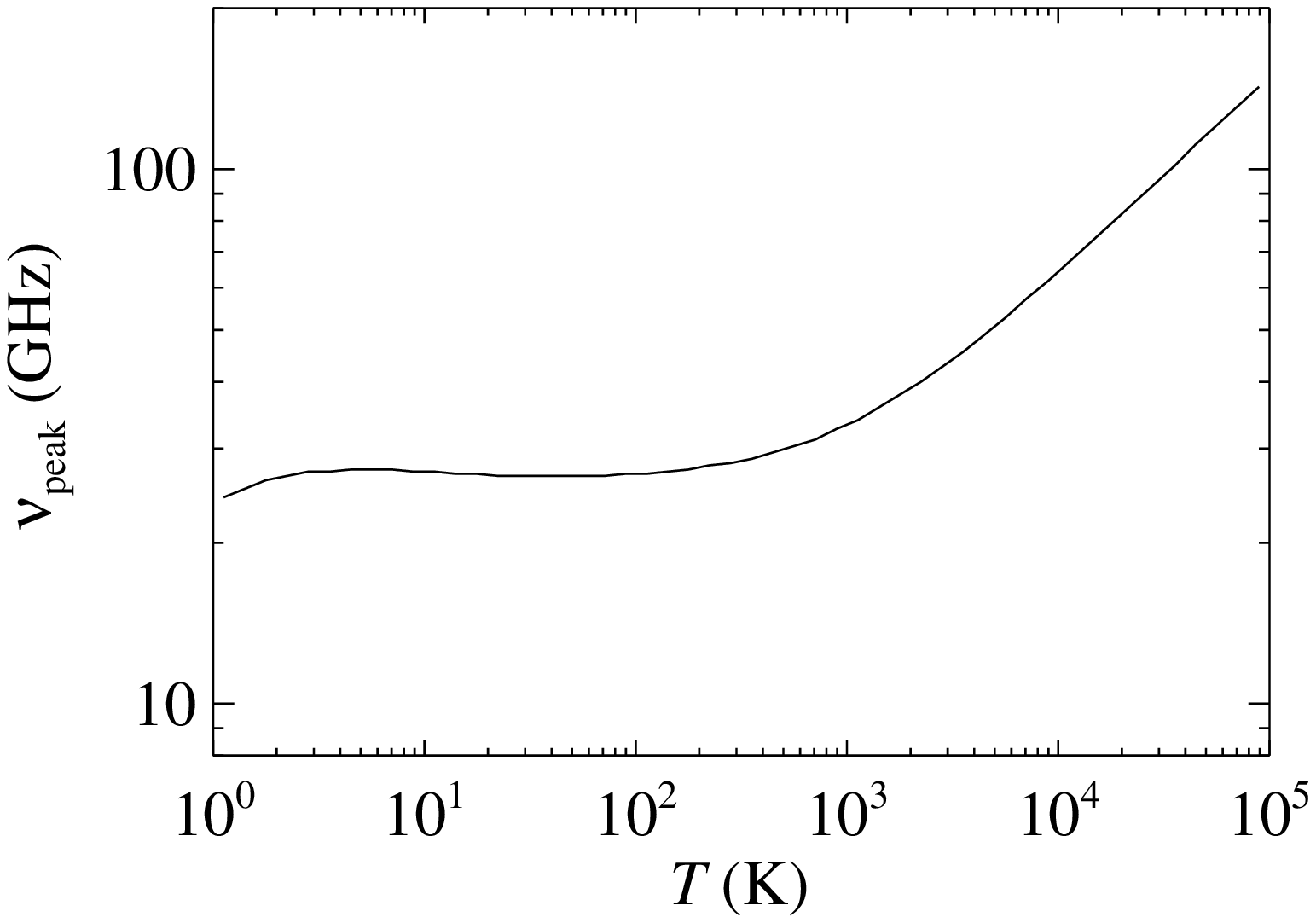}
  \includegraphics[width = 80mm]{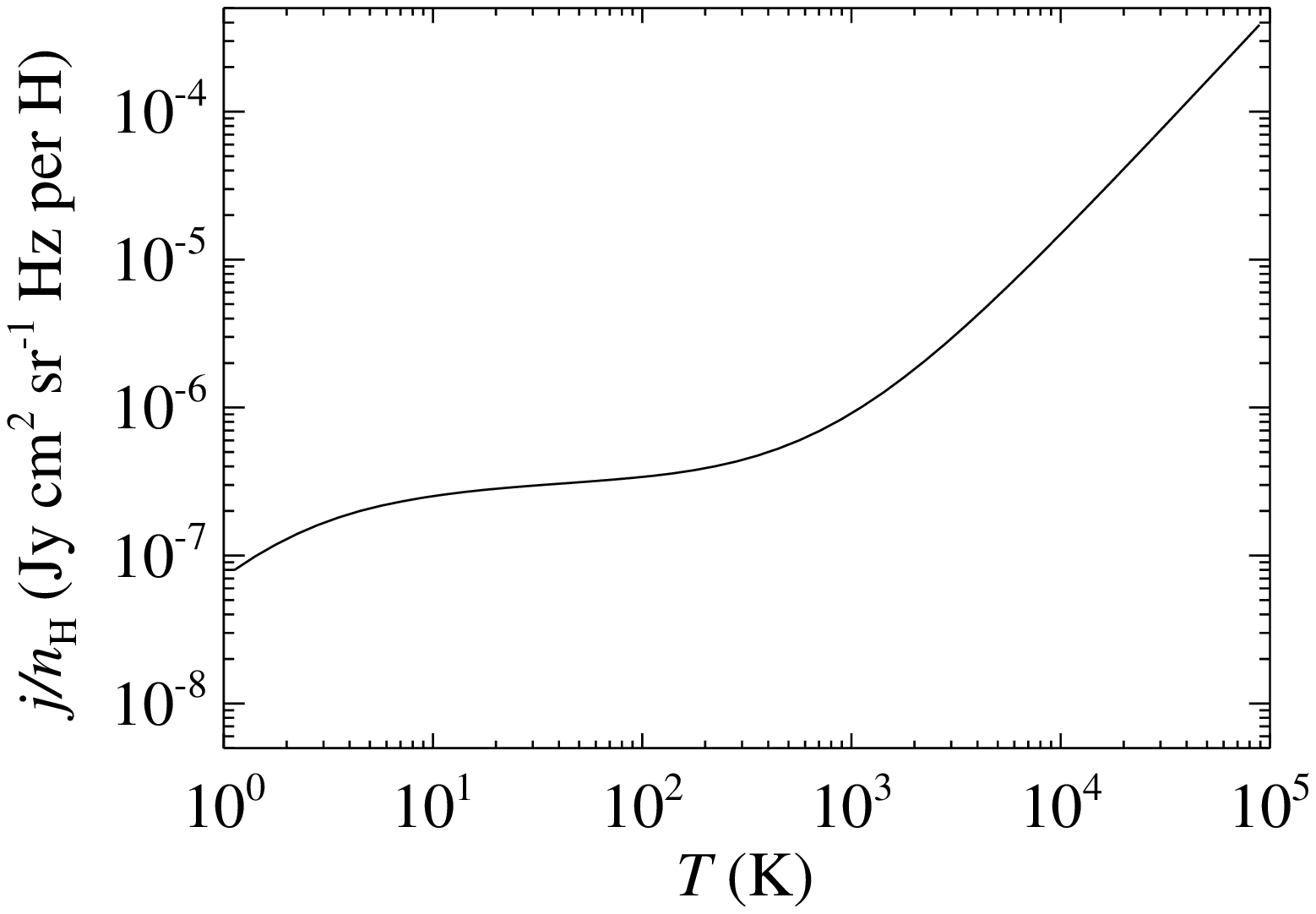} 
  \includegraphics[width = 80mm]{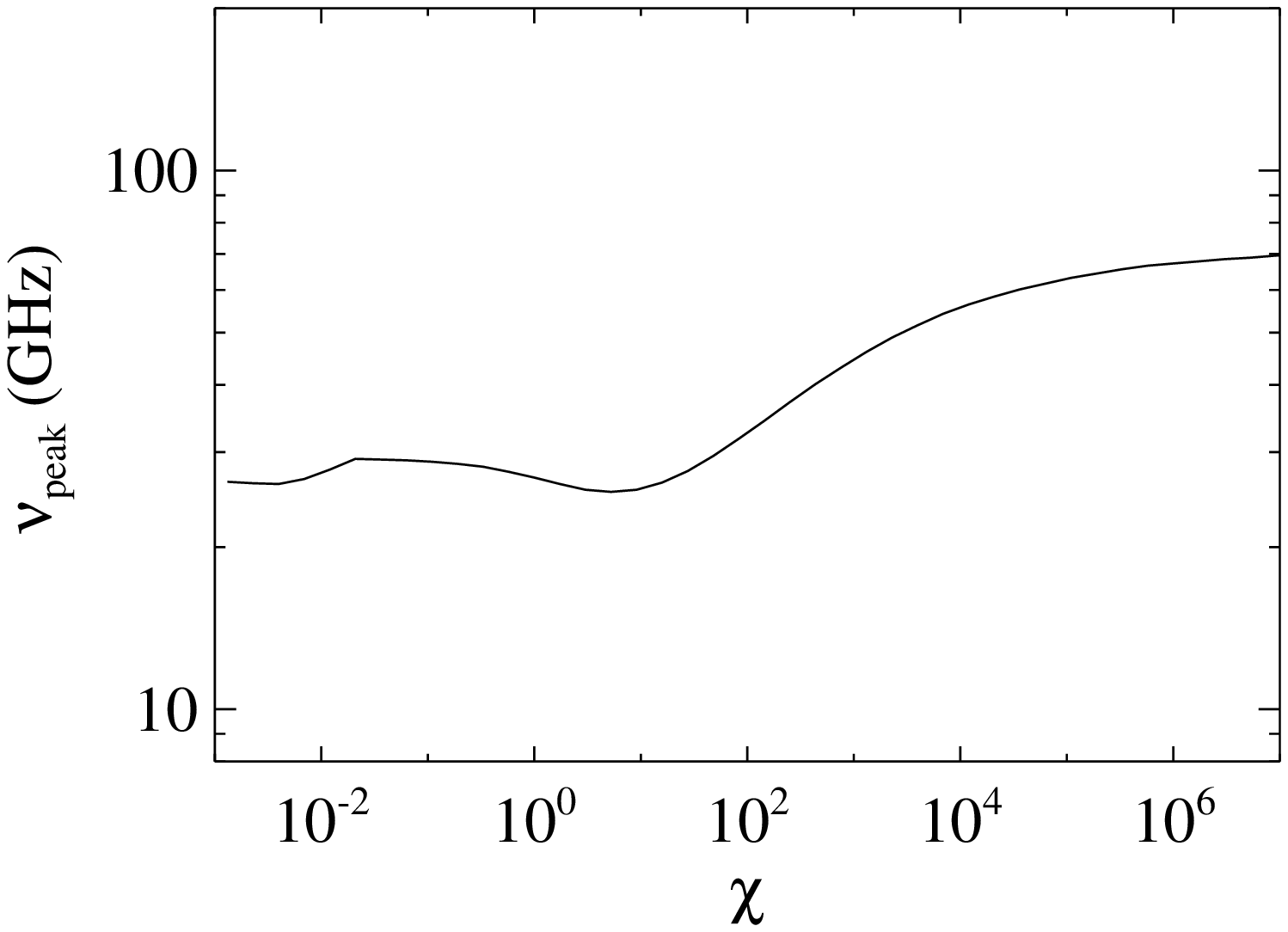}
  \includegraphics[width = 80mm]{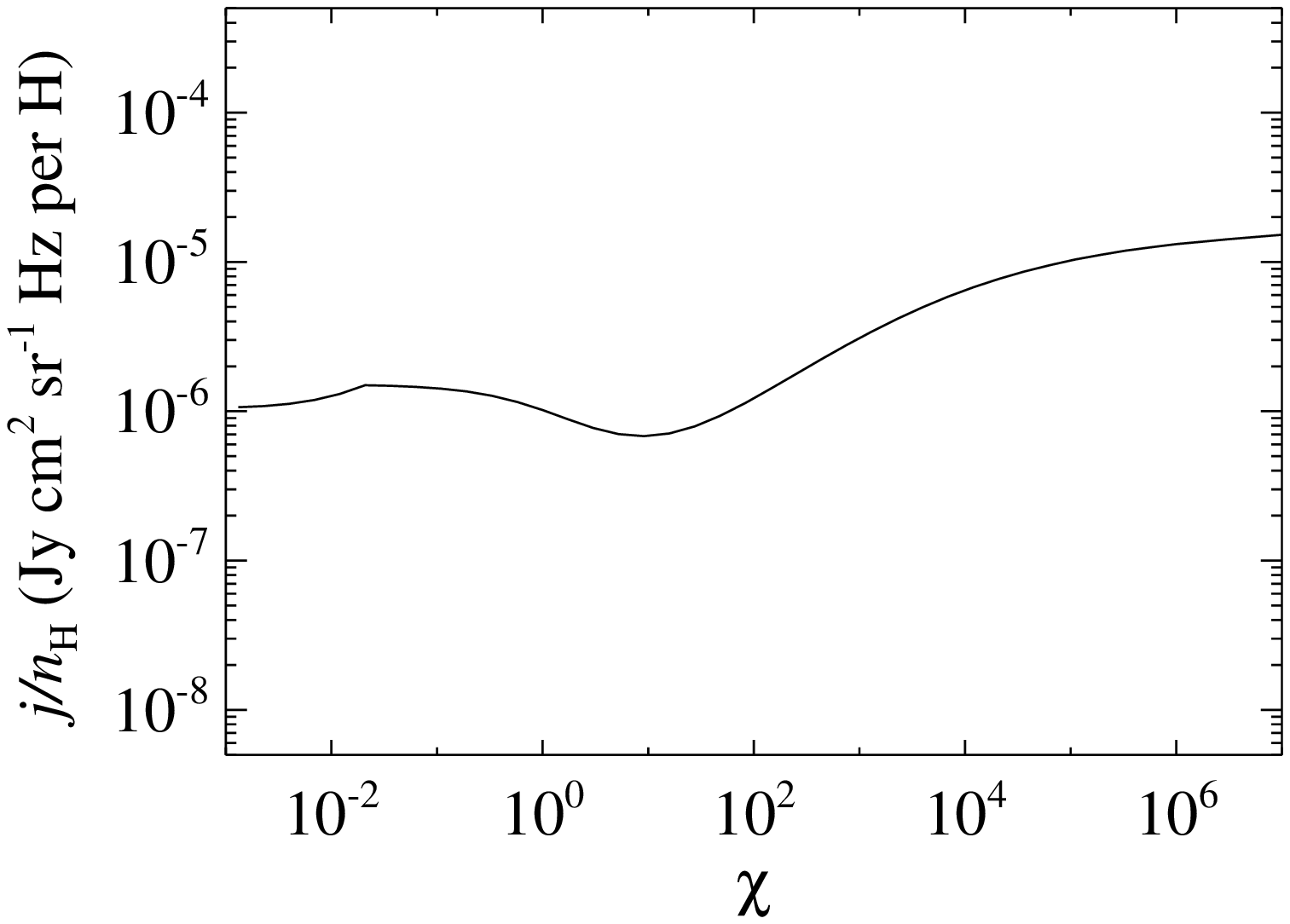}
  \includegraphics[width = 80mm]{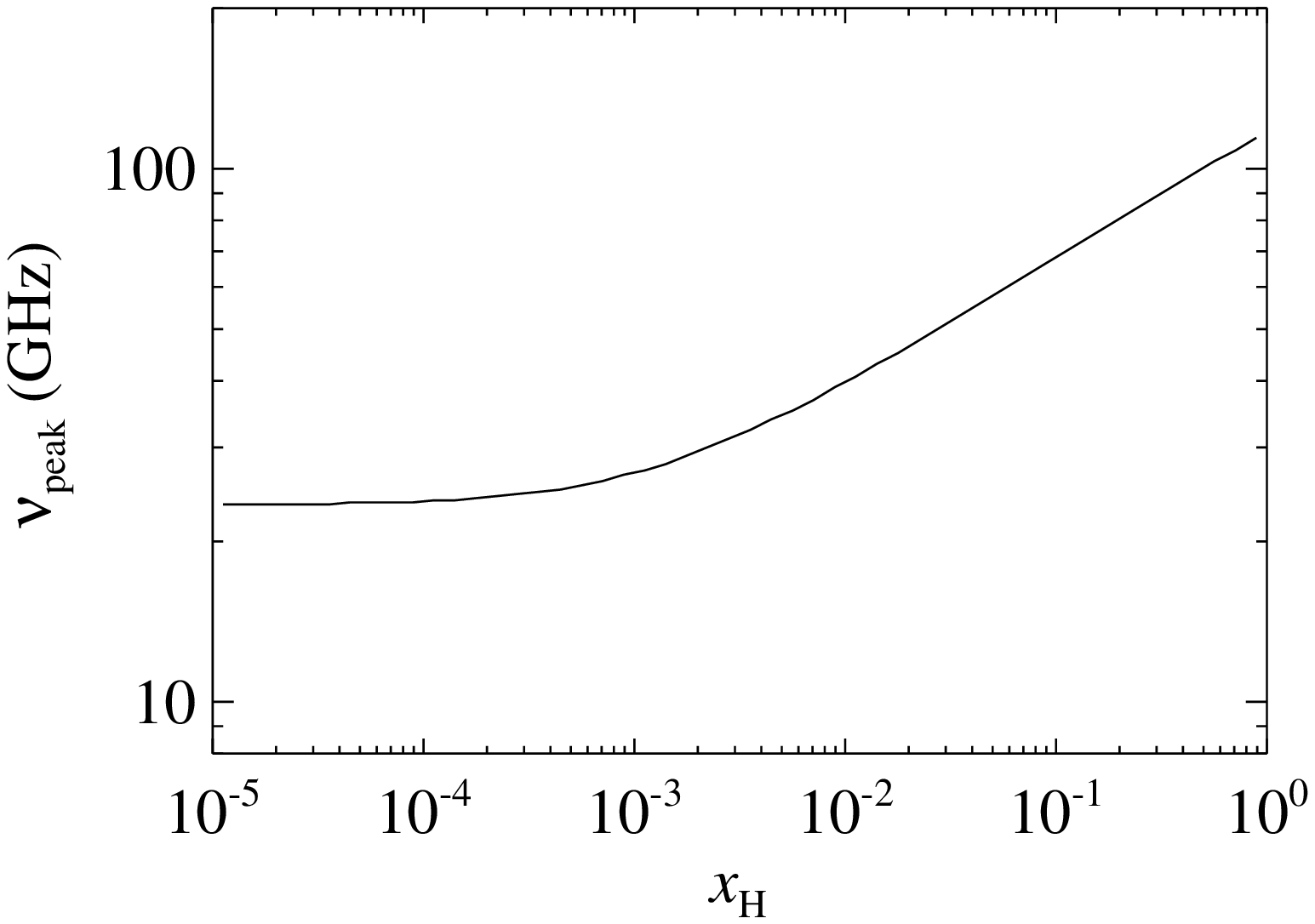}
  \includegraphics[width = 80mm]{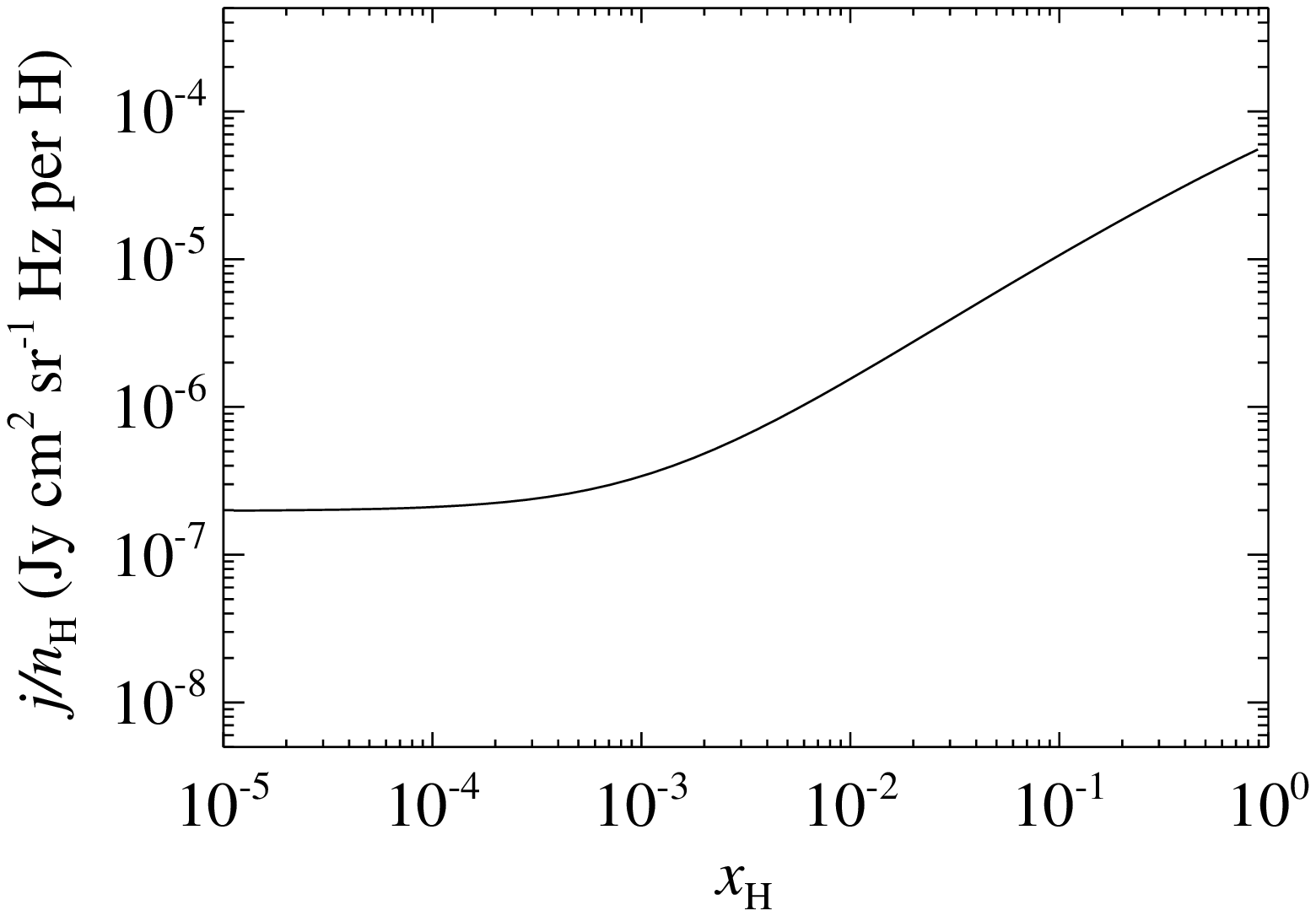}
  \caption{Effect of various environmental parameters on the peak frequency and the total spinning dust emission. When one parameter is varied, other environment parameters are set to CNM conditions defined in Eq. (\ref{eq : CNM}). See sections \ref{section nH} to \ref{section xH} for a detailed description.}
  \label{figure : nh_effect}
\end{figure*}

Temperature has a less obvious effect on the spectrum and we need to analyse in detail every damping and excitation process. It turns out the charge distribution of the smallest grains varies very little over the range of temperature considered $ 1 \ \K < T < 10^5 \ \K$ and they remain mostly neutral throughout this interval. The distribution remains strongly non-Maxwellian for $T$ greater than a few K. \\[5pt]
\emph{Low temperature limit}\\
At very low temperatures, the dominant excitation process is collisions with ions. Indeed, the grains being mostly neutral, the ions interact strongly with the electric dipole potential. As $\tilde \mu \propto T^{-1}$ and $\phi \propto T^{-1/2}$, one can see from Eq.(\ref{eq : Gi neutral}) and (\ref{h2}) that $G_i \propto T^{-2}$. Plasma drag has also $G_p \propto T^{-2}$ in principle but this becomes a shallower power law at low temperatures as the interaction timescale becomes longer than the rotation timescale. We find numerically, though, that roughly $G \propto T^{-1.5}$ as $G$ is not strictly equal to $G_i$ (collisions with neutrals are also significant at low temperatures). Using Eqs. (\ref{nu_peak, xi ll 1}), (\ref{power, xi ll 1}), and (\ref{eq : tau_dependence}) we find
\beq
\nu_{\rmn{peak}}\left(3 \ \rmn{K} \lesssim T \lesssim 10^2 \ \rmn{K} \right) \approx 35 \ \rmn{GHz}
\eeq
\beq
j/\nH\left( 3 \ \rmn{K} \lesssim T \lesssim 10^2 \ \rmn{K} \right) \approx \rmn{constant}
\eeq
Note that for extremely low temperatures, the distribution would become Maxwellian, and one would get, according to Eqs. (\ref{nu_peak, xi gg 1}), (\ref{power, xi gg 1}), 
\beq
\nu_{\rmn{peak}}\left(T \rightarrow 0 \right) \propto T^{1/2} 
\eeq
\beq
j/\nH\left(T \rightarrow 0 \right) \propto T^2 
\eeq
which can be guessed at the extreme low temperature end of Fig.~\ref{figure : nh_effect}. Temperatures below $\sim 3 \ \K$ are of course unphysical, but for other environmental conditions than those of Eq. (\ref{eq : CNM}), the behaviour discussed above could take place at higher, observed temperatures.\\[5pt]
\emph{High temperature limit}\\
At very high temperatures, collisions with neutrals are the dominant damping and excitation process. The CNM environment being mostly neutral, $F_n \rightarrow 1$ and $G_n \rightarrow 1/2$ at high temperatures ($G_n^{(ev)} \propto T_{ev}/T \rightarrow 0 $). Moreover, the distribution becomes strongly non-Maxwellian, as  $\xi \propto T^{1/2}$. We therefore obtain
\beq
\nu_{\rmn{peak}}(T \rightarrow \infty ) \approx 200 \ T_5^{3/8} \ \rmn{GHz} \label{equation : T_inf}
\eeq
\beq
j/\nH\left(  T \rightarrow \infty \right) \propto T^{3/2}
\eeq
Fig.~\ref{figure : nh_effect} shows that these power laws describes the behavior of the peak frequency and total power with very good accuracy.  

\subsection{Effect of the radiation field intensity $\chi$}
The radiation field affects the spectrum through only two ways. First of all, it changes the charge distribution of the grains as an increased radiation implies a higher photoemission rate. Second of all, it affects the rate of damping and excitation through infrared emission. (and photoelectric emission, but this is subdominant).\\[5pt] 
\emph{Low radiation intensity limit}\\
In a low radiation field, $F_{\rmn{rad}}$ and $G_{\rmn{rad}}$ become negligible. The photoemision charging rate becomes insignificant compared with collisional charging, and the charge distribution function depends only on other environment parameters. Thus, one expects the spectrum to reach an asymptotic shape for very low radiation fields. The distribution is strongly non-Maxwellian, and the dominant excitation mechanism is collisions with ions, whereas the dominant damping mechanisms are plasma drag and collisions with neutrals. Thus, we find
\beq
\nu_{\rmn{peak}}(\chi \rightarrow 0 ) \approx 35  \ \rmn{GHz} 
\eeq
\beq
j/\nH\left( \chi \rightarrow 0 \right) \rightarrow \rmn{constant}
\eeq
The kink around $\chi \sim 2 \times 10^{-2}$ is due to our discontinuous treatment of the evaporation temperature for low intensities of the radiation field.\\
Around $\chi \approx 1-10$, the grain becoming more and more positively charged, the collisions with ions start being less efficient, although still the dominant excitation mechanism. This results in a slight decrease in both $\nu_{\rmn{peak}}$ and $j/\nH$.\\[5pt]
\emph{High radiation intensity limit}\\
In a high radiation field, $F \approx F_{\rmn{IR}}$, and $G \approx G_{\rmn{IR}}$. Both $F_{\rmn{IR}}$ and $G_{\rmn{IR}}$ are approximately (although not strictly) linear in $\chi$, as shown in DL98b for the thermal spikes limit (see their equations (31) and (44)). Thus, $\xi \sim \chi^{-1}$ so the distribution becomes Maxwellian. The peak frequency and total emitted power are then given by Eqs. (\ref{nu_peak, xi ll 1}) and (\ref{power, xi ll 1}), which imply that
\beq
\nu_{\rmn{peak}}(\chi \rightarrow \infty) \approx \rmn{constant}
\eeq
\beq
j/\nH\left( \chi \rightarrow \infty \right) \approx \rmn{constant}
\eeq
These asymptotic forms are not strictly valid because $F_{\rmn{IR}}$ and $G_{\rmn{IR}}$ are not strictly linear in $\chi$, and do not have a simple dependence on that parameter.

\subsection{Effect of the ionization fraction $x_{\rmn H}$} \label{section xH}

The Hydrogen ionization fraction affects the charge distribution by modifying the contribution from collisions with protons. It also changes the contribution of collisions with ions, neutrals and plasma drag. Characteristic timescales are left invariant, and $\xi \gg 1$ for any ionization fraction in otherwise CNM conditions.\\[5pt] 
\emph{Low ionization fraction limit}\\ 
In that limit the rotational distribution function reaches an asymptotic form where collisions with protons and plasma drag due to protons can be neglected. However, there are still $C^+$ ions in the gas so collisions with ions and plasma drag may still be important, although the dominant excitation process is collisions with neutrals. We find
\beq
\nu_{\rmn{peak}}\left( x_{\rmn H}\rightarrow 0\right) \approx   30  \textrm{GHz}   
\eeq
\beq
j/\nH \left( x_{\rmn H}\rightarrow 0\right) \rightarrow \rmn{constant}
\eeq
\emph{High ionization fraction limit}\\ 
In that case collisions with ions are the dominant excitation process. Using  Eqs. (\ref{nu_peak, xi gg 1}) and (\ref{power, xi gg 1}) along with $G \approx G_i \propto x_{\rmn H}$, we find
\beq
\nu_{\rmn{peak}}( x_{\rmn H}\rightarrow 1) \approx  90  (x_{\rmn H}/0.1)^{1/4} \textrm{GHz}    
\eeq
\beq
j/\nH \left( x_{\rmn H}\rightarrow 1 \right) \propto x_{\rmn H}
\eeq

\subsection{Concluding remarks}

We remind the reader that all the estimates in the previous section were given by assuming that the peak frequency of the spinning dust spectrum is determined by that of the smallest grains, and that the total power follows the same dependence upon environmental parameters as the power emitted by the smallest grains. Therefore they should be taken as an aid to understand the physics of spinning dust, but not as an accurate description, which requires numerical calculations.

The overall conclusion of this section is that varying a single environmental parameter may change the peak frequency by up to an order of magnitude, and the total emitted power by several orders of magnitude. There is therefore a very large range of possible peak frequencies and total powers that can be produced by spinning dust radiation. Multiphase environments, in particular, could emit very broad spinning dust spectra. Deducing the environment parameters from en observed spectrum could therefore be a difficult task.

\begin{figure*}
\includegraphics{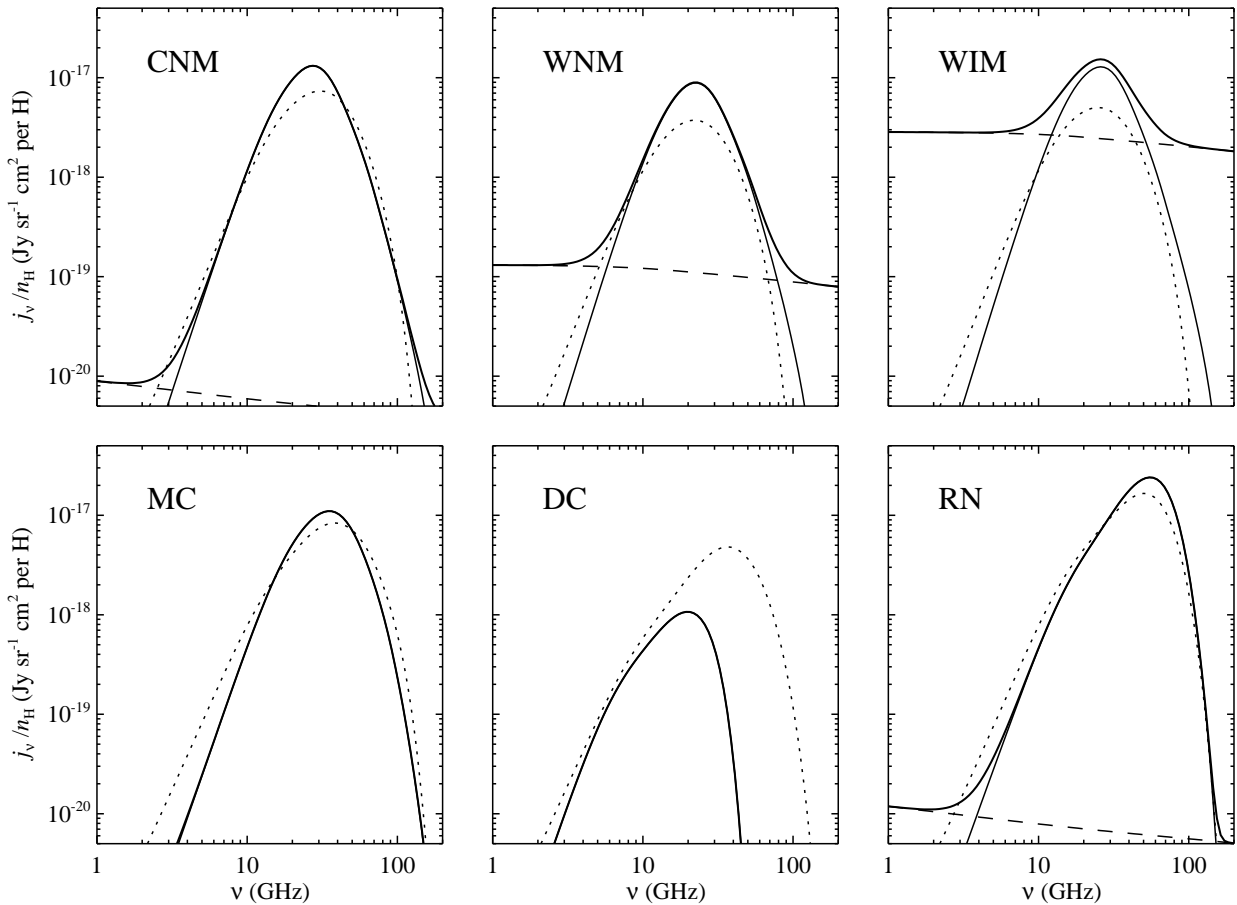}
\caption{Spinning dust spectra for several environment conditions : Cold Neutral Medium (CNM), Warm Neutral Medium (WNM), Warm Ionized Medium(WIM), Molecular Cloud (MC), Dark Cloud (DC) and Reflection Nebula (RN). The environments are defined in DL98b, Table 1. The thin solid line is the result of our calculation, the dotted line being DL98 prediction, and the dashed line is the free-free emission (the free-free gaunt factor were taken from \citet{Sutherland}). The parameters for the grain size distribution are : $R_V = 3.1,  \ b_C = 6 \times 10^{-5}$ for the diffuse CNM, WNM and WIM environments, and  $R_V = 5.5, \ b_C = 3 \times 10^{-5}$ for the dense MC, DC and RN environments. The apparent systematic increase of power around the peak frequency for our result is mainly due to the grain size distribution we use, which has an increased number of small grains compared to that used by DL98 (compare e.g. Fig.~2 from WD01a with Fig.~8 from DL98b). Note that for the DC environment, for which rotational excitation and damping is dominated by collisions with neutral species (mainly H$_2$ molecules), DL98 prediction largely overestimates the peak frequency and total power because they underestimate the damping rate (see Eq. (\ref{G_coll_ev}) and subsequent discussion).}
\label{environments}
\end{figure*}

We show the spinning dust spectrum for various environments and compare them to DL98 results in Fig.~\ref{environments}.

\section{Conclusion} \label{conclusion}

We have presented a detailed analysis of the rotational excitation and damping of small carbonaceuous grains. We have refined DL98b results in the case of collisions, accounting properly for the 
centrifugal potential which increases the net damping rate. In the case of collisions with ions, we accounted for the effect of the electric dipole potential on the collision cross section. We found that 
this is a small effect in the case of charged grains, but that it may significantly increase the damping and excitation rates in the case of neutral grains. We evaluated the contribution of ``plasma 
drag'' by considering hyperbolic trajectories and rotating grains in the case of charged grains, and straightlines in the case of neutral grains. We corrected DL98b results for the damping through 
infrared emission. Finally, we calculated the rotational distribution function by solving the Fokker-Planck equation.

We believe our model provides a much more accurate description of the spinning dust spectrum than previous work. However, we would like to remind the reader of its uncertainties and limitations. First, 
our model only computes the total intensity of the emitted radiation and not the polarization, which would require an additional study of the alignment mechanisms for PAHs.
Secondly, the dust grains properties are poorly known:
\begin{itemize}
\item
The size distribution and abundance of the smallest grains is uncertain, and in particular the nature of the cutoff at small grain sizes $a\sim a_{\rm min}$ can have a large effect on the spectrum.
\item
The permanent electric dipole moments of dust grains are not directly constrained by other dust observables.  Given that it cannot be computed from first principles, one may regard it as a free 
parameter (or parameters) of the spinning dust model.
\end{itemize}
Thirdly, we made some simplified calculations in some cases, as an accurate calculation would have been intractable numerically or substantially complicated the code:
\begin{itemize}
\item
We used the Fokker-Planck approximation, which starts to break down for our smallest grains because a single collision suffices to change the rotational state.  We expect that the main consequence of a 
full treatment would be a tail in the emission spectrum extending to high frequencies, because impulsive collisions would be able to increase the rotation velocities of the grains to $>2\nu_{\rm peak}$ 
before dissipative forces had time to act (an effect missed by the Fokker-Planck treatment).  Therefore one should not place too much confidence in the many order-of-magnitude falloff at $\sim 100\ $GHz 
seen in most of our models.  (In many cases this will be unimportant observationally since at high frequencies the vibrational dust contribution is dominant.)
\item
In the plasma drag calculation, we neglected the electric dipole potential when evaluating the trajectory of ions, taking the straight-line (neutral grain) or hyperbolic (charged grain) approximation. Relying on the study of collisions with ions, we may expect the dipole moment to have a small effect in the case of a charged grain. On the other hand, its effect in the case of a neutral grain may be more important, as in that case the electric dipole potential provides the dominant interaction.
\item
We assumed the evaporation temperature for the smallest grains was the ``temperature'' of the grain just after it has absorbed a UV photon.  This is a physically motivated assumption but its validity is not established. The evaporation temperature can have a significant effect on the spectrum, as can be seen from Fig. \ref{CNM Tev effect} and one should be aware of the uncertainty in this parameter. Also, we assumed that collisions transition from being sticking to elastic, as the density exceeds a given threshold. Our model is therefore inaccurate in the transition region.
\item
When calculating the infrared emission spectrum of the grains, we used DL01 ``thermal continuous'' approximation, which is not very accurate to describe the low energy part of the spectrum. Whereas 
these uncertainties are not important if one only wants the spectrum $F_{\nu}$ in the mid-infrared, they may lead to significant errors when calculating the corresponding damping and excitation rates, 
which are 
proportional to $\int \nu^{-2}F_{\nu} \rmd \nu$ and $\int \nu^{-1}F_{\nu} \rmd \nu$ respectively.
\item
We ignored systematic torques, although this may not be a major omission for the smallest dust grains.
\end{itemize}

Despite these uncertainties, we believe that this model is the most complete thus far, and will be a useful tool for comparison to observations and testing the spinning dust hypothesis for anomalous 
microwave emission in various ISM phases.

\section*{Acknowledgments}

We thank G. Dobler, B. Draine, D. Finkbeiner, and A. Lazarian for numerous conversations about the physics of grain rotation. Y. A.-H. also thanks T. Readhead and T. Pearson for useful discussions. YA-H. and CMH. are supported by the U.S. Department of Energy (DE-FG03-92-ER40701) and the National Science Foundation (AST-0807337). The early phase of this project was funded by the NSF grant AST-0607857. C.H. is supported by the Alfred P. Sloan Foundation. C.D. acknowledges support from the U.S. {\it Planck} project, which is funded by the NASA Science Mission Directoriate.

\appendix

\section{Plasma drag : numerical calculation of $\mathcal I$ in the general case}
\label{app:i}

The numerical calculation of $\mathcal I$ is tricky because it involves integrating an oscillating function which frequency goes to infinity at one limit of the integral, as $t(\alpha \rightarrow 
\alpha_e) \rightarrow \infty$.  Here we describe our implementation for both the positive and negative grain charges.

\subsection{Positively charged grains: $Z_g>0$}

We first make the change of variable
\beq
z = \sqrt{\gamma} \cot \frac{\alpha}{2},
\eeq
where $\gamma=(e-1)/(e+1)$.
The expression for the time is now:
\beq
\omega t(z) = \frac{\omega b }{v} \frac{1}{\sqrt{e^2 - 1}} \left( \ln \frac{z + 1}{z - 1} + 2 e \frac{z}{z^2- 1} \right).
\eeq
The $\mathcal I$-integral is then
\barr
\mathcal I &=& 4 \gamma\left[ \Re \int_1^{\infty} \rme^{i\omega t(z)} \frac{z^2 - \gamma}{(z^2 + \gamma)^2} \rmd z  \right]^2
\nonumber \\ &&
 + \  16 \gamma^2 \left[ \Im \int_1^{\infty} \rme^{i\omega t(z)} \frac{z}{(z^2 + \gamma)^2} dz \right]^2.
\earr

The functions inside the integrals are analytical on the complex plane, deprived from the branch cut $[-1, 1]$ on the real axis and the two poles $\pm \rmi\sqrt\gamma$.
The integrands are at least $\mathcal O(z^{-2})$ as $|z| \rightarrow \infty$. Moreover, for $y \rightarrow 0^{+}$,
\beq
\Re\left[\rmi \omega t(1 - \rmi y)\right] \propto \Re\left[\rmi \ln\left(-1 + \frac{2 \rmi}{y}\right) - \frac{e}{y}\right] < 0
\eeq
Thus, using the fact that the integral over the lower right part of the complex plane vanish, we can replace our integrals by integrals over the axis
\beq
z = 1 - \rmi y, \ \ \ 0<y<+\infty.
\eeq
Note that for $e \rightarrow 1$, $\mathcal I = \mathcal O(e-1)$, as one may expect from almost parabolic trajectories if the grain repels the ion. Also, in the limit $\omega b /v \rightarrow 0$, 
$\mathcal I \rightarrow (e^2-1)/e^2$.

\subsection{Negatively charged grains: $Z_g<0$}

This time we make the change of variable
\beq
z = \frac1{\sqrt\gamma} \tan \frac{\alpha}{2}.
\eeq
The expression for the time is now
\beq
\omega t(z) = \frac{\omega b }{v} \frac{1}{\sqrt{e^2 - 1}} \left( \ln \frac{z + 1}{z - 1} - 2 e \frac{z}{z^2- 1} \right).
\eeq
And we have:
\barr
\mathcal I &=& 4 \gamma\left[ \Re \int_1^{\infty} \rme^{\rmi\omega t(z)} \frac{1 - \gamma z^2}{(1 + \gamma z^2 )^2} \rmd z  \right]^2
\nonumber \\ &&
 + 16  \gamma^2 \left[ \Im\int_1^{\infty} \rme^{\rmi\omega t(z)} \frac{z}{( 1 + \gamma z^2 )^2} \rmd z \right]^2.
\earr
The functions inside the integrals are analytical on the complex plane, deprived from the branch cut $[-1, 1]$ on the real axis and the two poles $\pm \rmi/\sqrt\gamma$. This time 
$\Re(i \omega t)$ is negative for $z$ close to 1 when $\Im z >0$. Moreover, the two poles tend to infinity when $e \rightarrow 1$ so to avoid integrating too close to the poles, we 
integrate over the line
\beq
z = 1 + \rme^{\rmi \pi/4} y, \ \ \ 0<y<+\infty.
\eeq
In that case, the integrals are not simply bounded anymore for nearly parabolic trajectories. One can show, by making the previous change of variables, that
\beq
\mathcal I\left(\frac{\omega b}{v}, e, Z_g <0 \right) = \exp\frac{2 \pi {\omega b}}{{v}\sqrt{e^2-1}}\ \mathcal I\left(\frac{\omega b}{v}, e, Z_g > 0 \right).
\eeq
This expression is ill behaved for nearly parabolic trajectories, as the exponential factor diverges whereas the $\mathcal I$-integral vanishes. In order to avoid numerical problems, in the case of nearly parabolic trajectories, we make the change of variables
\beq 
u = \left(\tan\frac{\alpha}{2}\right)^{-1}
\eeq
The expression for the $\mathcal I$-integral is then, for $ e-1 \ll 1$:
\barr
\mathcal I  \approx  4 \left\{  \int_0^{\sqrt{\frac{2}{e-1}}} \cos\left[\frac{\omega b}{v}(e-1)(u + \frac{u^3}{3}) \right] \frac{u^2 - 1}{(u^2 + 1)^2} \rmd u \right\}^2
\nonumber \\ 
+ \ 16 \left\{  \int_0^{\sqrt{\frac{2}{e-1}}} \sin\left[\frac{\omega b}{v}(e-1)(u + \frac{u^3}{3}) \right] \frac{u}{(u^2 + 1)^2} \rmd u\right\}^2
\earr
Note that in terms of the true anomaly $f$ we have $u = -\tan f/2$ and the expression for the time can be found in \citet{orbits}, Eq. (2.3.9).\\
Here again we integrate along $u = \rme^{\rmi \pi/6} y$, $0<y<\infty$, which cancels the $\mathcal O (u^3)$ real part of the time and maximizes its positive imaginary part at infinity. Notice 
that for very small eccentricities, this is mainly a function of $(\omega b/ v)(e-1)$.

\section{Quantum treatment of infrared emission}
\label{app:q}

In Section~\ref{s:ir}, we computed the net angular momentum loss due to infrared emission using classical electrodynamics.  Here we reconsider the effect with a quantum calculation.  We assume a 
spherically symmetric grain for simplicity, and neglect vibration-rotation interaction.  We will recover the classical result in the limit $J\gg 1$, which is applicable to the dust grains considered 
in this paper.

The Hilbert space of the grain is characterized by the vibrational quantum numbers (generically denoted ${\bmath v}$) and the three rotation quantum numbers $J$, $K$, $M$, where $K$ is the projection of 
angular momentum onto 
the grain $z$-axis.  The energy levels are given by
\beq
E_{J,K,M,\bmath v} = E^0_{\bmath v} + \frac{\hbar^2J(J+1)}{2I},
\eeq
where $I$ is the grain moment of inertia and $E^0_{\bmath v}$ is the vibrational energy.  The rotational wave functions are
\beq
\Psi_{J,K,M}(\chi) = \sqrt{\frac{2J+1}{8\pi^2}} D^{J}_{K,M}(\chi),
\eeq
where $\chi=(\theta,\phi,\psi)\in\,$SO(3) is the set of Euler angles, $8\pi^2$ is the volume of SO(3), and ${\mathbfss D}^J$ is the passive rotation matrix in the spin-$J$ representation, i.e. 
$D^J_{M_1,M_2} = \langle JM_{1\,\rm grain}|JM_{2\,\rm lab}\rangle$.

Spontaneous infrared vibrational transitions are possible from vibrational state ${\bmath v}$ to ${\bmath v}'$; 
their rate is given by
\barr
A_{J,K,M,{\bmath v}\rightarrow J',K',M',{\bmath v}'} \!\!\!\!&=&\!\!\!\! \frac{4(E_{J,K,M,\bmath v}-E_{J',K',M',{\bmath v}'})^3}{3\hbar^4 c^3}
\nonumber \\ && \!\!\!\!\!\!\!\!\!\!\!\!\times
  \left| \langle J',K',M',{\bmath v}' | {\bmu} | J,K,M,{\bmath v} \rangle \right|^2,
\label{eq:spon}
\earr
where ${\bmu}$ is the electric dipole moment operator.  In the absence of vibration-rotation interaction, we may take the operator $\bmu$ to depend only on the vibrational quantum numbers and on the 
rotation matix ${\mathbfss R}(\chi)$ that converts grain-fixed to lab-fixed coordinates:
\barr
&&\langle J',K',M',{\bmath v}' | {\bmu} | J,K,M,{\bmath v} \rangle
\nonumber \\
&&  = \langle J',K',M'| {\mathbfss R}(\chi) | J,K,M \rangle \langle {\bmath v}'|\bmu^{(g)}|{\bmath v}\rangle.
\earr
Here $\bmu^{(g)}$ is the dipole moment in grain coordinates.

The transition rates can be determined by writing $\bmu^{(g)}$ in the polar basis,
\beq
\mu^{(g)}_0 = \mu^{(g)}_z {\rm ~~~and~~~}
\mu^{(g)}_{\pm 1} = \frac{\mp \mu^{(g)}_x + \rmi \mu^{(g)}_y}{\sqrt2},
\eeq
in which $\{\mu^{(g)}_m\}_{m=-1}^1$ transform in the $L=1$ representation of SO(3). Written in this basis, the rotation matrix ${\mathbfss R}(\chi)$ is the inverse of ${\mathbfss D}^1(\chi)$, which for 
unitary ${\mathbfss D}^1$ is the same as the Hermitian conjugate:
\barr
&&\langle J',K',M',{\bmath v}' | \mu_m | J,K,M,{\bmath v} \rangle
\nonumber \\
&&  = \langle J',K',M'| D^{1\ast}_{m',m}(\chi) | J,K,M \rangle \langle {\bmath v}'|\mu_{m'}^{(g)}|{\bmath v}\rangle.
\earr
The first matrix element can be evaluated by the three rotation matrix integral,
\barr
&&\langle J',K',M'| D^{1\ast}_{m',m}(\chi) | J,K,M \rangle
\nonumber \\
&& = \frac{\sqrt{(2J'+1)(2J+1)}}{8\pi^2} \nonumber \\ && \times
\int 
{D}^{J'\ast}_{K',M'}(\chi)
{D}^{1\ast}_{m',m}(\chi)
{D}^{J}_{K,M}(\chi)\,
\rmd^3\chi
\nonumber \\
&& = \sqrt{(2J'+1)(2J+1)} (-1)^{K'+m'+M'+m}
\nonumber \\ && \times
\left( \begin{array}{ccc} J' & 1 & J \\ -K' & -m' & K \end{array}\right)
\nonumber \\ && \times
\left( \begin{array}{ccc} J' & 1 & J \\ -M' & -m & M \end{array}\right).
\earr
This transforms the spontaneous decay rate, Eq.~(\ref{eq:spon}), into
\barr
&&
\frac{4(E_{J,K,M,\bmath v}-E_{J',K',M',{\bmath v}'})^3}{3\hbar^4 c^3}
(2J'+1)(2J+1)
\nonumber \\
&& \times
\sum_{m=-1}^1 \Bigl|
\sum_{m'=-1}^1 (-1)^{m'}
\langle {\bmath v}'|\mu_{m'}^{(g)}|{\bmath v}\rangle
\nonumber \\
&& \times
\left( \begin{array}{ccc} J' & 1 & J \\ -K' & -m' & K \end{array}\right)
\nonumber \\
&& \times
\left( \begin{array}{ccc} J' & 1 & J \\ -M' & -m & M \end{array}\right)
\Bigr|^2.
\earr
We would now like to find the net decay rates to states of different $J'$.  To do this, we assume the grain is randomly oriented, i.e. we average over initial projections $K$, and sum over final 
projections $K'$.  Using the $3j$ symbol orthonormality relations, one obtains
\barr
A_{J,M,{\bmath v}\rightarrow J',M',{\bmath v}'}
&=&\frac{4(E_{J,\bmath v}-E_{J',{\bmath v}'})^3}{3\hbar^4 c^3}
(2J'+1)
\nonumber \\ && \times
\sum_{m'} \left|\langle {\bmath v}'|\mu_{m'}^{(g)}|{\bmath v}\rangle\right|^2
\nonumber \\ && \times
\sum_m
\left(\begin{array}{ccc} J' & 1 & J \\ -M' & -m & M \end{array} \right)^2.
\label{eq:A-dist}
\earr
(The terms mixing different values of $m'$ are eliminated by orthogonality relations.)  The summation over $m$ of course has at most one term, with $m=\Delta M \equiv M'-M$.

We are interested in the net angular momentum loss, which is most easily obtained by taking an initial state with $M=J$.  There are then six possible values of $\Delta J$ and $\Delta M$, constrained by 
selection rules ($\Delta J,\Delta M=-1,0,+1$) and the restriction $\Delta M\le\Delta J$.  The branching ratios are constrained by (i) the energy difference factors in Eq.~(\ref{eq:A-dist}), (ii) the 
factor of $2J'+1$, and (iii) the $3j$ symbol.  We consider each.

The energy factors do not depend on $\Delta M$.  If we take natural frequency $\nu=(E^0_{\bmath v}-E^0_{{\bmath v}'})/h$, then the energies differences are given by
\beq
E_{J,\bmath v}-E_{J',{\bmath v}'} = h\nu - \frac{\hbar^2}{2I}(2J+1+\Delta J)\Delta J.
\eeq
The classical grain rotation rate is $\omega = \hbar J/I$.  In the limit of $J\gg 1$ and $\omega\ll\nu$, the energy difference is proportional to $1-\omega\Delta J/2\pi\nu$, so the cube of the energy 
difference is proportional to $1-3\omega\Delta J/2\pi\nu$.

The square of the $3j$ symbol, multiplied by $2J'+1$, can be directly evaluated for the six cases of interest.  It is:
\barr
\frac{2J-1}{2J+1} & & \Delta J = -1,\, \Delta M = -1,
\nonumber \\
\frac1{J+1} & & \Delta J = 0,\, \Delta M = -1,
\nonumber \\
\frac J{J+1} & & \Delta J = 0,\, \Delta M =0,
\nonumber \\
\frac1{(J+1)(2J+1)} & & \Delta J = +1,\, \Delta M = -1,
\nonumber \\
\frac1{J+1} & & \Delta J = +1,\, \Delta M = 0, {\rm ~~and}
\nonumber \\
1 & & \Delta J = +1,\, \Delta M = +1.
\label{eq:table}
\earr
By multiplying these relative probabilities by $1-3\omega\Delta J/2\pi\nu$, it is easily seen that the average $\langle \Delta M\rangle$ is exactly zero if $\omega=0$.  Therefore the leading contribution 
to $\langle \Delta M\rangle$ can be obtained by taking the large-$J$ limit of the $3j$ symbols.  Transitions with $\Delta J\neq\Delta M$ are suppressed by powers of $J$ in Eq.~(\ref{eq:table}), so one 
has three available transitions: $\Delta M=-1,0,+1$, $\Delta J=\Delta M$.  Since the factors in Eq.~(\ref{eq:table}) go to unity, the branching ratio for these three transitions is determined entirely by 
the energy factor:
\beq
P(\Delta M) = \frac13 - \frac{\omega}{2\pi\nu}\Delta M.
\eeq
This implies an average loss of $z$-component of angular momentum
\beq
\langle\Delta M\rangle = -\frac{\omega}{\pi\nu}.
\eeq
In particular, we may find the ratio of angular momentum loss to energy loss ($h\nu$), which is
\beq
\frac{\dot L_z}{\dot E} = \frac{\hbar\omega/\pi\nu}{h\nu} = \frac{\omega}{2\pi^2\nu^2}.
\eeq
With the normalization of Eq.~(\ref{eq:y-power}) and this ratio, one recovers Eq.~(\ref{eq:y-dlz}).

\label{lastpage}

\end{document}